\documentclass{article}

\usepackage{styling/arxiv}

\usepackage[utf8]{inputenc}
\usepackage{csquotes}
\usepackage{array}

\usepackage{amsthm}
\usepackage{newtxmath}

\usepackage{amsmath,amssymb,amsfonts}

\usepackage{xcolor}
\usepackage{graphicx}
\graphicspath{{./FIGURES/}}
\usepackage[caption=false,font=footnotesize]{subfig}

\usepackage{placeins}
\usepackage{float}
\usepackage{array,makecell,tabularx,ragged2e}
\newcolumntype{Y}{>{\RaggedRight\arraybackslash}X}
\usepackage{booktabs,multirow,arydshln}
\usepackage{caption}
\usepackage{subcaption}

\usepackage[inkscapelatex=false]{svg}
\svgsetup{
  inkscapeexe=inkscape,
  inkscapeopt=-o,
  inkscapeversion=1
}

\usepackage{algorithm}
\usepackage{algpseudocode}

\usepackage{url}
\usepackage{microtype}
\usepackage[unicode]{hyperref}
\usepackage{doi}

\usepackage[siunitx]{circuitikz}
\usetikzlibrary{intersections,decorations.pathreplacing,calc}

\usepackage{mdframed}

\usepackage{pifont}

\usepackage{cuted}
\newtheorem{definition}{Definition}[section]

\usepackage{hhline}
\usepackage{bm}

\usepackage{todonotes}

\renewcommand{\d}[1]{\operatorname{d}\!{#1}}

\newcommand{\vect}[1]{\bm{\mathrm{#1}}} 

\usepackage{amsmath}
\usepackage{dsfont}

\newmdtheoremenv{boxthm}{Theorem}
\newmdtheoremenv{boxlem}{Lemma}
\newmdtheoremenv{boxcor}{Corollary}
\newmdtheoremenv{boxdef}{Definition}                     

\newcommand{\cev}[1]{\reflectbox{\ensuremath{\vec{\reflectbox{\ensuremath{#1}}}}}}

\DeclareMathOperator{\Var}{Var}

\usetikzlibrary{positioning}
\usetikzlibrary{positioning,calc,shapes.geometric} 

\usepackage{tikz,colortbl}
\usepackage{xifthen}
\usetikzlibrary{intersections}
\usetikzlibrary{calc}
\usetikzlibrary{arrows.meta}
\usepackage{zref-savepos}
\usepackage{tkz-euclide}

\pgfdeclarelayer{bg}    
\pgfsetlayers{bg,main}  
\definecolor{beige}{RGB}{245, 245, 220}
\definecolor{darkgrey}{RGB}{65, 65, 65}
\definecolor{grey}{RGB}{240, 240, 240}
\definecolor{lightgrey}{RGB}{250, 250, 250}

\usetikzlibrary{calc, arrows, fit, positioning, patterns, decorations.pathreplacing, shapes}
\tikzstyle{dash} = [dashed]
\tikzstyle{line} = [draw]
\tikzstyle{box} = [draw, minimum size=.8cm]
\tikzstyle{detbox} = [draw, minimum size=.5cm] 
\tikzstyle{bigbox} = [draw, minimum size=13mm]
\tikzstyle{box_horizontal} = [draw, minimum width=3.5cm, minimum height=.8cm]
\tikzstyle{box_h} = [draw, minimum width=4.8cm, minimum height=.8cm]
\tikzstyle{smallbox} = [draw, minimum size=5mm]
\tikzstyle{roundbox} = [draw, circle, inner sep=0pt, minimum size=.8cm]
\tikzstyle{clamped} = [draw, fill=black, minimum size=0.35cm]
\tikzstyle{roundclamped} = [draw, circle, inner sep=0pt, fill=darkgrey, text=white, minimum size=0.35cm]
\tikzstyle{optim} = [draw, circle, inner sep=0pt, fill=white, minimum size=0.35cm]
\tikzstyle{msgcircle} = [shape=circle, draw, inner sep=0pt, minimum size=5mm, fill=white]
\tikzstyle{darkmsgcircle} = [shape=circle, draw, inner sep=0pt, minimum size=5mm, fill=darkgrey, text=white, font=\bfseries]
\tikzstyle{msgdoublecircle} = [shape=circle, double, double distance=1.5pt, draw, inner sep=0pt, minimum size=5mm, fill=white]
\tikzstyle{darkmsgdoublecircle} = [shape=circle, double, double distance=1.5pt, draw, inner sep=0pt, minimum size=5mm, fill=darkgrey, text=white, font=\bfseries]

\tikzstyle{q} = [draw, shape=circle, minimum size=3mm, fill=white, anchor=center, inner sep=0pt] 
\tikzstyle{p} = [draw, minimum size=2.8mm, fill=white, anchor=center, inner sep=0pt] 
\tikzstyle{v} = [circle split, rotate=90, anchor=center, minimum size=0mm, inner sep=0pt] 
\tikzstyle{h} = [circle split, anchor=center, minimum size=0mm, inner sep=0pt] 
\tikzstyle{s} = [draw, shape=semicircle, minimum size=1.5mm, fill=white, anchor=center, inner sep=0pt] 
\tikzstyle{f} = [draw, shape=circle, minimum size=3mm, fill=darkgrey, anchor=center, inner sep=0pt, text=white, font=\bfseries] 

\tikzstyle{BHDA} = [draw, minimum width=3.5cm, minimum height=2cm]
\tikzstyle{BHDA2} = [draw, minimum width=3.5cm, minimum height=2.5cm]
\tikzset{
  break mark/.pic={
    \draw[thick] (0,0) -- ++(0.8,0.2);
    \draw[thick] (0,0.2) -- ++(0.8,0.2);
  }
}

\tikzstyle{dot} = [circle, fill=black, inner sep=1.5pt]

\tikzstyle{rectangle_h} = [draw, minimum width=2.6cm, minimum height=.8cm]
\tikzstyle{rectangle_v} = [draw, minimum width=.8cm, minimum height=1.6cm]
\tikzset{
  declare function={
    atan3(\a,\b)=ifthenelse(atan2(0,1)==90, atan2(\a,\b), atan2(\b,\a));
  },
  jump radius/.initial=+.125cm,
  @jump/.initial=+, jumps/.is choice,
  jumps/left/.style={@jump=-},
  jumps/right/.style={@jump=+},
  jump/.style args={(#1)--(#2)}{
    to path={
      let \p{@jp@}=($(\tikztotarget)-(\tikztostart)$),
          \n{@jp@}={atan3(\p{@jp@})+180} in
      -- ($(intersection of \tikztostart--{\tikztotarget} and #1--#2)!%
             \pgfkeysvalueof{/tikz/jump radius}!(\tikztostart)$)
      arc [ radius     =\pgfkeysvalueof{/tikz/jump radius},
            start angle=\n{@jp@},
            delta angle=\pgfkeysvalueof{/tikz/@jump}180 ]
      -- (\tikztotarget)
    }
  }
}

\newcommand{\msg}[6]{
      \ifthenelse{\isin{#1}{left} \AND \isin{#2}{down}}{
            \coordinate (anchor) at ($({#3})!{#5}!({#4})$);
            \node[msgcircle, xshift=-5.5mm] at (anchor) {#6};
            \node[xshift=-1.5mm] at (anchor) {$\downarrow$};
      }{}
      \ifthenelse{\isin{#1}{right} \AND \isin{#2}{down}}{
            \coordinate (anchor) at ($({#3})!{#5}!({#4})$);
            \node[msgcircle, xshift=5.5mm] at (anchor) {#6};
            \node[xshift=1.5mm] at (anchor) {$\downarrow$};
      }{}

      \ifthenelse{\isin{#1}{down} \AND \isin{#2}{right}}{
            \coordinate (anchor) at ($({#3})!{#5}!({#4})$);
            \node[msgcircle, yshift=-6.0mm] at (anchor) {#6};
            \node[yshift=-2.0mm] at (anchor) {$\rightarrow$};
      }{}
      \ifthenelse{\isin{#1}{up} \AND \isin{#2}{right}}{
            \coordinate (anchor) at ($({#3})!{#5}!({#4})$);
            \node[msgcircle, yshift=6.0mm] at (anchor) {#6};
            \node[yshift=2.0mm] at (anchor) {$\rightarrow$};
      }{}

      \ifthenelse{\isin{#1}{down} \AND \isin{#2}{left}}{
            \coordinate (anchor) at ($({#3})!{#5}!({#4})$);
            \node[msgcircle, yshift=-6.0mm] at (anchor) {#6};
            \node[yshift=-2.0mm] at (anchor) {$\leftarrow$};
      }{}
      \ifthenelse{\isin{#1}{up} \AND \isin{#2}{left}}{
            \coordinate (anchor) at ($({#3})!{#5}!({#4})$);
            \node[msgcircle, yshift=6.0mm] at (anchor) {#6};
            \node[yshift=2.0mm] at (anchor) {$\leftarrow$};
      }{}

      \ifthenelse{\isin{#1}{left} \AND \isin{#2}{up}}{
            \coordinate (anchor) at ($({#3})!{#5}!({#4})$);
            \node[msgcircle, xshift=-5.5mm] at (anchor) {#6};
            \node[xshift=-1.5mm] at (anchor) {$\uparrow$};
      }{}
      \ifthenelse{\isin{#1}{right} \AND \isin{#2}{up}}{
            \coordinate (anchor) at ($({#3})!{#5}!({#4})$);
            \node[msgcircle, xshift=5.5mm] at (anchor) {#6};
            \node[xshift=1.5mm] at (anchor) {$\uparrow$};
      }{}
}

\newcommand{\darkmsg}[6]{
      \ifthenelse{\isin{#1}{left} \AND \isin{#2}{down}}{
            \coordinate (anchor) at ($({#3})!{#5}!({#4})$);
            \node[darkmsgcircle, xshift=-5.5mm] at (anchor) {#6};
            \node[xshift=-1.5mm] at (anchor) {$\downarrow$};
      }{}
      \ifthenelse{\isin{#1}{right} \AND \isin{#2}{down}}{
            \coordinate (anchor) at ($({#3})!{#5}!({#4})$);
            \node[darkmsgcircle, xshift=5.5mm] at (anchor) {#6};
            \node[xshift=1.5mm] at (anchor) {$\downarrow$};
      }{}

      \ifthenelse{\isin{#1}{down} \AND \isin{#2}{right}}{
            \coordinate (anchor) at ($({#3})!{#5}!({#4})$);
            \node[darkmsgcircle, yshift=-6.0mm] at (anchor) {#6};
            \node[yshift=-2.0mm] at (anchor) {$\rightarrow$};
      }{}
      \ifthenelse{\isin{#1}{up} \AND \isin{#2}{right}}{
            \coordinate (anchor) at ($({#3})!{#5}!({#4})$);
            \node[darkmsgcircle, yshift=6.0mm] at (anchor) {#6};
            \node[yshift=2.0mm] at (anchor) {$\rightarrow$};
      }{}

      \ifthenelse{\isin{#1}{down} \AND \isin{#2}{left}}{
            \coordinate (anchor) at ($({#3})!{#5}!({#4})$);
            \node[darkmsgcircle, yshift=-6.0mm] at (anchor) {#6};
            \node[yshift=-2.0mm] at (anchor) {$\leftarrow$};
      }{}
      \ifthenelse{\isin{#1}{up} \AND \isin{#2}{left}}{
            \coordinate (anchor) at ($({#3})!{#5}!({#4})$);
            \node[darkmsgcircle, yshift=6.0mm] at (anchor) {#6};
            \node[yshift=2.0mm] at (anchor) {$\leftarrow$};
      }{}

      \ifthenelse{\isin{#1}{left} \AND \isin{#2}{up}}{
            \coordinate (anchor) at ($({#3})!{#5}!({#4})$);
            \node[darkmsgcircle, xshift=-5.5mm] at (anchor) {#6};
            \node[xshift=-1.5mm] at (anchor) {$\uparrow$};
      }{}
      \ifthenelse{\isin{#1}{right} \AND \isin{#2}{up}}{
            \coordinate (anchor) at ($({#3})!{#5}!({#4})$);
            \node[darkmsgcircle, xshift=5.5mm] at (anchor) {#6};
            \node[xshift=1.5mm] at (anchor) {$\uparrow$};
      }{}
}

\newcommand{\bwmsg}[6]{
      \ifthenelse{\isin{#1}{left} \AND \isin{#2}{down}}{
            \coordinate (anchor) at ($({#3})!{#5}!({#4})$);
            \node[msgdoublecircle, xshift=-5.5mm] at (anchor) {#6};
            \node[xshift=-1.5mm] at (anchor) {$\downarrow$};
      }{}
      \ifthenelse{\isin{#1}{right} \AND \isin{#2}{down}}{
            \coordinate (anchor) at ($({#3})!{#5}!({#4})$);
            \node[msgdoublecircle, xshift=5.5mm] at (anchor) {#6};
            \node[xshift=1.5mm] at (anchor) {$\downarrow$};
      }{}

      \ifthenelse{\isin{#1}{down} \AND \isin{#2}{right}}{
            \coordinate (anchor) at ($({#3})!{#5}!({#4})$);
            \node[msgdoublecircle, yshift=-6.0mm] at (anchor) {#6};
            \node[yshift=-2.0mm] at (anchor) {$\rightarrow$};
      }{}
      \ifthenelse{\isin{#1}{up} \AND \isin{#2}{right}}{
            \coordinate (anchor) at ($({#3})!{#5}!({#4})$);
            \node[msgdoublecircle, yshift=6.0mm] at (anchor) {#6};
            \node[yshift=2.0mm] at (anchor) {$\rightarrow$};
      }{}

      \ifthenelse{\isin{#1}{down} \AND \isin{#2}{left}}{
            \coordinate (anchor) at ($({#3})!{#5}!({#4})$);
            \node[msgdoublecircle, yshift=-6.0mm] at (anchor) {#6};
            \node[yshift=-2.0mm] at (anchor) {$\leftarrow$};
      }{}
      \ifthenelse{\isin{#1}{up} \AND \isin{#2}{left}}{
            \coordinate (anchor) at ($({#3})!{#5}!({#4})$);
            \node[msgdoublecircle, yshift=6.0mm] at (anchor) {#6};
            \node[yshift=2.0mm] at (anchor) {$\leftarrow$};
      }{}

      \ifthenelse{\isin{#1}{left} \AND \isin{#2}{up}}{
            \coordinate (anchor) at ($({#3})!{#5}!({#4})$);
            \node[msgdoublecircle, xshift=-5.5mm] at (anchor) {#6};
            \node[xshift=-1.5mm] at (anchor) {$\uparrow$};
      }{}
      \ifthenelse{\isin{#1}{right} \AND \isin{#2}{up}}{
            \coordinate (anchor) at ($({#3})!{#5}!({#4})$);
            \node[msgdoublecircle, xshift=5.5mm] at (anchor) {#6};
            \node[xshift=1.5mm] at (anchor) {$\uparrow$};
      }{}
}

\newcommand{\bwdarkmsg}[6]{
      \ifthenelse{\isin{#1}{left} \AND \isin{#2}{down}}{
            \coordinate (anchor) at ($({#3})!{#5}!({#4})$);
            \node[darkmsgdoublecircle, xshift=-5.5mm] at (anchor) {#6};
            \node[xshift=-1.5mm] at (anchor) {$\downarrow$};
      }{}
      \ifthenelse{\isin{#1}{right} \AND \isin{#2}{down}}{
            \coordinate (anchor) at ($({#3})!{#5}!({#4})$);
            \node[darkmsgdoublecircle, xshift=5.5mm] at (anchor) {#6};
            \node[xshift=1.5mm] at (anchor) {$\downarrow$};
      }{}

      \ifthenelse{\isin{#1}{down} \AND \isin{#2}{right}}{
            \coordinate (anchor) at ($({#3})!{#5}!({#4})$);
            \node[darkmsgdoublecircle, yshift=-6.0mm] at (anchor) {#6};
            \node[yshift=-2.0mm] at (anchor) {$\rightarrow$};
      }{}
      \ifthenelse{\isin{#1}{up} \AND \isin{#2}{right}}{
            \coordinate (anchor) at ($({#3})!{#5}!({#4})$);
            \node[darkmsgdoublecircle, yshift=6.0mm] at (anchor) {#6};
            \node[yshift=2.0mm] at (anchor) {$\rightarrow$};
      }{}

      \ifthenelse{\isin{#1}{down} \AND \isin{#2}{left}}{
            \coordinate (anchor) at ($({#3})!{#5}!({#4})$);
            \node[darkmsgdoublecircle, yshift=-6.0mm] at (anchor) {#6};
            \node[yshift=-2.0mm] at (anchor) {$\leftarrow$};
      }{}
      \ifthenelse{\isin{#1}{up} \AND \isin{#2}{left}}{
            \coordinate (anchor) at ($({#3})!{#5}!({#4})$);
            \node[darkmsgdoublecircle, yshift=6.0mm] at (anchor) {#6};
            \node[yshift=2.0mm] at (anchor) {$\leftarrow$};
      }{}

      \ifthenelse{\isin{#1}{left} \AND \isin{#2}{up}}{
            \coordinate (anchor) at ($({#3})!{#5}!({#4})$);
            \node[darkmsgdoublecircle, xshift=-5.5mm] at (anchor) {#6};
            \node[xshift=-1.5mm] at (anchor) {$\uparrow$};
      }{}
      \ifthenelse{\isin{#1}{right} \AND \isin{#2}{up}}{
            \coordinate (anchor) at ($({#3})!{#5}!({#4})$);
            \node[darkmsgdoublecircle, xshift=5.5mm] at (anchor) {#6};
            \node[xshift=1.5mm] at (anchor) {$\uparrow$};
      }{}
}

\tikzset{mainstyle/.style={fill=white, draw=black, shape=rectangle, align=center}}

\tikzset{dstyle/.style={mainstyle, minimum size=5mm, inner sep=0pt, text width=4mm}}

\tikzset{astyle/.style={fill=black, draw=black, shape=diamond, minimum size=3mm, inner sep=0pt, text width=0mm}}

\tikzset{sstyle/.style={mainstyle, minimum size=7mm, inner sep=0pt, text width=5mm}}

\tikzset{ostyle/.style={fill=black, draw=black, shape=rectangle, minimum size=0.2cm, inner sep=0pt, text width=2mm}}

\tikzset{sophstyle/.style={fill=white,draw=black, shape=rectangle, minimum size=0.2cm, inner sep=0pt, text width=2mm}}

\tikzset{actionstyle/.style={fill=black, draw=black, shape=diamond, minimum size=0.3cm, inner sep=0pt, text width=2mm}}

\tikzset{estyle/.style={fill=white, draw=black, shape=rectangle, minimum height=0.7cm, minimum width=2.5cm, inner sep=3pt}}

\tikzset{cstyle/.style={fill=white, draw=black, shape=rectangle, minimum height=0.7cm, minimum width=2cm, inner sep=3pt}}

\tikzset{bstyle/.style={fill=white, draw=black, shape=circle, minimum size=0.7cm, inner sep=0pt, text width=4.5mm}}

\tikzstyle{observation}=[ostyle]
\tikzstyle{deterministic}=[dstyle]
\tikzstyle{stochastic}=[sstyle]
\tikzstyle{action}=[astyle]
\tikzstyle{environment}=[estyle]
\tikzstyle{agent}=[bstyle]
\tikzstyle{control}=[cstyle]
\tikzstyle{action}=[actionstyle]
\tikzstyle{soph}=[sophstyle]

\tikzstyle{filter}=[mainstyle, minimum width=1cm, minimum height=0.5cm]
\tikzstyle{selector}=[fill=white, draw=black, shape=trapezium, rotate=180, minimum width=1cm, minimum height=0.5cm]

\arrayrulecolor{black}

\tikzset{
  boxbigger/.style={
    draw,
    rectangle,
    minimum width=13mm,
    minimum height=13mm,
    inner sep=3pt,
    thick
  }
}
\tikzset{
  happyface/.style={
    draw,
    circle,
    minimum size=1.6cm, 
    inner sep=0pt,
    line width=0.5pt,
    fill=white,
    path picture={
      \pgftransformshift{\pgfpointanchor{path picture bounding box}{center}}
      \fill (-0.24cm, 0.24cm) circle (0.08cm);
      \fill ( 0.24cm, 0.24cm) circle (0.08cm);
      \draw[line width=0.4pt]
        (-0.30cm,-0.16cm) arc[start angle=200, end angle=340, radius=0.30cm];
    }
  }
}

\title{A Probabilistic Generative Model for Spectral Speech Enhancement}

\author{
  Marco Hidalgo-Araya\textsuperscript{1} \\
  \texttt{m.d.hidalgo.araya@tue.nl} \\
  \And
  Rapha{\"e}l Tr{\'e}sor\textsuperscript{1} \\
  \And
  Bart van Erp\textsuperscript{2} \\
  \And
  Wouter W.L. Nuijten\textsuperscript{1,2} \\
  \And
  Thijs van de Laar\textsuperscript{1} \\
  \And
  Bert de Vries\textsuperscript{1,2,3} \\
}

\date{}

\begin{document}
\maketitle

\begin{center}
\textsuperscript{1}Department of Electrical Engineering, Eindhoven University of Technology, the Netherlands \\
\textsuperscript{2}Lazy Dynamics B.V., Utrecht, the Netherlands \\
\textsuperscript{3}GN Advanced Science, Eindhoven, the Netherlands
\end{center}

\begin{abstract}
Speech enhancement in hearing aids remains a difficult task in nonstationary acoustic environments, mainly because current signal processing algorithms rely on fixed, manually tuned parameters that cannot adapt in situ to different users or listening contexts. This paper introduces a unified modular framework that formulates signal processing, learning, and personalization as Bayesian inference with explicit uncertainty tracking. The proposed framework replaces ad hoc algorithm design with a single probabilistic generative model that continuously adapts to changing acoustic conditions and user preferences. It extends spectral subtraction with principled mechanisms for in-situ personalization and adaptation to acoustic context. The system is implemented as an interconnected probabilistic state-space model, and inference is performed via variational message passing in the \texttt{RxInfer.jl} probabilistic programming environment, enabling real-time Bayesian processing under hearing-aid constraints. Proof-of-concept experiments on the \emph{VoiceBank+DEMAND} corpus show competitive speech quality and noise reduction with 85 effective parameters. The framework provides an interpretable, data-efficient foundation for uncertainty-aware, adaptive hearing-aid processing and points toward devices that learn continuously through probabilistic inference.
\end{abstract}

\keywords{Bayesian Inference \and Belief Propagation \and Forney-style Factor Graphs \and Hearing Aids \and Probabilistic Graphical Models \and Speech Enhancement \and Variational Inference \and Variational Message Passing}

\section{INTRODUCTION}
\label{sec:introduction}

Hearing loss affects an estimated 430~million people worldwide, a number projected to exceed 700~million by 2050; nearly one in ten individuals will require rehabilitation for disabling hearing loss \cite[Exec.~Summary, pp.~4–8]{worldhealthorganizationWorldReportHearing2021}. The global economic impact of unaddressed hearing loss has been estimated at US\$~980~billion annually \cite{mcdaidEstimatingGlobalCosts2021}, alongside substantial social costs such as increased risk of isolation, cognitive decline, and reduced quality of life \cite{kamilEffectsHearingImpairment2015,nachtegaalHearingAbilityWorking2012,linHearingLossCognition2011}. Despite the scale of the problem, Hearing Aid (HA) adoption remains low: only about 17\% of those who could benefit obtain them \cite{orjiGlobalRegionalNeeds2020a}, and even among owners, approximately 20\% report never using their devices, while another 30\% use them only occasionally \cite[Fig.~3]{dillonAdoptionUseNonuse2020b}.

One reason for this limited adoption is inadequate speech enhancement quality: HAs may improve audibility in quiet settings but often fall short in noisy, nonstationary environments where users need support most \cite{marcottiAssociationUnaidedSpeech2025,marcos-alonsoFactorsImpactingUse2023,fuentes-lopezAssociationHometohealthcareCenter2024}. This gap is especially apparent in everyday acoustic scenes such as restaurants, public transit, and busy public spaces where noise sources are diverse, nonstationary, and highly unpredictable. Conventional one-size-fits-all processing delivers limited benefit in such conditions, causing many users to underutilize their devices.

Fine-tuning of a HA's sound processing algorithm for one listener in a known acoustic environment is feasible, yet developing a solution that performs satisfactorily for many listeners across diverse acoustic scenes \emph{without} on-the-fly personal adjustments remains extremely challenging. Unfortunately, manual, real-time parameter tuning imposes a cognitive and practical burden that most users are unwilling to accept, particularly when such interactions do not yield immediate, perceivable improvements. In-situ personalization is therefore essential, but it must impose minimal burden on the HA user.

Variants of spectral subtraction currently remain the most popular noise reduction algorithm in commercial HAs, due to their simplicity and low computational cost (i.e. \cite{kamathMultibandSpectralSubtraction2002,martinNoisePowerSpectral2001}). Still, heuristic parameter tuning makes them vulnerable to audible artifacts in real-world conditions \cite{beroutiEnhancementSpeechCorrupted1979}. More recent noise reduction algorithms based on deep learning, such as Conv-TasNet \cite{luoConvTasNetSurpassingIdeal2019}, Demucs \cite{defossezRealTimeSpeech2020}, and diffusion-based models \cite{scheiblerUniversalScorebasedSpeech2024,luConditionalDiffusionProbabilistic2022}, outperform spectral subtraction on standard benchmark datasets. However, their large model sizes and computational demands strain the strict latency and energy budgets of HAs. Furthermore, the ICASSP~2023 Clarity Challenge highlighted performance degradation when deep learning systems trained on simulated data were evaluated on real acoustic recordings \cite{coxOverview2023Icassp2023}. Lightweight architectures such as TinyLSTM \cite{fedorovTinyLSTMsEfficientNeural2020}, FSPEN \cite{yangFspenUltraLightweightNetwork2024}, and Mamba-SEUNet \cite{kimMambabasedHybridModel2025} demonstrate that streaming operations are possible under hardware constraints, yet these solutions lack mechanisms for in-situ personalization.

In this paper, we present a novel framework for hearing aid algorithm design that explicitly tackles the challenges described above. Our long-term objective is to establish a framework for developing HA sound processing algorithms that can be trained both \emph{offline}, like deep neural networks, and \emph{online} through in-situ appraisals provided by HA users. In essence, we seek to design HA algorithms with a small computational footprint that can nonetheless adapt on-the-spot to unforeseen performance degradations in real-world listening environments.

In Section~\ref{sec:problem}, we identify the root cause of the problem, namely the absence of a principled approach for selecting and adapting tuning parameters in HA algorithms. We address this \emph{fitting} problem directly by requiring that parameter settings in an HA algorithm depend not on the skills or insights of an audiologist, but rather quantitatively on (input and/or output) audio signals and user appraisals, i.e., signals that can be observed directly in nature.

Our approach is based on a Bayesian modeling framework to enable continual (lifelong) learning, and a principled treatment of uncertainty throughout the entire chain from user feedback to HA output. In this framework, all processing, including offline and online learning as well as signal processing itself, is formulated as Bayesian inference in a probabilistic generative model.

To keep the study focused, we select speech enhancement as our first application task, noting that other HA processing functions can be addressed within the same framework. In Section~\ref{sec:architecture}, we derive a modular architecture for a generative speech enhancement model. Section~\ref{sec:model-specification} details the technical specifications of the generative model. To ensure a small computational footprint, we draw upon well-established mechanisms from the spectral subtraction literature.

Once the model is defined, all processing can be expressed as Bayesian inference tasks. In Section~\ref{sec:inference}, we discuss the state of the art in efficient automated Bayesian inference for dynamic models. We review message passing-based inference in factor graphs and describe how this framework can be used to automate both signal processing and parameter learning, in both offline and online settings, within our generative model.

Section~\ref{sec:experiments} presents validation experiments. Our goal is to establish feasibility: can the HA output signal be inferred automatically through message passing in a factor graph representation of a generative speech enhancement model? Where do the current weaknesses lie, and how can they be addressed in future work? We demonstrate that our model achieves performance in the same ballpark as comparable algorithms while requiring significantly fewer tuning parameters. This compact parameterization is essential for future in-situ personalization, and the results motivate continued development along this line of research.

A further review of related work is deferred to Section~\ref{sec:related-work}, where we place our approach in the context of existing methods.
Section~\ref{sec:discussion} discusses the broader implications for future HA design and outlines our forward-looking research agenda.

\section{PROBLEM STATEMENT}\label{sec:problem}

Consider a sound processing algorithm
\begin{equation}\label{eq:sound-processing-algo}
y_t = f_w(x^t),
\end{equation}
where $t$ is a discrete time index, and $f_w(\cdot)$ is a function, parameterized by weight vector $w$, that maps (the history of) an input signal $x^t \triangleq x_{1:t}$ to an output signal $y_t$. This algorithm can be executed in a HA if both $x^t$ and $w$ are given. The signal $x^t$ is measured by a microphone, but $w$ is not directly observable and must be chosen by a human specialist, such as an algorithm designer or an audiologist. In the HA industry, this problem is referred to as the \emph{fitting} problem.

Choosing appropriate values for $w$ is a challenging problem that often leads to suboptimal outcomes. First, HA users are exposed to diverse acoustic environments, and optimal parameter values for $w$ may vary across these conditions. Second, each user has a unique hearing loss profile, along with individual preferences and expectations regarding sound perception. Consequently, suitable parameter values should be both \emph{personalized} and \emph{adaptive to changing acoustic contexts}.

A large part of the audiological industry is devoted to the problem of fitting HA parameters for end users. While substantial progress has been made in personalizing fixed (non-adaptive) HA parameters, \emph{personalized adaptation} remains beyond the scope of traditional HA fitting. This is because problems arising under in-situ conditions must be addressed in situ, that is, in real-life listening environments where the audiologist is not present. 

To meet the requirements of both in-situ personalization and adaptation, we depart from treating \eqref{eq:sound-processing-algo} as a proper algorithmic design objective. Instead, our objective is to construct a probabilistic model of the form
\begin{equation}\label{eq:py|xr}
p(y_t \,|\, x^t, r^t),
\end{equation}
where $r^t \triangleq r_{1:t}$ denotes the history of all in-situ user appraisals. By user appraisals, we refer to signals provided by HA users, such as gestures or button presses, that indicate their evaluation of the current sound processing under in-situ conditions, for example, expressing “like” or “dislike”.

Note that, in contrast to \eqref{eq:sound-processing-algo}, the evaluation of \eqref{eq:py|xr} assumes access to signals $x^t$ and $r^t$ that both can be directly observed and recorded in nature. In other words, implementing a HA algorithm according to \eqref{eq:py|xr} circumvents the challenging problem of parameter fitting.

Note also that \eqref{eq:py|xr} is expressed as a probability distribution that should be read as a distribution over outputs $y_t$, for a given input signal $x^t$ and a given set of user appraisals $r^t$. Clearly, the true preferences of HA users are unobserved as we only have access to their appraisals $r^t$. To adequately capture how uncertainties about user preferences propagate from appraisals to the resulting HA output signal $y_t$, we consider a Bayesian modeling framework as principled approach to reasoning with uncertainty.

The need for \emph{personalization} of HAs is accommodated in model \eqref{eq:py|xr} through its dependence on in-situ recorded user appraisals $r^t$. \emph{Adaptation} to varying acoustic environments is achieved by inferring distinct acoustic conditions from $x^t$ and using the recognized conditions to modulate the output signal $y_t$. These mechanisms will be discussed in more detail in Sections~\ref{sec:architecture} and \ref{sec:model-specification}. 

The challenge addressed in the next section is to construct an architecture for model~\eqref{eq:py|xr} that decomposes it into a set of manageable modules.

\section{SYSTEM ARCHITECTURE}\label{sec:architecture}

\subsection{MODULARIZATION}\label{sec:modularization}

We begin by noting that there exists a rich body of literature on effective sound processing algorithms of the form \eqref{eq:sound-processing-algo}. Since this paper focuses on speech enhancement, the extensive research on spectral subtraction algorithms and their variants provides a valuable foundation that we aim to build upon.

In a Bayesian framework, detailed knowledge about an algorithm of the form \eqref{eq:sound-processing-algo} can be incorporated into a custom probabilistic model of the form
$p(y_t \,|\,x^t, w_t)$. Accordingly, we decompose \eqref{eq:py|xr} into two modules as follows:

\begin{subequations}
\begin{align}
p(y_t \,|\, x^t, r^t) &= \int p(y_t, w_t \,|\, x^t, r^t)\mathrm{d}w_t, \\
&= \int \underbrace{p(y_t \,|\, x^t, w_t)}_{\text{from }\eqref{eq:sound-processing-algo}} p(w_t \,|\, x^t, r^t)\mathrm{d}w_t. \label{eq:AIDA-decomp-1}
\end{align}    
\end{subequations}

\begin{figure*}[t!]\label{fig:eq:AIDA-architecture}
\begin{subequations}\label{eq:AIDA-architecture}
\begin{align}
 p(&y_t \,|\, x^t, r^t) = \int p(y_t \,|\, x^t, w_t) p(w_t \,|\, x^t, r^t) \mathrm{d}w_t, \label{eq:a-AIDA-architecture} \\
 &=  \int p(y_t \,|\, x^t, w_t) \bigg( \int p(w_t \,|\, x^t, \theta_t) p(\theta_t\,|\,x^t,r^t) \mathrm{d}\theta_t \bigg) \mathrm{d}w_t, \label{eq:b-AIDA-architecture} \\
 &=  \int p(y_t \,|\, x^t, w_t) \bigg( \int p(w_t \,|\, x^t, \theta_t) \Big(\int p(\theta_t\,|\,c_t,r^t) p(c_t|x^t) \mathrm{d}c_t \Big) \mathrm{d}\theta_t \bigg) \mathrm{d}w_t, \label{eq:c-AIDA-architecture} \\
&\propto  \int \underbrace{p(y_t \,|\, x^t, w_t)}_{\text{WFB}} \bigg( \int \underbrace{p(w_t \,|\, x^t, \theta_t)}_{\text{SEM}} \Big(\int \underbrace{p(\theta_t\,|\,c_t,r^t)}_{\text{EUM}}\underbrace{p(x^t\,|\,c_t) p(c_t)}_{\text{ACM}}\mathrm{d}c_t \Big) \mathrm{d}\theta_t \bigg) \mathrm{d}w_t.\label{eq:d-AIDA-architecture}
\end{align} 
\end{subequations}
\caption{Proposed decomposition of \eqref{eq:py|xr}; see Section~\ref{sec:modularization} for a detailed discussion.}
\end{figure*}

The second factor, $p(w_t\, |\, x^t, r^t)$, is a model that should predict personalized and context-adaptive filter coefficients $w_t$, given input signal $x^t$ and user appraisals $r^t$. 

Since $x^t$ and $r^t$ are observable in nature, we may interpret $p(w_t \,|\, x^t, r^t)$ as a \emph{data-driven fitting model}. To realize this objective, we decompose \eqref{eq:AIDA-decomp-1} further into submodules as shown in \eqref{eq:AIDA-architecture}.

We now discuss the decomposition \eqref{eq:AIDA-architecture} in more detail. Our objective is to construct the left-hand side of \eqref{eq:a-AIDA-architecture}, that is, a model $p(y_t \,|\,x^t, r^t)$ which predicts a HA output signal $y_t$ only from signals that are actually observed in nature, namely the HA input signal $x^t$ and user appraisals $r^t$. 

Equation~\eqref{eq:a-AIDA-architecture} mirrors the decomposition in \eqref{eq:AIDA-decomp-1}, separating the model into a front-end filter $p(y_t \,|\, x^t, w_t)$ and a personalized, adaptive filter-coefficient predictor $p(w_t \,|\, x^t, r^t)$. We adopt a warped-frequency filter bank (WFB) as the front-end module. Details of the WFB  are discussed in Section~\ref{sec:WFB}.

In \eqref{eq:b-AIDA-architecture}, the filter coefficient model $p(w_t \,|\, x^t, r^t)$ is further decomposed into a parameterized model $p(w_t \,|\, x^t, \theta_t)$ and a back-end model $p(\theta_t \,|\, x^t, r^t)$. We will draw on the spectral subtraction literature to develop a custom Bayesian Speech Enhancement Model (SEM), represented by $p(w_t \,|\, x^t, \theta^t)$. Personalization and adaptation of the SEM are handled by the module $p(\theta_t\,|\, x^t, r^t)$, referred to as the end-user model (EUM). 

Finally, we will not let the EUM depend directly on the input signal $x^t$, but instead on an acoustic context signal $c^t$ inferred from $x^t$. While the sound signal $x^t$ varies on a millisecond timescale, the acoustic context signal $c_t$ captures slower changes in the acoustic environment, such as transitions between scenes like “in the car” or “at home”. Equation~\eqref{eq:c-AIDA-architecture} introduces the context signal $c^t$, and \eqref{eq:d-AIDA-architecture} reformulates $p(c^t \,|\, x^t)$ in a generative model format using Bayes rule.

In summary, the targeted sound processing model $ p(y_t \,|\, x^t, r^t)$ is decomposed into (1) a front-end filter bank (WFB), (2) a Bayesian version of a spectral speech enhancement algorithm (SEM), (3) an end-user model (EUM) that supports \emph{personalization} (through $r^t$) and acoustic \emph{adaptation} (through $c^t$) of the SEM, and (4) an acoustic context (ACM) that infers relevant acoustic scenes from $x^t$.

\subsection{STATE-SPACE MODELING}

To complete the architectural design of the proposed model, we recognize that storing the complete histories $x^t$ or $r^t$ within the algorithms is undesirable, as the associated memory requirements would grow without bound.

Focusing on the front-end, we will instead store the relevant past of $x^t$ in a \emph{state vector} $z_t$ and update the representation $z_t$ with every observation $x_t$. Rather than using $p(y_t|x^t,w_t)$, we will consider a state-space model of the form

\begin{align}\label{eq:front-end-SSM}
     p(y_t,z_t\,|\,z_{t-1},x_t,w_t) = \underbrace{p(y_t\,|\,z_t,w_t)}_{\text{output}} \underbrace{p(z_t\,|\,z_{t-1},x_t)}_{\text{state transition}}.
\end{align}

Model~\eqref{eq:front-end-SSM} no longer depends on an ever-growing memory. The first factor, $p(y_t\,|\, z_t, w_t)$, generates the output signal $y_t$ from the current state $z_t$ and filter weights $w_t$, while the second factor updates the state, based on a new observation $x_t$. Similar conversions to state-space models are applied to the SEM, EUM, and ACM. 

Altogether, the full probabilistic generative model for situated personalized and context-adaptive sound processing is composed of the following modules:

\begin{subequations}\label{eq:AIDA-full-architecture}
 \begin{align}
    &p(y_t\,|\,z_t,w_t) p(z_t\,|\,z_{t-1},x_t) \quad &&\text{(WFB)} \label{eq:WFB-model}\\
   &p(w_t\,|\,\xi_t,\theta_t) p(\xi_t\,|\,\xi_{t-1},z_t) \quad &&\text{(SEM)} \label{eq:SEM-model}\\ 
   &p(\xi_t\,|\,c_t)p(c_t\,|\,c_{t-1}) \quad &&\text{(ACM)} \label{eq:ACM-model}\\
    &p(\theta_t\,|\,h_t,c_t) p(h_t\,|\,h_{t-1},r_t) \quad &&\text{(EUM)} \label{eq:EUM-model}
\end{align}   
\end{subequations}

In \eqref{eq:AIDA-full-architecture}, the WFB  employs $z_t$ as latent state variables to represent a history trace of $x_t$; the SEM  uses latent states $\xi_t$ to capture the relevant history of $z_t$; the EUM  maintains latent states $h_t$ to track the history of $r_t$, and the ACM  maintains a slowly varying scene state $c_t$. 

The complete generative model for a single time step $t$ is given by the product of the modules in \eqref{eq:AIDA-full-architecture}, while the model for a temporal segment is obtained by multiplying the models of the individual time steps. 

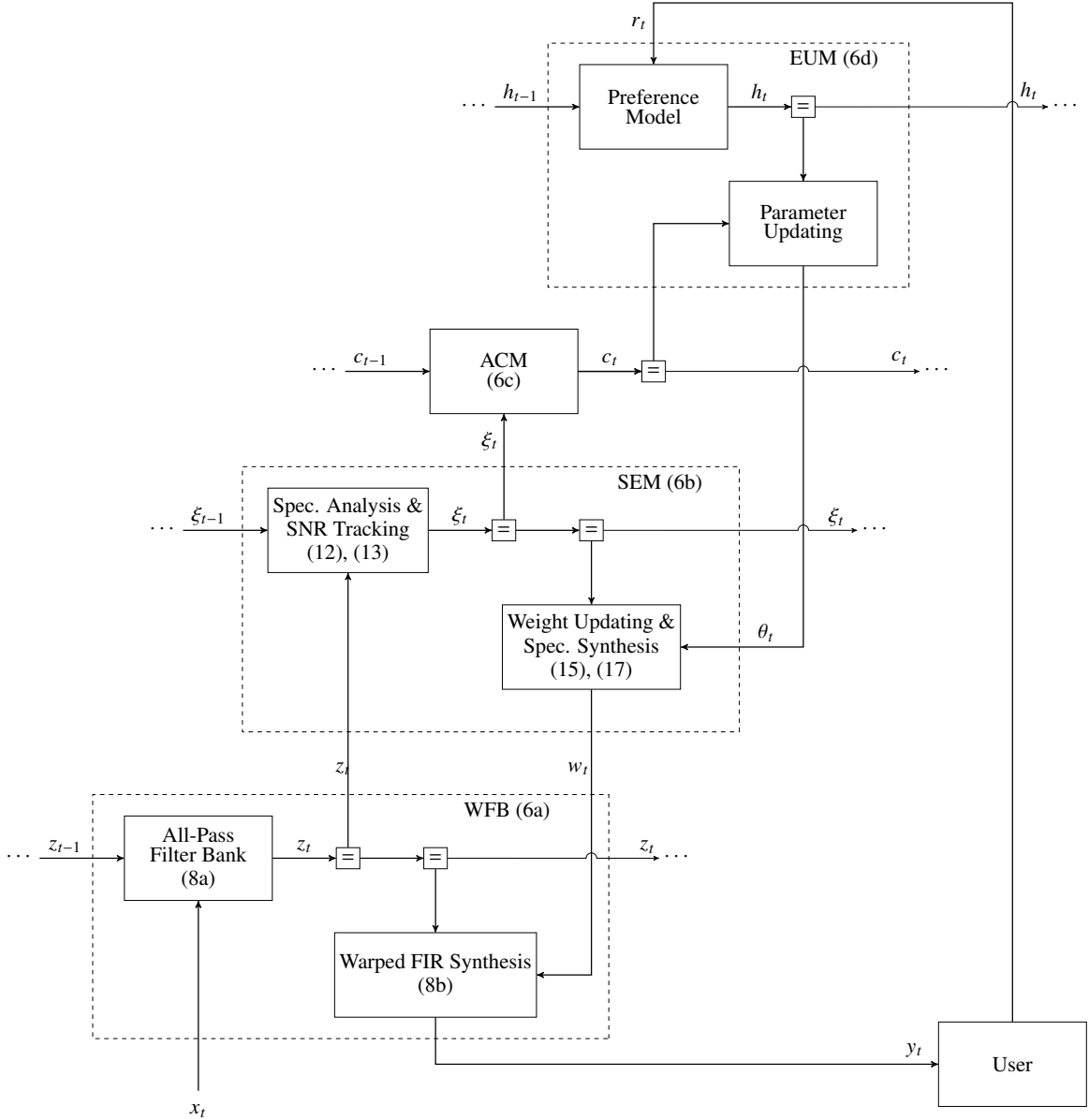
\begin{figure*}
  \centering
  \scalebox{0.6}{%
    \begin{tikzpicture}
[node distance=10mm,auto,>=stealth',
every node/.append style={font=\Large}]

\node[](input-x){};
\node[BHDA, above=45mm of input-x](AllPass-Bank){\shortstack{All-Pass \\ Filter Bank \\ \\\eqref{eq:all-pass-FB}}};
\node[left=20mm of AllPass-Bank](pre-dots-AllPass-Bank-states){$\cdots$};
\node[smallbox, right=15mm of AllPass-Bank](equality_AllPass-Bank_transition){$=$};
\node[smallbox, right=15mm of equality_AllPass-Bank_transition](equality_AllPass-Bank_transition2){$=$};

\node[BHDA, below=15mm of equality_AllPass-Bank_transition2](AllPass-Bank-FIR){\shortstack{Warped FIR Synthesis
 \\\eqref{eq:FIR-filter}}};

\node[right=50mm of equality_AllPass-Bank_transition2](post-dots-AllPass-Bank-states){$\cdots$};

\node[BHDA, below right =1mm and 95mm of AllPass-Bank-FIR] (user) {User};

\draw[->, thick](input-x)--node[above,yshift=-30mm] {$x_{t}$} (AllPass-Bank);
\draw[->, thick](pre-dots-AllPass-Bank-states)--node[xshift=-4mm]{$z_{t-1}$}(AllPass-Bank);
\draw[->, thick](AllPass-Bank)--node[]{$z_t$}(equality_AllPass-Bank_transition);
\draw[->, thick](equality_AllPass-Bank_transition)--node[]{}(equality_AllPass-Bank_transition2);
\draw[->, thick](equality_AllPass-Bank_transition2)--node[]{}(AllPass-Bank-FIR);
\draw[->, thick](AllPass-Bank-FIR.south)|-node[]{}(user.west);

\draw[dashed]
  ([xshift=41mm, yshift=-5mm] AllPass-Bank-FIR.south) rectangle 
  ([xshift=-25mm, yshift=5mm] AllPass-Bank.north);
  \node[align=center, xshift=45mm, yshift=25mm] at ($( AllPass-Bank-FIR)!0.5!(AllPass-Bank)$) {WFB \eqref{eq:WFB-model}};

\node[BHDA, above=65mm of equality_AllPass-Bank_transition](Analysis){\shortstack{Spec. Analysis \& \\ SNR Tracking  \\\eqref{eq:spectral-analysis-1}, \eqref{eq:SEM-full-architecture}}};
\node[left=20mm of Analysis](pre-dots-analysis-states){$\cdots$};
\node[smallbox, right=15mm of Analysis](equality_Analysis_transition){$=$};
\node[smallbox, right=15mm of equality_Analysis_transition](equality_Analysis_transition2){$=$};
\node[right=60mm of equality_Analysis_transition2](post-dots-Analysis){$\cdots$};

\node[BHDA,below=15mm of equality_Analysis_transition2](Wiener-Gain){\shortstack{Weight Updating \&\\ Spec. Synthesis\\\eqref{eq:weight-update-model}, \eqref{eq:weight-synthesis}}};

\node[below=40mm of Wiener-Gain](invisible_box_analysis){}; 

\node[above right=-10mm and 17mm of Wiener-Gain](theta){$\theta_t$};
\node[below right=16mm and -28mm of Wiener-Gain](weight){$w_t$};
\node[below right=16mm and -83mm of Wiener-Gain](z_updateSEM){$z_t$};

\node[below right=34.5mm and -11mm of Wiener-Gain](weight){$z_t$};

\draw[->, thick](equality_AllPass-Bank_transition)--node[]{}(Analysis);
\draw[->, thick](pre-dots-analysis-states)--node[xshift=-4mm]{$\xi_{t-1}$}(Analysis);
\draw[->, thick](Analysis)--node[]{$\xi_{t}$}(equality_Analysis_transition);
\draw[->, thick](equality_Analysis_transition)--(equality_Analysis_transition2);
\draw[->, thick](equality_Analysis_transition2)--(Wiener-Gain);
\draw[->, thick](Wiener-Gain)|-(AllPass-Bank-FIR.east);

\draw[->]
  (equality_AllPass-Bank_transition2)
  to[jump=(Wiener-Gain)--(invisible_box_analysis), jumps=left] (post-dots-AllPass-Bank-states);

\draw[dashed]
  ([xshift=35mm, yshift=-10mm] Wiener-Gain.south) rectangle 
  ([xshift=-25mm, yshift=5mm] Analysis.north);
  \node[align=center, xshift=45mm, yshift=25mm] at ($(Wiener-Gain)!0.5!(Analysis)$) {SEM \eqref{eq:SEM-model}};


\node[BHDA, above=25mm of equality_Analysis_transition](Context-transition){\shortstack{ACM \\\eqref{eq:ACM-model}}};
\node[left=20mm of Context-transition](pre-dots-context-states){$\cdots$};
\node[smallbox, right=15mm of Context-transition](equality_context_transition){$=$};
\node[right=60mm of equality_context_transition](post-dots-context){$\cdots$};

\node[below right=3mm and -24mm of Context-transition](snr-updateACM){$\xi_t$};
\node[below right=21mm and 58mm of Context-transition](snr-updateACM2){$\xi_t$};

\draw[->, thick](Context-transition)--node[]{}(equality_context_transition);
\draw[->, thick](pre-dots-context-states)--node[xshift=-4mm]{$c_{t-1}$}(Context-transition);
\draw[->, thick](Context-transition)--node[]{$c_{t}$}(equality_context_transition);

\draw[->,thick](equality_Analysis_transition)--node[]{}(Context-transition);

\node[BHDA, above=50mm of equality_context_transition](Preference-transition){\shortstack{Preference \\ Model}};
\node[left=20mm of Preference-transition](pre-dots-preference-states){$\cdots$};
\node[smallbox, right=15mm of Preference-transition](equality_preference_transition){$=$};
\node[right=55mm of equality_preference_transition](post-dots-preference){$\cdots$};
\node[BHDA,below=15mm of equality_preference_transition](Parameter){\shortstack{Parameter \\Updating}};

\node[below=50mm of Parameter](invisible_box_context){};

\draw[->, thick](pre-dots-preference-states)--node[xshift=-4mm]{$h_{t-1}$}(Preference-transition);
\draw[->, thick](Preference-transition)--node[]{$h_{t}$}(equality_preference_transition);
\draw[->, thick](equality_preference_transition)--(Parameter);
\draw[->, thick](equality_context_transition)|-(Parameter);

\draw[->, thick](Parameter)|-(Wiener-Gain);

\draw[->]
  (equality_Analysis_transition2)
  to[jump=(Parameter)--(invisible_box_context), jumps=left] (post-dots-Analysis);

  \draw[->]
  (equality_context_transition)
  to[jump=(Parameter)--(invisible_box_context), jumps=left] (post-dots-context);

\draw[dashed]
  ([xshift=25mm, yshift=-5mm] Parameter.south) rectangle 
  ([xshift=-25mm, yshift=5mm] Preference-transition.north);
  \node[align=center, xshift=25mm, yshift=25mm] at ($(Parameter)!0.5!(Preference-transition)$) {EUM \eqref{eq:EUM-model}};

  \node[above=240 mm of user](inv_userup){};

  \draw[->, thick]
  (user.north) |- (inv_userup.center) -| (Preference-transition.north);
  \node[above right=7mm and -24mm of Preference-transition](rating){$r_t$};

  \draw[->]
  (equality_preference_transition)
  to[jump=(inv_userup)--(user), jumps=left] (post-dots-preference);

  \node[above right=0.5mm and 50mm of equality_preference_transition.center](rating){$h_t$};

  \node[above right=0.5mm and 55mm of equality_context_transition.center](rating){$c_t$};
  \node[above left=0.5mm and 20mm of user.center](rating){$y_t$};
\end{tikzpicture}
  }
 \caption{%
Forney-style factor graph of the AIDA-2 framework defined in~\eqref{eq:AIDA-full-architecture}. 
The figure illustrates the conditional-dependency structure among the four modules: 
the Warped-frequency Filter Bank (WFB), the Bayesian Speech Enhancement Model (SEM), 
the Acoustic Context Model (ACM), and the End-User Model (EUM). 
Arrows indicate conditional dependencies and information flow within the generative framework (see Section~\ref{sec:modularization}).}
\label{fig:AIDA-2-architecture}
\end{figure*}

For reference, we refer to the full model in \eqref{eq:AIDA-full-architecture} as the AIDA-2 model. AIDA-2 represents an updated version of the original AIDA sound processing model introduced in \cite{podusenkoAIDAActiveInferenceBased2022a}. The architecture of the AIDA-2 model is illustrated in Figure~\ref{fig:AIDA-2-architecture}.

\subsection{SCOPE AND IMPLEMENTATION FOCUS}

In this work, we focus on the technical specifications of the sound-processing modules, WFB and SEM, and demonstrate how Bayesian inference enables efficient computation of the output signal $y_t$. We further present validation experiments that illustrate the feasibility and promise of this framework. The complementary components, namely parameter learning mechanisms and the full integration of the EUM and ACM, will be addressed in forthcoming work focused on personalization and contextual adaptation.

\section{MODEL SPECIFICATION}
\label{sec:model-specification}

In this section, we specify the technical details of the submodels in \eqref{eq:AIDA-full-architecture}. Throughout this section, Figure~\ref{fig:AIDA-2-architecture} serves as a reference for contextual orientation. The system architecture is described bottom-up. We begin with the front-end module, continue with the SEM, and conclude with the ACM and EUM. 

\subsection{THE WARPED-FREQUENCY FILTER BANK}\label{sec:WFB}

Following \eqref{eq:AIDA-full-architecture}, the front-end of the model comprises the state-space model
\begin{align}\label{eq:WFB-model-2}
    p(y_t|z_t,w_t) p(z_t|z_{t-1},x_t)\,.
\end{align}

We employ a warped-frequency filter bank, a common sound processing front-end in the HA industry \cite{katesDynamicrangeCompressionUsing2003}. 

In a WFB, the output signal $y_t$ is generated by  
\begin{subequations}\label{eq:WFB}
    \begin{align}
    z_{tj} &= \begin{cases} 
        x_t & \text{for }j=1 \\
        A(q^{-1}) z_{t,j-1} & \text{for }j=2,3,\ldots,J 
    \end{cases} \label{eq:all-pass-FB}\\
    y_t &= \sum_{j=1}^J w_{mj} z_{tj} \,, \label{eq:FIR-filter}
\end{align}
\end{subequations}

where $m$ indexes blocks of $M$ samples, $J$ is the number of taps in the delay line, and $q^{-1}$ represents the unit delay operator.

The state vector $z_t$ in \eqref{eq:WFB} is generated by passing $x_t$ through a tapped delay line of first-order all-pass filters

\begin{equation}\label{eq:AP-coeff}
    A(q^{-1}) = \frac{q^{-1}-\alpha}{1-\alpha q^{-1}}\,,
\end{equation}

where $\alpha$ is the warping parameter that controls the frequency resolution characteristics. If $\alpha = 0$, the WFB reduces to a Finite Impulse Response (FIR) filter. However, by setting the all-pass coefficient to a value $0 < \alpha < 1$, the frequency resolution of the WFB can be warped to more closely align with human auditory perception \cite{laineWarpedLinearPrediction1994, smithBarkERBBilinear1999}.

The system \eqref{eq:WFB}–\eqref{eq:AP-coeff} constitutes the WFB, illustrated in Figure~\ref{fig:front-end-architecture}. In simple terms, the proposed sound processing system is a warped-frequency FIR filter with time-varying filter weights $w_m$, which are updated every $M$ samples (i.e., once per block). All remaining modules, SEM, ACM, and EUM, serve to convey personalization and adaptation information into the computation of $w_m$.

For clarity, it is straightforward to reformulate deterministic equations such as \eqref{eq:WFB} into probabilistic models, as in \eqref{eq:WFB-model-2}, by introducing a Dirac delta function. For example, the deterministic relation in \eqref{eq:FIR-filter} can be expressed as
\begin{equation}
p(y_t | z_t, w_t) = \delta\bigg(y_t - \sum_{j=1}^J w_{mj} z_{tj}\bigg)\,.
\end{equation}
Analogous reformulations apply to all deterministic equations. 
More details about the WFB module are provided in Appendix~\ref{app:wfb-details}.

\subsection{THE SPEECH ENHANCEMENT MODEL}

Our goal here is to specify the details of \eqref{eq:SEM-model}, a probabilistic generative model in which processing of observations $z_t$ by Bayesian inference gives rise to a spectral speech enhancement algorithm.

\subsubsection{Block Processing} Before we proceed with the technical details of \eqref{eq:SEM-model}, some remarks on sampling rates and block processing. Processing in the SEM serves a single purpose: to infer weights $w_m$ every $M$ samples (i.e., once per processing block), where $m$ denotes the block index. To update $w_m$, SEM draws on two sources of information. Once every $M$ samples, SEM receives the signal $z_m$, representing a history trace of $x_t$, from the WFB module. At the same block rate, SEM reads the value of $\theta_m$, representing personalized adaptation information from the EUM. Hence, the SEM updates itself only once per block of $M$ samples and therefore runs at a significantly slower rate than the front-end. 

In fact, each module runs independently at its own sampling rate. The front-end updates itself at the audio sampling rate (in our case: $f_s = 16$ [kHz]); the SEM runs at block rate $f_s/M$; the EUM runs at $1$ [Hz] since it only needs to process user interactions; and the ACM runs at $1$ [Hz]. Keeping track of the distinct sampling rates for each module would complicate the notation more than it would aid clarity. Therefore, in the following, we distinguish only between the audio sample index $t$ and the block index $m$. 

\subsubsection{A Model Refinement}

Since spectral speech enhancement operates in the spectral domain, we introduce spectral analysis and synthesis modules. Together with block index notation, this extension refines the respective analysis and synthesis models from \eqref{eq:SEM-model} into

\begin{subequations}\label{eq:SEM-model-1}
\begin{align}
    p(\xi_m \,|\, \xi_{m-1}, z_m) &= \int \overbrace{p(\xi_m \,|\, \xi_{m-1}, \tilde{z}_m)}^{\text{\shortstack{SNR\\ Tracking}}} \overbrace{p(\tilde{z}_m | z_m)}^{\smash{\text{\shortstack{Spectral\\ Analysis}}}} \d{\tilde{z}_m},\\
    p(w_m \,|\, \xi_m, \theta_m) &= \int \underbrace{p(w_m \,|\, \tilde{w}_m)}_{\text{\shortstack{Spectral\\ Synthesis}}} \underbrace{p(\tilde{w}_m \,|\, \xi_m, \theta_m)}_{\text{\shortstack{Weight \\Updating}}}  \d{\tilde{w}_m}.
\end{align}
\end{subequations}
The variables $\tilde{w}_m$ and $\tilde{z}_m$ hold spectral representations of the filter coefficients $w_m$ and the WFB state $z_m$, respectively. 

Next, we discuss each of these submodules in \eqref{eq:SEM-model-1}, beginning with spectral analysis, proceeding through SNR tracking and weight updating, and concluding with spectral synthesis.

\subsubsection{Spectral Analysis}

Once per block, SEM receives a $J$-dimensional vector $z_m$ representing a history trace of the acoustic input signal $x_t$. The spectral analysis stage transforms $z_m$ into a spectral vector $\tilde{z}_m$ by

\begin{subequations}\label{eq:spectral-analysis-1}
\begin{align}
Z_{m} &= \mathrm{DFT}[\text{hann} \odot z_{m}] \label{eq:spectral-analysis-DFT-1} \\
\tilde{z}_{mj} &= \log( |Z_{mj}|^2) \quad \text{for } j = 1,\ldots,J \label{eq:log-spectral-analysis-Power-1}
\end{align}   
\end{subequations}

In \eqref{eq:spectral-analysis-DFT-1}, $\text{hann}$ denotes a Hanning window of length $J$, DFT refers to the discrete Fourier transformation, and $\odot$ indicates element-wise multiplication. 

The signal $\tilde{z}_m$ holds a $J$-dimensional vector containing spectral logarithmic power coefficients of the signal block $z_m$. 

Each power coefficient $\tilde{z}_{mj}$ in frequency band $j \in \{1,2,\ldots,J\}$ is processed independently, ultimately resulting in a corresponding spectral weight update $\tilde{w}_{mj}$. For notational simplicity, the index $j$ will be omitted in the following.

\subsubsection{SNR Tracking}\label{sec:snr-tracking}

Signal-to-noise ratio (SNR) tracking is a key component of many spectral enhancement algorithm \cite{martinNoisePowerSpectral2001, ephraimSpeechEnhancementUsing1984, ephraimSpeechEnhancementUsing1985, wienerExtrapolationInterpolationSmoothing1964, limEnhancementBandwidthCompression1979,cohenNoiseEstimationMinima2002,cohenSpeechEnhancementNonstationary2001}. Our aim is to estimate the latent log-SNR $\xi_m = s_m - n_m$ by jointly tracking the speech and noise log-powers, $s_m$ and $n_m$, within a probabilistic state-space model. In the spectral enhancement literature, this power tracking is typically achieved using a leaky integrator filter, whose forgetting factor is dynamically adjusted to achieve the desired tracking behavior. Our goal is to reproduce this behavior by formulating a probabilistic model in which Bayesian inference over observations $\tilde{z}_m$ yields equivalent leaky-integrator dynamics.

Our formulation is inspired by the switching Kalman filter framework in \cite{murphySwitchingKalmanFilters1998}, adapted to voice-aware SNR tracking. In particular, tracking the SNR in the log-power domain can be described as Bayesian filtering in the following model:

\begin{subequations}\label{eq:SEM-full-architecture}
 \begin{align}
    p(\xi_m \,|\,s_m,n_m) &= \delta(\xi_m -  (s_m - n_m))\label{eq:SNR-model}\\
   p(s_m \,|\, s_{m-1}) &= \mathcal{N}(s_m \,|\,s_{m-1},\sigma^2_{s})\label{eq:Speech-model}\\ 
   p(n_m \,|\, n_{m-1}) &= \mathcal{N}(n_m \,|\,n_{m-1},\sigma^2_{n})\label{eq:Noise-model}\\ 
   p(\pi_m\,|\,\xi_m) &= \mathrm{Bernoulli}\big(\pi_m \,|\, \sigma(\xi_m- \kappa_m)) \label{eq:Voice-model}\\
   p(\tilde{z}_m \,|\, s_m, n_m, \pi_m) &=
   \begin{cases}
       \mathcal{N}(\tilde z_m \,|\, s_m,\, 1.0) & \text{if }\pi_m=1\\
       \mathcal{N}(\tilde z_m \,|\, n_m,\, 1.0) & \text{if }\pi_m=0\,.
   \end{cases}\label{eq:Observation-model}
\end{align}   
\end{subequations}

The model \eqref{eq:SEM-full-architecture} specifies two linear Gaussian trackers with different adaptation speeds, given by the evolution models in \eqref{eq:Speech-model} and \eqref{eq:Noise-model}.
A binary voice-activity variable $\pi_m$ governed by \eqref{eq:Voice-model} determines which tracker is updated. When $\pi_m = 1$, the observation $\tilde z_m$ contributes to the speech tracker, and when $\pi_m = 0$, it contributes to the noise tracker, as specified in \eqref{eq:Observation-model}.

Each tracker behaves as a Bayesian leaky integrator, as detailed in Appendix~\ref{app:bayesian-leaky-integrator}. The speech tracker follows rapid fluctuations in the observed power, and the noise tracker captures slower background variations. This behavior agrees with established principles in decision-directed SNR estimation and minimum-statistics noise tracking \cite{ephraimSpeechEnhancementUsing1984,martinNoisePowerSpectral2001}. The resulting structure provides a probabilistic formulation of classical SNR-tracking strategies, with the switching variable $\pi_m$ coordinating the updates of the two trackers in a voice-aware manner.

\subsubsection{Weight Updating}

Spectral enhancement gains are commonly based on the foundations of Wiener filtering \cite{martinNoisePowerSpectral2001, cohenNoiseEstimationMinima2002, cohenNoiseSpectrumEstimation2003, upadhyaySingleChannelSpeech2016, scalartSpeechEnhancementBased1996}. In Appendix~\ref{app:wiener}, we show that the Wiener gain can be written as a logistic function of the log-spectral SNR,
\begin{align}
    \tilde{w}_m = \sigma(\xi_m) \triangleq \frac{1}{1+e^{-\xi_m}}\,,
\end{align}
where $\xi_m$ is the SNR ratio in log-domain and $\sigma(\cdot)$ is known as the logistic function. A large positive SNR drives $\tilde{w}_m \rightarrow 1$, leaving the frequency bin effectively unprocessed. In contrast, a large negative SNR (indicating more noise than signal) drives $\tilde{w}_m \rightarrow 0$, thereby suppressing the signal.

Following this rationale, we specify the spectral weight update model as
\begin{align}\label{eq:weight-update-model}
    p(\tilde{w}_m\,|\,\xi_m,\theta_m) &= \mathrm{Bernoulli}\big(\tilde{w}_m \,|\, \sigma(\xi_m- \theta_m) \big) \,. 
\end{align}

In this model, $\tilde{w}_m$ represents a binary spectral mask variable, modeled as a Bernoulli random variable. Equation~\eqref{eq:weight-update-model} can thus also be read as
\begin{equation}
p(\tilde{w}_m = 1 \,|\, \xi_m, \theta_m) = \sigma(\xi_m - \theta_m)\,.
\end{equation}
The external control parameter $\theta_m$ introduces a “context-aware personalization shift,” enabling the gain computation to be adapted to individual user preferences. The SEM reads $\theta_m$ from the EUM, which learns $\theta_m$ from user appraisals and acoustic context.

\subsubsection{Spectral Synthesis}
The final stage of the SEM is to convert spectral filter weights $\tilde{w}_m$ back to (time-domain) filter coefficients $w_m$. We reverse the spectral analysis process here by
\begin{align}\label{eq:weight-synthesis}
    w_m= \mathrm{IDFT}\big[ \text{hann} \odot  \mathbb{E}[\tilde{w}_m]\big] \,,
\end{align}
where IDFT refers to the inverse discrete Fourier transform. 

\subsubsection{Summary of SEM Specification}

The SEM, technically specified by \eqref{eq:SEM-model-1}--\eqref{eq:weight-synthesis}, integrates and provides a compact generative description of how filter coefficients $w_m$ are inferred from observations $z_m$, conditioned on the control parameter $\theta_m$ supplied by the EUM. When the SEM processes its inputs $z_m$ and $\theta_m$ through Bayesian filtering, it generates probabilistic predictions for the filter coefficients $w_m$, which, when applied within the WFB, realize speech enhancement in accordance with established principles from the spectral enhancement literature.

\subsection{THE ACOUSTIC CONTEXT MODEL}

The ACM captures contextual information from the acoustic input signal $x_t$ and represents the acoustic scene over a multiple-second time scale. Its generative structure, given by
\begin{align}
p(\xi_m\,|\,c_m)p(c_m\,|\,c_{m-1}) 
\end{align}
where $\xi_m$ denotes a latent SNR variable and $c_m$ captures the prevailing acoustic regime.

The latent variable $c_m$ describes the acoustic environment and evolves gradually over time. By distinguishing coarse SNR conditions (e.g., low, medium, or high), it conveys contextual information that guides the EUM's adaptive selection of the control parameter $\theta_m$.

The slow evolution of $c_m$ ensures that environmental context estimates change smoothly over time, maintaining smooth transitions over the inferred acoustic scenes. This behavior provides a reliable context representation for the EUM, enabling context-aware adaptation. Detailed technical specification and integration with other modules are beyond the scope of the present work and will be addressed in future work.

\subsection{THE END USER MODEL}
The task of the EUM is to translate the set of user appraisals $r^m$ into acoustic context-dependent proposals for $\theta_m$, a control variable within the SEM that regulates the spectral gains through \eqref{eq:weight-update-model}.

The generative model for the EUM is of the form (copied from \eqref{eq:EUM-model}, with block index notation)
\begin{align}\label{eq:EUM-1}
\underbrace{p(\theta_m\,|\,h_m,c_m)}_{\substack{\text{personalized,} \\ \text{context-aware,} \\ \text{control}}} \underbrace{p(h_m\,|\,h_{m-1},r_m)}_{\substack{\text{``happiness''} \\ \text{state tracking} }}\,.
\end{align}

The essential feature of \eqref{eq:EUM-1} is that predictions for $\theta_m$ depend both on acoustic context $c_m$ for \emph{adaptation} to environmental scenes and on \emph{personalized} user feedback $r_m$. 

To apply this model in an in-situ setting with HA users, it is necessary to model user interactions in the submodel $p(h_m\,|\,h_{m-1},r_m)$. Ideally, users should be able to indicate their appraisal of the current sound processing performance at any time. However, such interactions will only occur if (1) users experience that their feedback leads rapidly to perceptible improvements, and (2) the interaction procedure is extremely lightweight. In other words, the benefits of providing in-situ appraisals must clearly outweigh the associated effort.

Designing such interaction protocols is a challenging problem that requires careful research into the cognitive load of interaction protocols and their computational modeling aspects. For example, simple pairwise comparison protocols are unsuitable for in-situ scenarios, as they demand excessive short-term memory (i.e., cognitive load) and yield limited information per interaction.

For these reasons, the specification and implementation of EUM for in-situ elicitation of user preferences are deferred to a companion paper that is currently in preparation.

In this paper, we proceed with describing how observations $x^t$ are processed to produce a Bayesian speech enhancement algorithm.

\section{FITTING AND SIGNAL PROCESSING AS INFERENCE}
\label{sec:inference}

\subsection{OVERVIEW}
A key feature of defining a probabilistic generative model \eqref{eq:AIDA-full-architecture} for sound processing is that any relevant task can be expressed as a Bayesian inference process on this model, and inference itself is an automatable procedure. In practice, this means that tasks such as \emph{signal processing} (producing $y_t$) and \emph{parameter learning}, both offline from a given dataset $D = \{x_i^T, y_i^T\}_{i=1}^N$ and online from user appraisals $r^t$, can be automated by a Bayesian inference toolbox.

Technically, Bayesian inference in a probabilistic model entails a full commitment to using the sum and product rules of probability theory to compute
\begin{equation}
p(\text{what-I-want-know} \,|\, \text{what-has-been-observed})\,.
\end{equation}
In principle, this computation is always possible, yet often intractable. Consider a joint probability distribution $p(a, b, c)$ over three sets of variables, where $a$ denotes the variables of interest, $b$ represents the observed variables, and $c$ comprises latent (unobserved) variables that are not of direct interest. The desired posterior distribution $p(a|b)$ can then be obtained from the joint distribution through
\begin{equation}\label{eq:Bayesian-inference}
p(a|b) = \frac{p(a,b)}{p(b)} = \frac{\int p(a,b,c)\, \mathrm{d}c}{\iint p(a,b,c)\, \mathrm{d}a \mathrm{d}c}\,.
\end{equation}

In this framework, signal processing can be expressed as computing $p(y_t \,|\, x^t, r^t)$ by \eqref{eq:Bayesian-inference} in the AIDA-2 model. Assuming a set of tuning parameters $\phi$ within the AIDA-2 model, offline HA fitting corresponds to inferring $p(\phi \,|\, D)$, while online personalization involves further updating these parameters to $p(\phi \,|\, D, r^t)$.

The implications of the above could represent a pivotal shift for the HA industry for two main reasons. First, signal processing itself becomes an automatable inference process. Conventional HA algorithms are typically complex, with extensive codebases that are difficult to maintain, modify, and verify. In contrast, a probabilistic generative model, such as the one specified in Section~\ref{sec:model-specification}, provides a concise and transparent description of the underlying assumptions, typically expressible in only a few pages of code. This clarity not only simplifies model maintenance but also enhances interpretability and reproducibility.

More importantly, within the proposed framework, both HA fitting and online personalization are reformulated as automatable inference processes. Given that HA fitting remains a largely suboptimal procedure and that true online HA personalization is nearly absent in current practice, automating these processes within a logically consistent computational framework could have a transformative impact on the field. 

All of these promises depend on the ability to automate Bayesian inference. In Sections~\ref{sec:var-inference} and \ref{sec:MP-in-FFG}, we review the current state of Bayesian inference automation for real-time dynamic models in general, and in Section~\ref{sec:MP-for-SEM}, we discuss specific adaptations of this framework for inference in the SEM module.

\subsection{VARIATIONAL INFERENCE}\label{sec:var-inference}
Following Bayes rule given by \eqref{eq:Bayesian-inference}, we now formulate the inference problem for the SEM module. The aim is to infer the filter weights $w_m$ by processing the coefficients $z_m$ from the WFB together with the control parameter $\theta_m$ from the EUM, such that denoising arises naturally as a direct outcome of Bayesian inference.

Let $\nu_m = (\tilde{w}_m, \tilde{z}_m,\pi_m, s_m, n_m, \xi_m)$ denote the set of latent state variables in the SEM. The inference task for the SEM can then be described as computing 
\begin{equation}\label{eq:posterior-w}
    p(\underbrace{w_m}_{\substack{ \text{inferred} \\ \text{gains}}},\underbrace{\nu_m\,|\,\nu_{m-1}}_{\substack{\text{state} \\ \text{update}}},\underbrace{z_m,\theta_m}_{\substack{ \text{new} \\ \text{observations} }} )\,,
\end{equation}
by executing the (Bayesian inference) procedure described in \eqref{eq:Bayesian-inference}.

Unfortunately, computing the posterior in \eqref{eq:posterior-w} in closed form is infeasible for two main reasons.
First, marginalizing over the discrete voice-activity variable $\pi_m$ produces a Gaussian mixture whose number of components grows exponentially with time.
Second, the logistic nonlinearities in both the VAD prior \eqref{eq:Voice-model} and the gain mapping \eqref{eq:weight-update-model} break conjugacy with the Gaussian priors over the latent SNR $\xi_m$. As no conjugate prior exists for the resulting likelihood, Bayes rule cannot be evaluated in closed form, and an analytic solution for \eqref{eq:posterior-w} is not achievable. Given these intractabilities, we adopt a variational inference (VI) approach \cite{jordanIntroductionVariationalMethods1998}, which approximates the Bayesian posterior in \eqref{eq:posterior-w} by a tractable family of distributions $\mathcal{Q}$ by minimizing a variational free-energy (VFE) functional

\begin{align}\label{eq:VFE-SEM}
F[q] &= \mathbb{E}_{q} \bigg[\log  \frac{\overbrace{q(w_m,\nu_m \,|\,\nu_{m-1},z_m,\theta_m )}^{\text{variational posterior}}}{\underbrace{p(z_m,\theta_m,w_m,\nu_m \,|\,\nu_{m-1} )}_{\text{SEM generative model}}}\bigg] 
\end{align}

This VFE functional equals the Kullback–Leibler divergence between the approximate (variational) posterior $q(w_m,\nu_m \,|\,\nu_{m-1},z_m,\theta_m )$ and the exact Bayesian posterior $p(w_m,\nu_m \,|\,\nu_{m-1},z_m,\theta_m )$, up to an additive constant.

The variational inference task thus becomes an optimization problem, yielding the optimal variational posterior
\begin{equation}
q^*(w_m,\nu_m \,|\, \nu_{m-1}, z_m, \theta_m ) = \arg\min_{q\in\mathcal Q} F[q] \,.
\end{equation}

Having defined the variational objective in \eqref{eq:VFE-SEM}, we next describe how inference is carried out. Owing to the factorized structure of the SEM generative model, inference can be implemented as localized message passing on a Forney-style factor graph.

Before describing how inference proceeds in our model, we briefly review Forney-style factor graphs, which provide the foundation for automated message passing in the SEM module.

\subsection{AUTOMATED MESSAGE PASSING-BASED INFERENCE IN A FACTOR GRAPH}\label{sec:MP-in-FFG}

A Forney-style factor graph (FFG) represents the factorization of a multivariate function by a bipartite graph $\mathcal{G}=(\mathcal{V},\mathcal{E})$ whose nodes $a\in\mathcal{V}$ are factors and whose edges $j\in\mathcal{E}$ are variables. A factorized function over the variables $s$ has the form

\begin{equation}
  f(s)\;=\;\prod_{a\in\mathcal{V}} f_a(s_a),
  \label{eq:ffg_factorization}
\end{equation}

where $s_a$ collects the arguments of factor $a$. 
Variables that appear in more than two factors are branched by an equality node, 
$f_{=}(s, s', s'') \triangleq \delta(s-s')\,\delta(s-s'')$, 
while additive constraints are implemented by an addition node, 
$f_{+}(s_1,s_2,s_3) \triangleq \delta(s_1-(s_2+s_3))$. 
Observed variables are represented by clamping, i.e., replacing the corresponding incoming message with 
$\delta(s_1-\hat{s_1})$, which collapses the associated local integral(s)~\cite{loeligerIntroductionFactorGraphs2004}. Note that edge directions in FFGs are purely notational and do not reflect the causal directionality of the generative model.

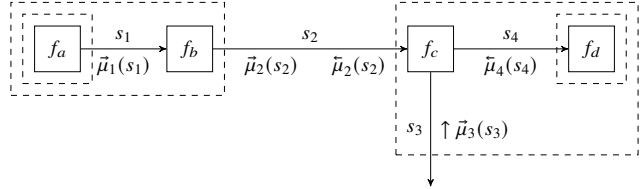
\begin{figure}[ht!]
  \centering
  \scalebox{0.76}{

\begin{tikzpicture}
    [node distance=15mm,auto,>=stealth']


    \node[box] (f_a) {$f_a$};
    \node[box, right =15 mm of f_a] (f_b) {$f_b$};
    
    \node[box, right =34mm of f_b] (f_c) {$f_c$};
    \node[ below =20mm of f_c] (f_d) {};

    \node[ below =10mm of f_c] (f_d_inv) {};
    \node[box, right =20mm of f_c] (f_e) {$f_d$};

\path[line] (f_c) edge[->] node[anchor=south]{$s_4$} node[anchor=north]{$\cev{\mu}_{4}(s_4)$}(f_e);

    \node[fit=(f_e), draw, inner sep=2.0mm,dashed](f_d_box){};

    \node[fit=(f_a), draw, inner sep=2.0mm,dashed](f_a_box){};
    \node[fit=(f_a_box)(f_b), draw, inner sep=2.0mm,dashed](f_ab_box){};

     \node[fit=(f_c)(f_d_inv)(f_d_box), draw, inner sep=2.0mm,dashed](f_abc_box){};

    \path[line] (f_a) edge[->] node[anchor=south]{$s_1$} node[anchor=north]{$\vec{\mu}_{1}(s_1)$} (f_b);
    
    \path[line] (f_b) edge[->] node[anchor=south]{$s_2$} node[anchor=north, pos=0.3]{$\vec{\mu}_{2}(s_2)$} node[anchor=north, pos=0.75]{$\cev{\mu}_{2}(s_2)$}(f_c);
    
    \path[line] (f_c) edge[->] node[anchor=east, pos=0.5]{$s_3$} node[anchor=east, pos=0.5,xshift=15mm]{$\uparrow {\vec \mu}_{3}(s_3)$}(f_d);
\end{tikzpicture}}
  \caption{Illustration of message passing on a Forney-style factor graph. Each node $f_a$ represents a factor, and each edge $s_j$ a variable shared between factors. Forward messages $\vec{\mu}_{j}(s_j)$ and backward messages $\cev{\mu}_{j}(s_j)$ propagate in opposite directions along edges, summarizing information from their respective subgraphs.The posterior marginal on edge $s_j$ is obtained as the normalized product of the two colliding messages, as in \eqref{eq:tree_marginal}.} 
  \label{fig:ffg_example}
\end{figure}

As a simple example consider,
\begin{equation}
  f(s_1,s_2,s_3,s_4) = f_a(s_1)\, f_b(s_1,s_2)\, f_c(s_2,s_3,s_4)\, f_d(s_4),
  \label{eq:example_f}
\end{equation}
whose FFG is shown in Figure~\ref{fig:ffg_example}. 
Observing $s_3=\hat{s}_3$ multiplies the graph by the clamping factor $\delta(s_3-\hat{s}_3)$ and terminates the corresponding half-edge~\cite{loeligerIntroductionFactorGraphs2004}.

Global integrals, such as marginalizations, decompose into local \emph{messages} that summarize the contributions of subgraphs and propagate along edges.  
For the example in \eqref{eq:example_f}, the unnormalized marginal over $s_2$ is

\begin{equation}
    \label{eq:example_marginalization}
    f(s_2) = \iiint f(s_1,s_2,s_3,s_4)\,\mathrm{d}s_1\,\mathrm{d}s_3\,\mathrm{d}s_4,
\end{equation}
which factorizes into two colliding messages on the edge $s_2$ 
\begin{align}
\label{eq:message_decomposition}
&f(s_2) =\\
&\overbrace{\int f_b(s_1,s_2)\underbrace{f_a(s_1)}_{\vec{\mu}_{1}(s_1)}\mathrm{d}s_1}^{\vec{\mu}_{2}(s_2)}
\overbrace{\iint f_c(s_2,s_3,s_4)\underbrace{f_d(s_4)}_{\cev{\mu}_{4}(s_4)}\mathrm{d}s_3\mathrm{d}s_4}^{\cev{\mu}_{2}(s_2)}. \notag
\end{align}


Forney-style factor graphs provide a unified computational framework that supports automated message-passing algorithms for probabilistic inference.
Different algorithmic variations correspond to different interpretations of the local update rules applied along the edges of the graph.
In this work, we focus on two main algorithms: the \emph{sum–product} algorithm \cite{kschischangFactorGraphsSumproduct2001a}, which performs exact inference in conjugate and tree-structured models, and \emph{variational message passing (VMP)} \cite{dauwelsVariationalMessagePassing2007}, which enables approximate inference in non-conjugate or loopy models by minimizing the variational free energy.
Both algorithms can operate within the same graphical representation, differing only in how messages are computed and combined at each node.

\subsubsection{The Sum–Product Algorithm (Belief-Propagation)}

Among the algorithms supported by Forney-style factor graphs, 
the \emph{sum–product algorithm} performs exact inference in tree-structured models with conjugate dependencies.

Consider a probabilistic model factorized as
\begin{equation}
\label{eq:model_factorization}
p(y,s)\;\propto\;\prod_{a\in\mathcal V} f_a(y_a,s_a),
\end{equation}
where \(y\) denotes observed variables and \(s\) latent variables.   For a factor \(a\in\mathcal V\) and an incident edge \(j\in\mathcal E(a)\), the factor-to-edge (forward) message in the sum–product algorithm is given by~\cite{loeligerIntroductionFactorGraphs2004}

\begin{equation}
\vec{\mu}_{j}(s_j)\;=\;
\int f_a(y_a{=}\hat{y}_a,\,s_a)\;
\prod_{i\in \mathcal{E}(a)\setminus\{j\}}\vec{\mu}_{i}(s_i)\;\mathrm{d}s_{a\setminus j}.
\label{eq:sp}
\end{equation}

The corresponding backward message $\cev{\mu}_{j}(s_j)$ completes the bidirectional flow.
For trees, the posterior marginal on edge $j$ is given by
\begin{equation}
p(s_j\,|\,y{=}\hat{y})=
\frac{\vec{\mu}_{j}(s_j)\,\cev{\mu}_{j}(s_j)}
{\int \vec{\mu}_{j}(s_j)\,\cev{\mu}_{j}(s_j)\,\mathrm{d}s_j}.\,
\label{eq:tree_marginal}
\end{equation}

\subsubsection{Variational Message Passing}
Variational inference can likewise be expressed as message passing on a Forney-style factor graph. In this formulation, local message updates correspond to coordinate descent steps that minimize the global variational free energy functional.
Under the mean-field assumption
\begin{equation}
\label{eq:mf_factorization}
q(s) \;=\; \prod_{j\in\mathcal{E}} q_j(s_j),
\end{equation}
the approximate posterior $q(s)$ factorizes over the variables indexed by $\mathcal{E}$, allowing each variable to be updated using expectations over its local neighborhood in the graph.

For a factor $a$ and an incident edge $j\in\mathcal{E}(a)$, the forward variational message is~\cite[Sec.~3, Eqs.~(15-17)]{dauwelsVariationalMessagePassing2007}

\begin{equation}
\label{eq:mean_field_update}
\vec{v}_{j}(s_j) \;=\;
\exp\!\Bigg(\int 
\!\!\!\!\prod_{i\in \mathcal{E}(a)\setminus j}\!\!\!\!
q_i(s_i)\,
\ln f_a(y_a{=}\hat{y}_a, s_a)\,
\mathrm{d}s_{a\setminus j}\Bigg).
\end{equation}

The approximate marginal on $s_j$ is obtained by multiplication of incoming messages,
\begin{equation}
q_j(s_j)\propto \vec{v}_{j}(s_j)\,\cev{v}_{j}(s_j),
\end{equation}
followed by normalization. 
These updates are iterated until the approximate posterior $q(s)$ converges. Convergence is not guaranteed in general, although the variational free energy is guaranteed to decrease at each update step when all expectations and normalization constants are computed exactly~\cite{dauwelsVariationalMessagePassing2007}.

A key advantage of the factor-graph formulation is that providing additional local constraints allow \emph{hybrid inference}, 
where different subgraphs employ distinct update rules (e.g., VMP for non-conjugate parts and BP for conjugate parts) while exchanging compatible messages across shared interface variables~\cite{coxFactorGraphApproach2019,senozVariationalMessagePassing2021}.

The preceding review summarized the framework for message passing on Forney-style factor graphs, including both the sum–product and variational message passing formulations.
We now apply these principles to the SEM. Specifically, we represent the SEM as an FFG by connecting its main submodels (for speech, noise, SNR, gain, and VAD tracking) into a graphical structure.
The FFG representation enables localized inference updates for each node and provides the computational foundation for automated inference within AIDA-2.

The factor-graph formulation also enables \emph{automated} derivation and implementation of message passing algorithms.
Given a probabilistic model and a set of factorization and inference constraints, toolboxes such as \texttt{ForneyLab.jl} and its successor \texttt{RxInfer.jl} can compile the corresponding Forney-style factor graph into an executable schedule of local message updates, including hybrid combinations of sum–product, VMP, and related rules~\cite{coxForneyLabJlFast2018,bagaevRxInferJuliaPackage2023}.
In this way, Bayesian signal processing algorithms are obtained directly from the model specification, rather than being hand-crafted, which improves transparency, maintainability, and rapid exploration of alternative model and inference choices.

\subsection{MESSAGE PASSING FOR THE SEM MODULE}\label{sec:MP-for-SEM}

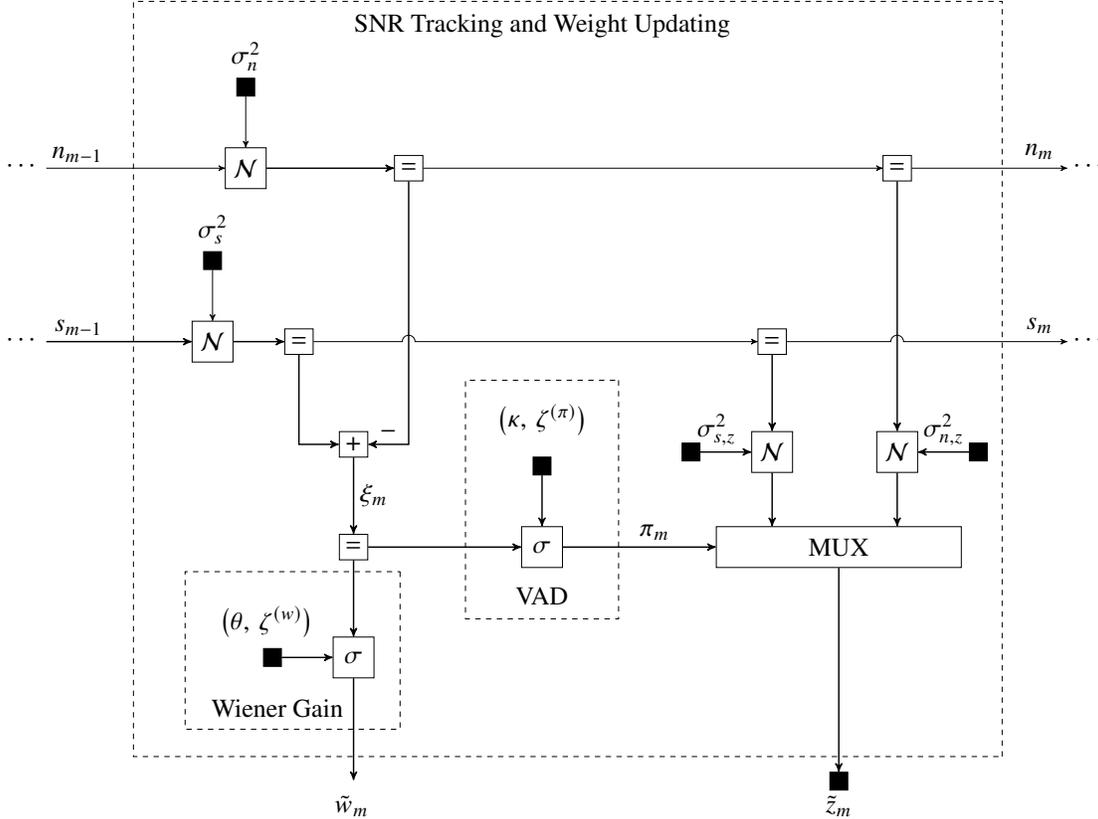
\begin{figure*}[ht!]
  \centering
  \resizebox{0.9\linewidth}{!}{\begin{tikzpicture}
[node distance=10mm,auto,>=stealth',
every node/.append style={font=\Large}]

\node[](dots_noise_prior){$\cdots$};
\node[box, right=35mm of dots_noise_prior](transition_noise){$\mathcal{N}$};

\node[clamped, above=10mm of transition_noise](noise_state_variance){};

\node[smallbox, right=25mm of transition_noise](equality_noise_transition){$=$};

\node[below=20mm of equality_noise_transition](equality_noise_transitioninv){};

\node[smallbox, right=90mm of equality_noise_transition](equality_noise_transitionobs){$=$};

\draw[->,thick] (transition_noise) -- node[xshift=-13mm, above] {} (equality_noise_transition);

\node[below=30mm of dots_noise_prior](dots_speech_prior){$\cdots$};
\node[box, right=28.5mm of dots_speech_prior](transition_speech){$\mathcal{N}$};
\node[clamped, above=10mm of transition_speech](speech_state_variance){};
\draw[->,thick] (dots_speech_prior) -- node[xshift=-8mm, above] {$s_{m-1}$} (transition_speech);
\node[smallbox, right=10mm of transition_speech](equality_speech_transition){$=$};

\node[smallbox, right=87mm of equality_speech_transition](equality_speech_transitionobs){$=$};

\node[box, below=15mm of equality_speech_transitionobs](transition_obs_s){$\mathcal{N}$};

\node[box, below =49mm of equality_noise_transitionobs](transi†ion_n_OBSERVATION){$\mathcal{N}$};

\draw[->,thick] (transition_speech) -- node[] {} (equality_speech_transition);

\draw[->] (dots_noise_prior) -- (transition_noise)node[xshift=-33mm, above]{$n_{m-1}$} ;
\draw[<-](transition_speech)--(speech_state_variance)node[above,yshift=2mm]{$\sigma^2_s$};
\draw[<-](transition_noise)--(noise_state_variance)node[above, yshift=2mm]{$\sigma^2_n$};

\node[smallbox, below right=15mm and 5mm of equality_speech_transition](SNR){$+$};

\node[smallbox, below =15mm  of SNR](equality_SNR){$=$};

\node[box, right=30mm of equality_SNR](VAD){$\sigma$};
\node[clamped, above=10mm of VAD](params_VAD){};
\node[box_h,right=30 mm of VAD](MUX){MUX};

\node[clamped,below=40mm of MUX](observation){};


\node[box, below=15mm of equality_SNR](Wiener){$\sigma$};

\node[below=20mm of Wiener](w){};
\node[clamped, left=10mm of Wiener](params_Wiener){};

\draw[->,thick] (equality_SNR) -- node[] {} (Wiener);
\draw[->,thick] (Wiener) -- node[xshift =-5mm,yshift=-15mm] {$\tilde w_{m}$} (w);

\draw[->,thick](params_Wiener.east)--(Wiener.west)node[xshift=-13mm, yshift = 7mm]{$\left(\theta,\,\zeta^{(w)}\right)$};

\draw[->,thick](params_VAD)--(VAD)node[yshift = 25mm]{$\left( \kappa,\,\zeta^{(\pi)}\right)$};
\draw[->,thick] (equality_SNR) -- node[] {} (VAD);


\draw[->,thick] (equality_speech_transition.south) |- node[xshift=-13mm, above] {} (SNR.west);

\draw[->,thick] (equality_noise_transition.south) |- node[xshift=-4mm, above] {$-$} (SNR.east);

\draw[->,thick] (SNR) -- node[] {$\xi_{m}$} (equality_SNR);

\draw[->,thick] (VAD) -- node[above,xshift=3mm] {$\pi_{m}$} (MUX.west);
\draw[->,thick] (MUX) -- node[above,yshift=-30mm] {$\tilde z_{m}$} (observation);

\draw[->,thick]
  (equality_noise_transitionobs.south) -- (transi†ion_n_OBSERVATION.north);

\draw[->,thick]
  (transi†ion_n_OBSERVATION.south) -- ($(transi†ion_n_OBSERVATION.south |- MUX.north)$);

\draw[->,thick]
  (equality_speech_transitionobs.south) -- (transition_obs_s.north);

\draw[->,thick]
  (transition_obs_s.south) -- ($(transition_obs_s.south |- MUX.north)$);

\node[clamped,  left=10mm of transition_obs_s](ps){};
\node[clamped, right=10mm of transi†ion_n_OBSERVATION](pn){};

\draw[->,thick]
  (ps.west) -- node[above] {$\sigma^2_{s,z}$} (transition_obs_s);

\draw[->,thick]
  (pn) -- node[above] {$\sigma^2_{n,z}$} (transi†ion_n_OBSERVATION.east);

\node[right=200mm of dots_noise_prior](dots_noise_prior_post){$\cdots$};
\node[right=200mm of dots_speech_prior](dots_speech_prior_post){$\cdots$};

\node[above=0.5mm of dots_noise_prior_post,yshift=-2mm, xshift=-10mm](dots_noise_prior_post2){$n_{m}$};
\node[above=0.5mm of dots_speech_prior_post,yshift=-2mm, xshift=-10mm](dots_speech_prior_post2){$s_{m}$};

\draw[->]
  (equality_speech_transitionobs) to [jump=(equality_noise_transitionobs)--(transi†ion_n_OBSERVATION), jumps=left]
  (dots_speech_prior_post){};

\draw[->] (equality_noise_transitionobs) --  (dots_noise_prior_post){};

\coordinate (mid_lambda) at ($(equality_speech_transition)!0.5!(equality_speech_transitionobs)$);

\draw[->]
  (equality_speech_transition)
  to[jump=(equality_noise_transition)--(equality_noise_transitioninv), jumps=left]  
  (equality_speech_transitionobs);

\draw[->,thick] (equality_noise_transition) (equality_noise_transitionobs){};

\draw[->] (equality_noise_transition) --(equality_noise_transitionobs){};

   \draw[dashed]
  ([xshift=32mm, yshift=3mm] observation.north) rectangle 
  ([xshift=-22mm, yshift=15mm] noise_state_variance.north);
  \node[align=center, yshift=80mm] at ($(observation)!0.5!(noise_state_variance)$) {SNR Tracking and Weight Updating};

     \draw[dashed]
  ([xshift=15mm, yshift=15mm] params_VAD.north) rectangle 
  ([xshift=-15mm, yshift=-10mm] VAD.south);

  \node[align=center] at ($(params_VAD)!1.6!(VAD)$) {VAD};

       \draw[dashed]
  ([xshift=25mm, yshift=15mm] params_Wiener.north) rectangle 
  ([xshift=-33mm, yshift=-10mm] Wiener.south);
  \node[align=center, xshift=-7mm, yshift=-10mm] at ($(params_Wiener)!0.5!(Wiener)$) {Wiener Gain};

\end{tikzpicture}}
  \caption{Forney-style factor graph representation of the SEM, shown without the analysis module, which transforms $z_m$ into the log-spectral coefficients $\tilde{z}_m$, and the synthesis module, which maps the spectral weights $\tilde{w}_m$ back to the time-domain filter coefficients $w_m$. Each node corresponds to a probabilistic factor, and edges denote shared latent variables.}
  \label{fig:SEM}
\end{figure*}

Inference in the SEM follows directly from the message-passing framework described above. 
The generative model in \eqref{eq:SEM-model-1} defines the factor graph structure.
Each node in the graph corresponds to a submodule of the SEM, namely the speech, noise, SNR, gain, and VAD trackers, and exchanges messages with its neighboring nodes according to the sum-product and variational message passing rules described in the previous section.
This structure yields an efficient approximate Bayesian inference procedure suitable for real-time implementation in embedded hearing aid systems.

The Bayesian Leaky Integrator (BLI) nodes for speech $s_m$ and noise $n_m$ tracking perform temporal smoothing of the observed power coefficients $\tilde z_m$, each with its own adaptation rate.
Each BLI propagates Gaussian beliefs over the corresponding log-power variable. The resulting variational posterior marginals are

\begin{subequations}
\begin{align}
    q(s_m) &= \mathcal{N}\!\big(s_m\,|\,\mu^{(s)}_m, v^{(s)}_m\big) \\
q(n_m) &= \mathcal{N}\!\big(n_m\,|\,\mu^{(n)}_m, v^{(n)}_m\big),
\end{align}    
\end{subequations} 

which jointly define the instantaneous SNR posterior marginal
\begin{equation*} 
q(\xi_m)=\mathcal{N}\!\big(\mu^{(s)}_m-\mu^{(n)}_m,\; v^{(s)}_m+v^{(n)}_m\big).
\end{equation*}
This SNR distribution acts as a coupling variable that drives the VAD and spectral weight-update nodes in the SEM.

The observation node $p(\tilde z_m\,|\,s_m,n_m,\pi_m)$ in \eqref{eq:Observation-model} connects the speech and noise nodes to the observations $\tilde z_m$ through the discrete switch variable $\pi_m$. Marginalizing over the switch variable yields a two-component Gaussian mixture,

\begin{equation}
    p(\tilde z_m \,|\, s_m, n_m) = \sum_{\pi_m \in \{0,1\}} p(\tilde z_m \,|\, s_m, n_m, \pi_m)\,p(\pi_m),
\end{equation}

which represents the probabilistic mixing between speech-active and speech-inactive states.
Under the mean-field approximation, this mixture is represented by separate variational marginals $q(\pi_m)$, $q(s_m)$, and $q(n_m)$ that exchange Bernoulli and Gaussian messages.
In practice, this approximation replaces the exact Gaussian mixture with a single Normal message whose precision is weighted by the posterior speech-activity marginal $q(\pi_m)$, providing an efficient surrogate for the full mixture.

The VAD in \eqref{eq:Voice-model} and the spectral weight update in \eqref{eq:weight-update-model} are both parameterized by logistic functions, which are non-conjugate with the Gaussian posterior $q(\xi_m)$.
To enable tractable message passing, the log-sigmoid term $\log \sigma(x)$ is approximated using the Jaakkola–Jordan (JJ) quadratic bound \cite{jaakkolaVariationalApproachBayesian1997a}. This bound introduces a local variational parameter that is updated analytically from the current moments of $q(\xi_m)$.
The resulting approximation converts the non-conjugate Bernoulli–logistic links into conditionally Gaussian factors, permitting closed-form Gaussian and Bernoulli message updates.
The corresponding derivation and update equations are given in Appendix~\ref{app:logit} and summarized in Table~\ref{tab:LogitNodeRules}.
The JJ bound thus provides a computationally efficient surrogate for the VAD node while maintaining a tight local free-energy bound.

The expectation $\mathbb{E}[\tilde w_m]$ provides the spectral gain applied by the enhancement stage through the WFB. This expectation is taken with respect to the variational marginal $q(\tilde w_m)$, which is obtained from the message-passing updates in the SEM.
Under the proposed generative model, this gain corresponds to a user-modulated Wiener gain, adapted by the control parameter $\theta$.

Together, Fig.~\ref{fig:SEM} and Table\ref{tab:LogitNodeRules} summarize the computational structure of the inference procedure in the SEM module.

\vspace{0.5em}
\begin{center}
\begin{table}[b!]
\captionsetup{font=normalsize,justification=centering}
\caption{Variational message passing updates for the logistic node using the Jaakkola–Jordan bound.
The variable $y\in\{0,1\}$ denotes either the gain $\tilde w$ or the VAD $\pi$; $\xi \in\mathbb{R}$ is the logit input; and $\zeta$ is an auxiliary variational parameter.}
\label{tab:LogitNodeRules}
\centering
\renewcommand{\arraystretch}{1.4}
\begin{tabular}{p{\linewidth}}
\toprule
\textbf{LOGIT NODE} \\ \midrule
\textbf{Forney-style Factor Graph} \\[4pt]
{\centering\scalebox{0.9}{\begin{tikzpicture}
[node distance=10mm,auto,>=stealth',
every node/.append style={font=\large}]

\node[](x){$\zeta$};
\node[boxbigger, below=20mm of x](logit){$\sigma$};
\node[left=20mm of logit](zeta){$\xi$};
\node[left=15mm of zeta](zetainv){};
\node[right=20mm of logit](y){$y$};

\draw[->,thick] (x) -- (logit)node[pos=0.5, left,scale=0.9] {$q(\vect{\zeta})$}
        node[pos=0.15, left,scale=0.9] {$\downarrow$}
        node[pos=0.2, left=2mm,scale=0.9] {$\vec{\nu}(\vect{\zeta})$}
        node[pos=0.8, left, scale=0.9] {$\uparrow$}
        node[pos=0.82, left=2mm,scale=0.9] {$\cev{\nu}(\vect{\zeta})$};
\draw[->,thick] (logit) -- (y)node[pos=0.5, above,scale=0.9] {$q(\vect{y})$}
        node[pos=0.2,below,scale=0.9] {$\rightarrow$}
        node[pos=0.25, below=0.15,scale=0.9] {$\vec{\nu}(\vect{y})$}
        node[pos=0.8,below,scale=0.9] {$\leftarrow$}
        node[pos=0.82, below=0.15,scale=0.9] {$\cev{\nu}(\vect{y})$};
\draw[->,thick] (zeta) --(logit)node[pos=0.5, above,scale=0.9] {$q(\vect{\xi})$}
        node[pos=0.2,below,scale=0.9] {$\rightarrow$}
        node[pos=0.25, below=0.15,scale=0.9] {$\vec{\nu}(\vect{\xi})$}
        node[pos=0.8,below,scale=0.9] {$\leftarrow$}
        node[pos=0.82, below=0.15,scale=0.9] {$\cev{\nu}(\vect{\xi})$};
\end{tikzpicture}}\par} \\[2pt] \midrule
\textbf{Functional form} \\[2pt]
$\begin{aligned}
f(y,\xi;\zeta) &= \operatorname{Ber}\!\big(y\,|\,\sigma(\xi)\big), \qquad
\sigma(\xi)    = \dfrac{1}{1+\exp(-\xi)}
\end{aligned}$ \\[2pt] \midrule

\textbf{Lower bound} \\[2pt]
$\begin{aligned}
\ln\sigma(\xi)\ &\ge\ \ln\sigma(\zeta)+\dfrac{\xi-\zeta}{2}
   -\lambda(\zeta)\bigl(\xi^2-\zeta^2\bigr), \qquad
\lambda(\zeta)\ =\ \dfrac{1}{2\zeta}\Bigl(\sigma(\zeta)-\tfrac{1}{2}\Bigr)
\end{aligned}$ \\[2pt] \midrule

\textbf{Factorization} \\[2pt]
$q(y,\xi,\zeta)=q(y)\,q(\xi)\,q(\zeta)$ \\[2pt] \midrule

\textbf{Variational posteriors} \\[2pt]
$\begin{aligned}
q(y) &\propto \operatorname{Ber}(y\,|\,\phi), \qquad
q(\xi) \propto \mathcal{N}(\xi \,|\, m_x, v_x), \qquad
q(\zeta) \propto \delta(\zeta - \hat{\zeta})
\end{aligned}$ \\[2pt] \midrule

\textbf{Update messages} \\[2pt]
$\begin{aligned}
\vec{\nu}(y) &\propto \operatorname{Ber}\!\big(y\,|\,\sigma(\overline{\xi})\big), \qquad
\cev{\nu}(\xi) \propto \mathcal{N}\!\left(\xi \,\Bigg|\,
\dfrac{\bar{y}-\tfrac{1}{2}}{2\lambda(\hat{\zeta})},\
(2\lambda(\hat{\zeta}))^{-1}\right), \qquad
\hat{\zeta} = \sqrt{\,\overline{\xi^2} + \Var[\xi]}
\end{aligned}$ \\[2pt] \midrule

\textbf{Average energy} \\[2pt]
$\mathcal{U} = -\overline{xy} - \ln\sigma(\hat{\zeta})
   + \dfrac{\bar{\xi}+\hat{\zeta}}{2}
   + \lambda(\hat{\zeta})\Bigl(\overline{\xi^{2}}+\Var[\xi]-\hat{\zeta}^{2}\Bigr)$\\[2pt]
\bottomrule
\end{tabular}
\end{table}

\end{center}
\vspace{0.5em}

\section{EXPERIMENTAL VALIDATION}
\label{sec:experiments}

\subsection{OBJECTIVES}

As discussed, this research report is part of a broader program aimed at transferring sound processing in HAs to an automated, data-driven inference process. At this stage, our primary goal is to validate the feasibility of this approach. The main question we address is whether current open-source technology for automated Bayesian inference \texttt{RxInfer.jl}\footnote{\url{https://github.com/ReactiveBayes/RxInfer.jl}} \cite{bagaevRxInferJuliaPackage2023} is capable of performing signal processing inference within a complex generative model for sound processing. A second question concerns whether the model specifications of the SEM and WFB modules yield acceptable noise reduction performance, even in the absence of offline or online parameter learning. A positive outcome would lend support to the adequacy of the proposed model specification. With these objectives in mind, we designed and conducted the experiments described below.  

\subsection{DATASET}
We evaluate the proposed SEM on the public \emph{VoiceBank+DEMAND} \cite{valentini-botinhaoc.NoisySpeechDatabase2017} corpus, a widely used benchmark for monaural speech enhancement. We perform \emph{no} training; all results are reported on the official test set, which comprises 824 noisy utterances from two held-out speakers mixed with five unseen DEMAND environments (bus, café, living room, office, public square) at four SNRs: 17.5, 12.5, 7.5, and 2.5~dB. Single-channel audio is resampled to 16~kHz. While the framework admits learning, this proof-of-concept evaluates SEM in a training-free setting on the official test set.

\subsection{EXPERIMENTAL SETUP}

The WFB front end employs a 32-tap warped filter bank with a warping coefficient $\alpha = 0.5$, operating at 16~kHz.  
This stage provides a perceptually motivated subband decomposition that serves as the input to the SEM.  

Following the WFB analysis, the SEM performs spectral analysis via the DFT to obtain complex spectral coefficients.  
Because the input signals are real-valued, inference in the SEM operates only on the non-redundant portion of the spectrum, i.e., 17 warped-frequency bands including the DC and Nyquist components. Across these bands, message passing-based inference maintains approximately $85$ variational posteriors.   

Table~\ref{tab:parameters} summarizes the used parameter settings in the SEM module.
\vspace{0.5em}
\begin{center}
\begin{table}[ht]
\centering
\caption{Module configuration and hyperparameters used in the experiments.
The symbols match the model specification. All values are clamped and fixed (no learning).}
\label{tab:parameters}
\begin{tabular*}{\linewidth}{@{\extracolsep{\fill}} l l l}
\toprule
\textbf{Module} & \textbf{Quantity} & \textbf{Value / Setting} \\
\midrule
\multirow{3}{*}{WFB}
  & Sampling rate & 16 kHz \\
  & Bands, taps, warp $\alpha$ & 17, 32, 0.5 \\
  & Analysis/synthesis & Reconstruction WFB \\
\midrule
\multirow{5}{*}{SEM}
  & $\tau_{90}(s)$, $\tau_{90}(n)$
      & $5$ ms, $700$ ms\\
  & $\lambda(\cdot)$ from $\tau_{90}(\cdot)$ & via \eqref{eq:tau_to_lambda} \\
  & $\sigma^2(\cdot)$ from $\lambda(\cdot)$ & via \eqref{eq:sigma_s-constraint_bli} (clamped) \\
  &$\kappa$  & $2$ dB  (clamped)\\
\midrule
\multirow{1}{*}{EUM}
  &  $\theta$ &  $12$ dB (clamped) \\
\bottomrule
\end{tabular*}
\end{table}

\end{center}
\vspace{0.5em}

\subsection{EXPERIMENTAL RESULTS}
We compare the proposed SEM against representative generative and discriminative baselines to assess how a minimal, training-free probabilistic approach fares relative to modern high-capacity models. Objective metrics include PESQ (ITU–T P.862) \cite{internationaltelecommunicationunionP8622WidebandExtension2005} with implementation from \cite{miaoPesqPythonWrapper}, and the composite DSMOS predictors SIG (signal distortion), BAK (background intrusiveness), and OVL (overall quality) \cite{reddyDnsmosP835NonIntrusive2022}. We also report parameter counts when available.

Table~\ref{tab:comparison_with_params} compares the proposed model against published baselines in terms of performance metrics and model size. Baseline scores and sizes are quoted from the cited publications. 

\vspace{0.5em}
\begin{table}[t]
\centering
\caption{VoiceBank+DEMAND results. Best scores are bold.
T–F = time–frequency domain, T = time domain.
The Wiener gain is taken from the results reported in WaveCRN~\cite{hsiehWaveCRNEfficientConvolutional2020}.}
\label{tab:comparison_with_params}
\begin{tabular*}{\linewidth}{@{\extracolsep{\fill}} l c r r r r r}
\toprule
System & Dom. & PESQ & SIG & BAK & OVL & Params \\
\midrule
Unprocessed        & T–F & 1.95 & 3.32 & 3.10 & 2.68 & N/A \\
Wiener~\cite{hsiehWaveCRNEfficientConvolutional2020}  & T–F & 2.22 & 3.23 & 2.68 & 2.67 & — \\
\textbf{SEM (ours)}& T–F & 2.17 & 3.29 & 3.47 & 2.78 & $\sim$85 \\
\midrule
MAMBA-SENet~\cite{kimMambabasedHybridModel2025} & T & \textbf{3.62} & \textbf{4.79} & \textbf{4.01} & \textbf{4.34} & 0.99\,M \\
DSEGAN~\cite{pascualSEGANSpeechEnhancement2017} & T & 2.39 & 3.46 & 3.11 & 2.90 & 43.2\,k \\
CDiffuSE~\cite{luConditionalDiffusionProbabilistic2022} & T & 2.52 & 3.72 & 2.91 & 3.10 & — \\
WaveCRN~\cite{hsiehWaveCRNEfficientConvolutional2020} & T & 2.64 & 3.94 & 3.37 & 3.29 & 4.65\,M \\
MOSE~\cite{chenMetricOrientedSpeechEnhancement2023b} & T & 2.54 & 3.72 & 2.93 & 3.06 & — \\
\bottomrule
\end{tabular*}
\end{table}
\vspace{0.5em}

Per-environment breakdowns (noise type $\times$ SNR) are provided in Appendix~\ref{app:results}. Audio examples and configuration files are included in the supplementary material to facilitate exact reproducibility (Section~\ref{sec:reproducibility}).

\vspace{0.5em}
\subsection{DISCUSSION}
\label{sec:discussion}

We conclude that the two main objectives of the validation experiments have been achieved. After a few additions to the message update rules in \texttt{RxInfer} (detailed in Appendix~\ref{app:logit}), we found it feasible to use \texttt{RxInfer} to infer an output signal $y_t$ solely through automated message passing in the AIDA-2 model. Moreover, using an ``untrained'' SEM model yields noise reduction performance that is in the same ballpark as state-of-the-art noise reduction algorithms that rely on far more tuning parameters. Since our longer-term goal is to further develop AIDA-2 to enable situated personalization based on sparse in-situ appraisals, we consider AIDA-2 particularly suitable for HA applications that require a very small computational footprint.

\section{RELATED WORK}
\label{sec:related-work}

Most single-channel enhancement algorithms used in hearing aids can be written as $y_t = f_w(x^t)$ (cf. \eqref{eq:sound-processing-algo}). As discussed in Section~\ref{sec:problem}, selecting a fixed $w$ is challenging because acoustic conditions and user preferences vary over time.
Prior work has focused on designing algorithmic architectures to estimate $w$, primarily in the spectral domain. These approaches have shaped the speech enhancement methods currently used in hearing aids.
The SEM in AIDA-2 extends this line: it builds on the classical spectral-subtraction foundation and reframes enhancement as probabilistic inference from observed signals only, integrating personalization and acoustic-context directly into the processing loop.

In many of these approaches \cite{kamathMultibandSpectralSubtraction2002,martinNoisePowerSpectral2001,beroutiEnhancementSpeechCorrupted1979,ephraimSpeechEnhancementUsing1984,ephraimSpeechEnhancementUsing1985, wienerExtrapolationInterpolationSmoothing1964,limEnhancementBandwidthCompression1979,cohenNoiseEstimationMinima2002,bollSuppressionAcousticNoise1979, wangIdealBinaryMask2005,wangTrainingTargetsSupervised2014,yongOptimizationEvaluationSigmoid2013}, the estimation of $w$ follows a two-stage structure. First, the algorithm tracks the time-varying relation of speech and noise, typically via an SNR estimate or proxy thereof. Second, it maps that estimate to a spectral gain, controlling the level of noise suppression. These two components, tracking SNR and mapping SNR to gain, define a common computational structure across classical spectral subtraction, Wiener filtering, and modern data-driven methods.

In the SEM, this pattern is expressed probabilistically: both tracking and mapping arise from inference on a generative model. The prior models for speech and noise can be designed to capture realistic acoustic behavior while remaining compatible with hearing-aid constraints on latency, memory, and computational load.

Early systems estimated noise during presumed silence and subtracted it from the noisy spectrum to obtain a positive residual spectrum \cite{bollSuppressionAcousticNoise1979}. In this approach, any negative values after subtraction were set to zero through a technique called \emph{half-wave rectification}. It ensured that spectral magnitudes remained non-negative. This operation produced residual tones known as musical noise, defined as short, random tonal bursts caused by frame-to-frame variability in the subtraction process. To reduce these artifacts, \emph{over-subtraction} scaled the noise estimate by a factor greater than one before subtraction, and \emph{spectral flooring} imposed a minimum spectral magnitude to prevent bins from being fully suppressed \cite{beroutiEnhancementSpeechCorrupted1979}. Multiband and perceptually weighted variants later introduced frequency-dependent control and \emph{auditory-masking-based} weighting to improve robustness in the face of colored noise and uneven speech spectra \cite{kamathMultibandSpectralSubtraction2002,viragSingleChannelSpeech1999}. The persistence of musical noise and the limited flexibility of heuristic fixes such as \emph{over-subtraction} and \emph{spectral flooring} motivated the use of estimators derived from statistical principles, where gains are computed as continuous functions of estimated SNR rather than by hard spectral subtraction.

Wiener filtering provided a soft gain that minimizes mean-square error under Gaussian assumptions \cite{wienerExtrapolationInterpolationSmoothing1964,limEnhancementBandwidthCompression1979}. Ephraim and Malah extended this to minimum mean-square error (MMSE) estimators in both spectral and log-spectral domains and introduced the decision-directed (DD) approach for estimating the a priori SNR \cite{ephraimSpeechEnhancementUsing1984,ephraimSpeechEnhancementUsing1985}.
In the DD method, the a priori SNR is computed recursively by combining the previous frame’s a priori estimate with the current a posteriori SNR through a fixed smoothing coefficient. This update behaves like a leaky integrator, stabilizing gain fluctuations and reducing musical noise, but the fixed smoothing factor imposes a constant adaptation rate that can lag in rapidly changing acoustic scenes.
These limitations motivated adaptive noise-tracking methods with data-dependent update rates, including minimum-statistics noise estimation and speech-presence-probability formulations \cite{martinNoisePowerSpectral2001,cohenNoiseEstimationMinima2002}.

Adaptive trackers replaced fixed smoothing with data-driven updates. The \emph{minimum-statistics} method estimated the noise floor via sliding minima and avoided an explicit VAD \cite{martinNoisePowerSpectral2001}, while other schemes adjusted  adjusted the forgetting factors using a voice-activity probability \cite{cohenNoiseEstimationMinima2002}. In the SEM, we unify these ideas with a Bayesian Leaky Integrator (BLI): speech and noise log-powers are treated as latent states in a linear–Gaussian model, and posterior uncertainty dynamically sets the effective forgetting (update) factors. This enables \emph{maximum-tracking} for speech peaks and \emph{minimum-tracking} for background noise, all without manual tuning. (Section ~\ref{sec:model-specification}, Appendix ~\ref{app:bayesian-leaky-integrator}).

Several works sought parametric and perceptually grounded formulations for estimating w that could explicitly balance noise suppression and speech distortion. Parametric sigmoid gain functions were introduced to provide additional degrees of freedom through slope and offset parameters, motivated by the need to control enhancement aggressiveness and reduce musical noise under different noise conditions \cite{yongOptimizationEvaluationSigmoid2013,damOptimizedSigmoidFunctions2024}. In parallel, work in computational auditory scene analysis defined the Ideal Binary Mask (IBM) as an intelligibility-maximizing target, though its hard decisions introduced perceptual artifacts \cite{wangIdealBinaryMask2005}. The Ideal Ratio Mask (IRM) was proposed to overcome these artifacts by assigning continuous gains proportional to the local SNR, yielding smoother reconstructions and improved perceptual scores \cite{wangTrainingTargetsSupervised2014,srinivasanComputationalAuditoryScene2006,kjemsRoleMaskPattern2009,liOptimalityIdealBinary2008}.

Taken together, prior work established enhancement as a mapping from an estimated SNR to a spectral gain: classical methods fixed this mapping, while later formulations introduced parametric or perceptually motivated control. The SEM module builds on this foundation by expressing both the user-preferred gain and the VAD as probabilistic functions of the inferred log-SNR, as specified in~\eqref{eq:weight-update-model} and~\eqref{eq:Voice-model}. In this formulation, both variables are modeled through Bernoulli–logistic links, which generalize classical gain rules and introduce controllable offsets through the parameters $\theta$ and $\kappa$.

Both the VAD and weight-update nodes in the SEM follow logistic forms that generalize classical mask-based approaches. Marginalizing uncertainty in the latent SNR $\xi$ yields IRM-like soft masks, while a maximum a posteriori decision reproduces the binary behavior of the IBM. Under Gaussian assumptions for the speech and noise priors, the posterior mean of $\tilde w$ corresponds to a Wiener-type gain whose parametric form reduces to a sigmoid modulated by the control parameter $\theta$. This formulation unifies the classical IBM, IRM, and Wiener perspectives within a single probabilistic framework that embeds personalization and acoustic context directly into the inference process.

Recent research and implementations rely heavily on deep learning architectures for single-channel speech enhancement. Representative model families include GAN-based and adversarial frameworks (e.g., DSEGAN \cite{pascualSEGANSpeechEnhancement2017}), diffusion and score-based generative models (e.g., CDiffuSE \cite{luConditionalDiffusionProbabilistic2022}), convolutional–recurrent architectures that capture spectro-temporal dependencies (e.g., WaveCRN \cite{hsiehWaveCRNEfficientConvolutional2020}), state-space and hybrid time–frequency systems (e.g., MAMBA-SENet \cite{kimMambabasedHybridModel2025}), and metric-oriented formulations that directly optimize perceptual measures such as PESQ or composite MOS predictors (e.g., MOSE \cite{chenMetricOrientedSpeechEnhancement2023b}). Many of these systems operate directly on the waveform. They achieve high perceptual scores on matched benchmarks but require computational and memory resources beyond hearing-aid constraints. Diffusion methods, in particular, involve iterative multi-step inference. In general, performance often declines when evaluated on real acoustic recordings rather than simulated mixtures, as reported in the Clarity Challenge \cite{coxOverview2023Icassp2023}, due to limited generalization to unseen conditions. A complementary research line targets embedded feasibility through architectural compression and multiply–accumulate (MAC) reduction, which minimize arithmetic operations to lower latency and power consumption. Examples include TinyLSTM and FSPEN \cite{fedorovTinyLSTMsEfficientNeural2020,yangFspenUltraLightweightNetwork2024}. These approaches move toward hearing-aid constraints on latency, battery, and memory but generally lack mechanisms for online adaptation, explicit uncertainty propagation, or user/context-aware control. In contrast, SEM performs online Bayesian inference with a small parameter set, propagates uncertainty through tracking and mapping, and incorporates user controls directly in the gain computation via the offset parameter defined in \eqref{eq:weight-update-model}. The design is aligned with hearing-aid latency, power, and memory constraints while enabling in-situ personalization through the EUM. Rather than aiming for benchmark maximization, SEM prioritizes interpretability, robustness across scenes, and energy-efficient operation, making it suitable for embedded HA applications.

\section{DISCUSSION}\label{sec:discussion}

\subsection{MAIN FINDINGS}

In the proposed generative probabilistic modeling approach, we employ compact models inspired by established spectral subtraction algorithms to maintain a low computational load. The framework also combines the offline trainability characteristic of deep neural network methods with the potential for online personalization, which introduces a new direction in HA research. A key advantage of the approach is that all processing tasks can be formulated within a unified framework of automated Bayesian inference. Through validation experiments, we also confirmed that the current reactive message passing technology in RxInfer is capable of executing such complex inference tasks efficiently.

In short, our initial findings are encouraging and demonstrate the feasibility of the proposed approach. However, several limitations and open challenges remain, which we outline next together with possible directions for future research.

\subsection{LIMITATIONS} 

Our approach has not yet been fully tested in several important respects. First, we have so far designed only the generative model for spectral speech enhancement, while omitting the crucial specification of the EUM module. Consequently, online personalization of the speech enhancement algorithm has not yet been evaluated. Furthermore, although RxInfer demonstrated encouraging inference performance for the signal processing task, challenges remain in handling complex non-conjugate prior–likelihood pairs, which prevented us from performing full parameter learning in the AIDA-2 model. Fortunately, research on inference mechanisms and research on generative model design can proceed independently, allowing both limitations to be addressed in parallel.

\subsection{FUTURE DIRECTIONS}

Our future work focuses on alleviating the above-mentioned limitations. 

A first priority for future work is to develop an end user model (EUM) that supports an interaction protocol users are willing to engage with in practice. This effort is being pursued in parallel with the ongoing work on the sound processing modules (WFB, SEM, and AC) under the same research program umbrella (see Acknowledgments section). The objective of this research thread is to pave the way toward online personalization of HA algorithms. The results of this study will be reported in a forthcoming publication.

A second research thread focuses on advancing automated inference through message passing. Ongoing work on RxInfer aims to refine message update computations in complex models where priors and likelihoods are non-conjugate \cite{lukashchuk2025quotient}. This line of research is expected to enable full automation of offline parameter learning for all components of the AIDA-2 model.

Finally, we believe that the current generative sound processing models can be further improved. In particular, future work will focus on refining the hierarchical model structures embedded in the SEM and ACM modules. Advancements in these components are expected to yield more accurate estimation of signal and noise spectra, ultimately leading to enhanced user experiences in challenging acoustic environments.    
\section{CONCLUSIONS}\label{sec:conclusions}

This paper presented a generative probabilistic modeling framework for speech enhancement in hearing aids, unifying signal enhancement, and both online, context-aware personalization and offline training as Bayesian inference tasks. We demonstrated that speech enhancement can be realized satisfactorily through inference, using efficient reactive message passing.

The results confirm two central hypotheses. First, a compact generative model composed of interpretable modules for spectral sound processing can achieve speech enhancement performance on par with more complex data-driven systems. Second, the \texttt{RxInfer.jl} framework provides a practical platform for automated low-latency inference in such dynamic models.

Two major research directions now emerge. The first is the development of an end user model that enables intuitive, in-situ personalization through lightweight user interactions. The second is the advancement of message passing update rules for inference with non-conjugate factors, which will allow complete offline learning of all model parameters. Further refinement of the hierarchical structure in the SEM and ACM modules is also expected to enhance noise estimation and improve user experience in diverse listening environments.

Taken together, these developments suggest a path toward a new class of hearing aids: devices that are not merely tuned by human experts, but that are completely data-driven and adapt through continual probabilistic inference.

\section{REPRODUCIBILITY}
\label{sec:reproducibility}

All code for the model specifications, variational message passing, evaluation scripts, and audio demos will be released under an MIT license at:
\url{https://github.com/biaslab/Publication_Spectral_Subtraction/releases/tag/v1.0.0}; (\texttt{commit: 3f80595 }).
Artifacts include: (i) reference implementations of the BLIs and logit-node updates, (ii) configuration files matching Table~\ref{tab:comparison_with_params} and the parameter table in Section~\ref{sec:experiments}, and (iii) scripts to reproduce VoiceBank+DEMAND scores and generate listening examples.

\section*{Acknowledgments}
The authors thank the members of the BIASlab team and Dr. Tanya Ignatenko from GN Advanced Science for their support and valuable discussions throughout this project.

This publication is part of the project ``ROBUST: Trustworthy AI-based Systems for Sustainable Growth'' with project number KICH3.LTP.20.006, which is (partly) financed by the Dutch Research Council (NWO), GN Hearing, and the Dutch Ministry of Economic Affairs and Climate Policy (EZK) under the program LTP KIC 2020--2023.

\bibliographystyle{unsrt}
\bibliography{BIBLIOGRAPHY/biaslab}

@article{bagaevRxInferJuliaPackage2023,
  title = {{{RxInfer}}: {{A Julia}} Package for Reactive Real-Time {{Bayesian}} Inference},
  shorttitle = {{{RxInfer}}},
  author = {Bagaev, Dmitry and Podusenko, Albert and De Vries, Bert},
  year = 2023,
  journal = {Journal of Open Source Software},
  volume = {8},
  number = {84},
  pages = {5161},
  issn = {2475-9066},
  doi = {10.21105/joss.05161},
  langid = {english}
}

@inproceedings{beroutiEnhancementSpeechCorrupted1979,
  title = {Enhancement of Speech Corrupted by Acoustic Noise},
  booktitle = {{{ICASSP}} '79. {{IEEE International Conference}} on {{Acoustics}}, {{Speech}}, and {{Signal Processing}}},
  author = {Berouti, M. and Schwartz, R. and Makhoul, J.},
  year = 1979,
  volume = {4},
  pages = {208--211},
  doi = {10.1109/ICASSP.1979.1170788},
  urldate = {2015-11-16}
}

@book{bishopPatternRecognitionMachine2006,
  title = {Pattern {{Recognition}} and {{Machine Learning}} ({{Information Science}} and {{Statistics}})},
  author = {Bishop, Christopher M.},
  year = 2006,
  publisher = {Springer-Verlag},
  urldate = {2014-04-10},
  isbn = {0-387-31073-8}
}

@article{bollSuppressionAcousticNoise1979,
  title = {Suppression of Acoustic Noise in Speech Using Spectral Subtraction},
  author = {Boll, S.},
  year = 1979,
  month = apr,
  journal = {IEEE Transactions on Acoustics, Speech and Signal Processing},
  volume = {27},
  number = {2},
  pages = {113--120},
  issn = {0096-3518},
  doi = {10.1109/TASSP.1979.1163209}
}

@book{butcherNumericalMethodsOrdinary2016,
  title = {Numerical {{Methods}} for {{Ordinary Differential Equations}}},
  author = {Butcher, John C.},
  year = 2016,
  edition = {3},
  publisher = {John Wiley \& Sons, Ltd},
  doi = {10.1002/9781119121534},
  urldate = {2025-10-28},
  isbn = {978-1-119-12153-4}
}

@inproceedings{chenMetricOrientedSpeechEnhancement2023b,
  title = {Metric-{{Oriented Speech Enhancement Using Diffusion Probabilistic Model}}},
  booktitle = {{{ICASSP}} 2023 - 2023 {{IEEE International Conference}} on {{Acoustics}}, {{Speech}} and {{Signal Processing}} ({{ICASSP}})},
  author = {Chen, Chen and Hu, Yuchen and Weng, Weiwei and Chng, Eng Siong},
  year = 2023,
  month = jun,
  pages = {1--5},
  issn = {2379-190X},
  doi = {10.1109/ICASSP49357.2023.10095046},
  urldate = {2025-09-05}
}

@article{cohenNoiseEstimationMinima2002,
  title = {Noise Estimation by Minima Controlled Recursive Averaging for Robust Speech Enhancement},
  author = {Cohen, Israel and Berdugo, Baruch},
  year = 2002,
  journal = {IEEE Signal Processing Letters},
  volume = {9},
  number = {1},
  doi = {10.1109/97.988717},
  langid = {english}
}

@article{cohenNoiseSpectrumEstimation2003,
  title = {Noise Spectrum Estimation in Adverse Environments: Improved Minima Controlled Recursive Averaging},
  shorttitle = {Noise Spectrum Estimation in Adverse Environments},
  author = {Cohen, I.},
  year = 2003,
  month = sep,
  journal = {IEEE Transactions on Speech and Audio Processing},
  volume = {11},
  number = {5},
  pages = {466--475},
  issn = {1558-2353},
  doi = {10.1109/TSA.2003.811544},
  urldate = {2025-06-26}
}

@article{cohenSpeechEnhancementNonstationary2001,
  title = {Speech Enhancement for Non-Stationary Noise Environments},
  author = {Cohen, Israel and Berdugo, Baruch},
  year = 2001,
  journal = {Signal Processing},
  issn = {0165-1684},
  doi = {10.1016/S0165-1684(01)00128-1},
  langid = {english}
}

@article{coxFactorGraphApproach2019,
  title = {A Factor Graph Approach to Automated Design of {{Bayesian}} Signal Processing Algorithms},
  author = {Cox, Marco and {van de Laar}, Thijs and {de Vries}, Bert},
  year = 2019,
  month = jan,
  journal = {International Journal of Approximate Reasoning},
  volume = {104},
  pages = {185--204},
  issn = {0888-613X},
  doi = {10.1016/j.ijar.2018.11.002},
  urldate = {2018-11-16}
}

@inproceedings{coxForneyLabJlFast2018,
  title = {{{ForneyLab}}.Jl: {{Fast}} and Flexible Automated Inference through Message Passing in {{Julia}}},
  booktitle = {International {{Conference}} on {{Probabilistic Programming}}},
  author = {Cox, Marco and {van de Laar}, Thijs and {de Vries}, Bert},
  year = 2018,
  month = oct,
  address = {Boston, MA}
}

@misc{coxOverview2023Icassp2023,
  title = {Overview {{Of The}} 2023 {{Icassp Sp Clarity Challenge}}: {{Speech Enhancement For Hearing Aids}}},
  shorttitle = {Overview {{Of The}} 2023 {{Icassp Sp Clarity Challenge}}},
  author = {Cox, Trevor J. and Barker, Jon and Bailey, Will and Graetzer, Simone and Akeroyd, Michael A. and Culling, John F. and Naylor, Graham},
  year = 2023,
  month = nov,
  number = {arXiv:2311.14490},
  eprint = {2311.14490},
  primaryclass = {cs},
  publisher = {arXiv},
  doi = {10.48550/arXiv.2311.14490},
  urldate = {2025-07-29},
  archiveprefix = {arXiv}
}

@article{damOptimizedSigmoidFunctions2024,
  title = {Optimized {{Sigmoid Functions}} for {{Speech Presence Probability}} and {{Gain Function}} in {{Speech Enhancement}}},
  author = {Dam, Hai Huyen and Nordholm, Sven and Yong, Pei Chee and Low, Siow Yong},
  year = 2024,
  month = may,
  journal = {Circuits, Systems, and Signal Processing},
  volume = {43},
  number = {5},
  pages = {2891--2908},
  issn = {1531-5878},
  doi = {10.1007/s00034-023-02549-2},
  urldate = {2025-06-27},
  langid = {english}
}

@inproceedings{dauwelsVariationalMessagePassing2007,
  title = {On {{Variational Message Passing}} on {{Factor Graphs}}},
  booktitle = {{{IEEE International Symposium}} on {{Information Theory}}},
  author = {Dauwels, Justin},
  year = 2007,
  month = jun,
  pages = {2546--2550},
  address = {Nice, France},
  doi = {10.1109/ISIT.2007.4557602}
}

@inproceedings{defossezRealTimeSpeech2020,
  title = {Real {{Time Speech Enhancement}} in the {{Waveform Domain}}},
  booktitle = {Interspeech 2020},
  author = {D{\'e}fossez, Alexandre and Synnaeve, Gabriel and Adi, Yossi},
  year = 2020,
  month = oct,
  pages = {3291--3295},
  publisher = {ISCA},
  doi = {10.21437/Interspeech.2020-2409},
  urldate = {2025-09-09},
  langid = {english}
}

@article{dillonAdoptionUseNonuse2020b,
  title = {Adoption, Use and Non-Use of Hearing Aids: A Robust Estimate Based on {{Welsh}} National Survey Statistics},
  shorttitle = {Adoption, Use and Non-Use of Hearing Aids},
  author = {Dillon, Harvey and Day, John and Bant, Sarah and Munro, Kevin J},
  year = 2020,
  month = jul,
  journal = {International Journal of Audiology},
  volume = {59},
  number = {8},
  pages = {567--573},
  publisher = {Taylor \& Francis},
  issn = {1499-2027},
  doi = {10.1080/14992027.2020.1773550},
  urldate = {2025-07-29},
  pmid = {32530329}
}

@article{ephraimSpeechEnhancementUsing1984,
  title = {Speech Enhancement Using a Minimum-Mean Square Error Short-Time Spectral Amplitude Estimator},
  author = {Ephraim, Yariv and Malah, David},
  year = 1984,
  journal = {IEEE Transactions on Acoustics, Speech, and Signal Processing},
  volume = {32},
  number = {6},
  pages = {1109--1121},
  doi = {10.1109/TASSP.1984.1164453},
  urldate = {2015-11-23}
}

@article{ephraimSpeechEnhancementUsing1985,
  title = {Speech Enhancement Using a Minimum Mean-Square Error Log-Spectral Amplitude Estimator},
  author = {Ephraim, Yariv and Malah, David},
  year = 1985,
  month = apr,
  journal = {IEEE Transactions on Acoustics, Speech, and Signal Processing},
  volume = {33},
  number = {2},
  pages = {443--445},
  issn = {0096-3518},
  doi = {10.1109/TASSP.1985.1164550},
  urldate = {2025-06-26}
}

@inproceedings{fedorovTinyLSTMsEfficientNeural2020,
  title = {{{TinyLSTMs}}: {{Efficient Neural Speech Enhancement}} for {{Hearing Aids}}},
  shorttitle = {{{TinyLSTMs}}},
  booktitle = {Interspeech 2020},
  author = {Fedorov, Igor and Stamenovic, Marko and Jensen, Carl and Yang, Li-Chia and Mandell, Ari and Gan, Yiming and Mattina, Matthew and Whatmough, Paul N.},
  year = 2020,
  month = oct,
  pages = {4054--4058},
  publisher = {ISCA},
  doi = {10.21437/Interspeech.2020-1864},
  urldate = {2025-07-29},
  langid = {english}
}

@article{fuentes-lopezAssociationHometohealthcareCenter2024,
  title = {Association between the Home-to-Healthcare Center Distance and Hearing Aid Abandonment among Older Adults},
  author = {{Fuentes-L{\'o}pez}, Eduardo and {Galaz-Mella}, Javier and Ayala, Salvador and {De la Fuente}, Carlos and {Luna-Monsalve}, Manuel and Nieman, Carrie and Marcotti, Anthony},
  year = 2024,
  month = may,
  journal = {Frontiers in Public Health},
  volume = {12},
  publisher = {Frontiers},
  issn = {2296-2565},
  doi = {10.3389/fpubh.2024.1364000},
  urldate = {2025-07-29},
  langid = {english}
}

@article{hsiehWaveCRNEfficientConvolutional2020,
  title = {{{WaveCRN}}: {{An Efficient Convolutional Recurrent Neural Network}} for {{End-to-End Speech Enhancement}}},
  shorttitle = {{{WaveCRN}}},
  author = {Hsieh, Tsun-An and Wang, Hsin-Min and Lu, Xugang and Tsao, Yu},
  year = 2020,
  journal = {IEEE Signal Processing Letters},
  volume = {27},
  pages = {2149--2153},
  issn = {1558-2361},
  doi = {10.1109/LSP.2020.3040693},
  urldate = {2025-09-09}
}

@techreport{internationaltelecommunicationunionP8622WidebandExtension2005,
  title = {P.862.2: {{Wideband}} Extension to {{Recommendation P}}.862 for the Assessment of Wideband Telephone Networks and Speech Codecs},
  author = {{International Telecommunication Union}},
  year = 2005,
  institution = {International Telecommunication Union}
}

@inproceedings{jaakkolaVariationalApproachBayesian1997a,
  title = {A {{Variational Approach}} to {{Bayesian Logistic Regression Models}} and Their {{Extensions}}},
  booktitle = {Sixth {{International Workshop}} on {{Artificial Intelligence}} and {{Statistics}}},
  author = {Jaakkola, Tommi S. and Jordan, Michael I.},
  year = 1997,
  month = jan,
  pages = {283--294},
  publisher = {PMLR},
  issn = {2640-3498},
  urldate = {2025-12-02},
  langid = {english}
}

@incollection{jordanIntroductionVariationalMethods1998,
  title = {An {{Introduction}} to {{Variational Methods}} for {{Graphical Models}}},
  booktitle = {Learning in {{Graphical Models}}},
  author = {Jordan, Michael I. and Ghahramani, Zoubin and Jaakkola, Tommi S. and Saul, Lawrence K.},
  editor = {Jordan, Michael I.},
  year = 1998,
  pages = {105--161},
  publisher = {Springer Netherlands},
  address = {Dordrecht},
  doi = {10.1007/978-94-011-5014-9_5},
  urldate = {2025-12-01},
  isbn = {978-94-010-6104-9 978-94-011-5014-9},
  langid = {english}
}

@inproceedings{kamathMultibandSpectralSubtraction2002,
  title = {A Multi-Band Spectral Subtraction Method for Enhancing Speech Corrupted by Colored Noise},
  booktitle = {2002 {{IEEE International Conference}} on {{Acoustics}}, {{Speech}}, and {{Signal Processing}}},
  author = {Kamath, Sunil and Loizou, Philipos},
  year = 2002,
  month = may,
  volume = {4},
  pages = {IV-4164-IV-4164},
  issn = {1520-6149},
  doi = {10.1109/ICASSP.2002.5745591},
  urldate = {2025-06-26},
  isbn = {0-7803-7402-9}
}

@article{kamilEffectsHearingImpairment2015,
  title = {The {{Effects}} of {{Hearing Impairment}} in {{Older Adults}} on {{Communication Partners}}: {{A Systematic Review}}},
  shorttitle = {The {{Effects}} of {{Hearing Impairment}} in {{Older Adults}} on {{Communication Partners}}},
  author = {Kamil, Rebecca J. and Lin, Frank R.},
  year = 2015,
  month = feb,
  journal = {Journal of the American Academy of Audiology},
  volume = {26},
  number = {2},
  pages = {155--182},
  issn = {1050-0545},
  doi = {10.3766/jaaa.26.2.6},
  urldate = {2025-07-29},
  langid = {english}
}

@inproceedings{katesDynamicrangeCompressionUsing2003,
  title = {Dynamic-Range Compression Using Digital Frequency Warping},
  booktitle = {The {{Thrity-Seventh Asilomar Conference}} on {{Signals}}, {{Systems}} \& {{Computers}}, 2003},
  author = {Kates, J.M.},
  year = 2003,
  month = nov,
  volume = {1},
  pages = {715-719 Vol.1},
  doi = {10.1109/ACSSC.2003.1292007},
  urldate = {2024-06-11}
}

@inproceedings{kimMambabasedHybridModel2025,
  title = {Mamba-Based {{Hybrid Model}} for {{Speech Enhancement}}},
  booktitle = {Proc. {{Interspeech}} 2025},
  author = {Kim, Se-Ha and Kim, Tae-Gyeong and Chun, Chang-Jae},
  year = 2025,
  pages = {5163--5167},
  doi = {10.21437/Interspeech.2025-1476},
  urldate = {2025-09-09},
  langid = {english}
}

@article{kjemsRoleMaskPattern2009,
  title = {Role of Mask Pattern in Intelligibility of Ideal Binary-Masked Noisy Speech},
  author = {Kjems, Ulrik and Boldt, Jesper B. and Pedersen, Michael S. and Lunner, Thomas and Wang, DeLiang},
  year = 2009,
  month = sep,
  journal = {The Journal of the Acoustical Society of America},
  volume = {126},
  number = {3},
  pages = {1415--1426},
  issn = {0001-4966},
  doi = {10.1121/1.3179673},
  urldate = {2025-10-03}
}

@article{kschischangFactorGraphsSumproduct2001a,
  title = {Factor Graphs and the Sum-Product Algorithm},
  author = {Kschischang, F.R. and Frey, B.J. and Loeliger, H.-A.},
  year = 2001,
  month = feb,
  journal = {IEEE Transactions on Information Theory},
  volume = {47},
  number = {2},
  pages = {498--519},
  issn = {1557-9654},
  doi = {10.1109/18.910572},
  urldate = {2025-12-01}
}

@inproceedings{laineWarpedLinearPrediction1994,
  title = {Warped Linear Prediction ({{WLP}}) in Speech and Audio Processing},
  booktitle = {Proceedings of {{ICASSP}} '94. {{IEEE International Conference}} on {{Acoustics}}, {{Speech}} and {{Signal Processing}}},
  author = {Laine, U.K. and Karjalainen, M. and Altosaar, T.},
  year = 1994,
  month = apr,
  volume = {iii},
  pages = {III/349-III/352 vol.3},
  issn = {1520-6149},
  doi = {10.1109/ICASSP.1994.390018},
  urldate = {2025-12-01}
}

@article{limEnhancementBandwidthCompression1979,
  title = {Enhancement and Bandwidth Compression of Noisy Speech},
  author = {Lim, J.S. and Oppenheim, A.V.},
  year = 1979,
  month = dec,
  journal = {Proceedings of the IEEE},
  volume = {67},
  number = {12},
  pages = {1586--1604},
  issn = {1558-2256},
  doi = {10.1109/PROC.1979.11540},
  urldate = {2025-06-27}
}

@article{linHearingLossCognition2011,
  title = {Hearing Loss and Cognition among Older Adults in the {{United States}}},
  author = {Lin, Frank R.},
  year = 2011,
  month = oct,
  journal = {The Journals of Gerontology. Series A, Biological Sciences and Medical Sciences},
  volume = {66},
  number = {10},
  pages = {1131--1136},
  issn = {1758-535X},
  doi = {10.1093/gerona/glr115},
  langid = {english},
  pmcid = {PMC3172566},
  pmid = {21768501}
}

@inproceedings{liOptimalityIdealBinary2008,
  title = {On the Optimality of Ideal Binary Time-Frequency Masks},
  booktitle = {2008 {{IEEE International Conference}} on {{Acoustics}}, {{Speech}} and {{Signal Processing}}},
  author = {Li, Yipeng and Wang, DeLiang},
  year = 2008,
  month = mar,
  pages = {3501--3504},
  issn = {2379-190X},
  doi = {10.1109/ICASSP.2008.4518406},
  urldate = {2025-10-03}
}

@article{loeligerIntroductionFactorGraphs2004,
  title = {An Introduction to Factor Graphs},
  author = {Loeliger, Hans-Andrea},
  year = 2004,
  month = jan,
  journal = {Signal Processing Magazine, IEEE},
  volume = {21},
  number = {1},
  pages = {28--41},
  doi = {10.1109/MSP.2004.1267047},
  urldate = {2014-04-10}
}

@inproceedings{luConditionalDiffusionProbabilistic2022,
  title = {Conditional {{Diffusion Probabilistic Model}} for {{Speech Enhancement}}},
  booktitle = {{{ICASSP}} 2022 - 2022 {{IEEE International Conference}} on {{Acoustics}}, {{Speech}} and {{Signal Processing}} ({{ICASSP}})},
  author = {Lu, Yen-Ju and Wang, Zhong-Qiu and Watanabe, Shinji and Richard, Alexander and Yu, Cheng and Tsao, Yu},
  year = 2022,
  month = may,
  pages = {7402--7406},
  issn = {2379-190X},
  doi = {10.1109/ICASSP43922.2022.9746901},
  urldate = {2023-12-27}
}

@article{luoConvTasNetSurpassingIdeal2019,
  title = {Conv-{{TasNet}}: {{Surpassing Ideal Time}}--{{Frequency Magnitude Masking}} for {{Speech Separation}}},
  shorttitle = {Conv-{{TasNet}}},
  author = {Luo, Yi and Mesgarani, Nima},
  year = 2019,
  month = aug,
  journal = {IEEE/ACM Transactions on Audio, Speech, and Language Processing},
  volume = {27},
  number = {8},
  pages = {1256--1266},
  issn = {2329-9304},
  doi = {10.1109/TASLP.2019.2915167},
  urldate = {2025-09-05}
}

@article{marcos-alonsoFactorsImpactingUse2023,
  title = {Factors {{Impacting}} the {{Use}} or {{Rejection}} of {{Hearing Aids}}---{{A Systematic Review}} and {{Meta-Analysis}}},
  author = {{Marcos-Alonso}, Susana and {Almeida-Ayerve}, Cristina Nicole and {Monopoli-Roca}, Chiara and {Coronel-Touma}, Guillermo Salib and {Pacheco-L{\'o}pez}, Sof{\'i}a and {Pe{\~n}a-Navarro}, Paula and {Serradilla-L{\'o}pez}, Jos{\'e} Manuel and {S{\'a}nchez-G{\'o}mez}, Hortensia and {Pardal-Refoyo}, Jos{\'e} Luis and {Batuecas-Caletr{\'i}o}, {\'A}ngel},
  year = 2023,
  month = jan,
  journal = {Journal of Clinical Medicine},
  volume = {12},
  number = {12},
  pages = {4030},
  publisher = {Multidisciplinary Digital Publishing Institute},
  issn = {2077-0383},
  doi = {10.3390/jcm12124030},
  urldate = {2025-07-29},
  copyright = {http://creativecommons.org/licenses/by/3.0/},
  langid = {english}
}

@article{marcottiAssociationUnaidedSpeech2025,
  title = {Association {{Between Unaided Speech Perception}} in {{Noise}} and {{Hearing Aid Use Mediated}} by {{Perceived Benefit}}},
  author = {Marcotti, Anthony and {Silva-Letelier}, Catherine and {Galaz-Mella}, Javier and Ianiszewski, Alejandro and Vargas, Nicole B. and {Fuentes-L{\'o}pez}, Eduardo},
  year = 2025,
  month = jun,
  journal = {Audiology Research},
  volume = {15},
  number = {3},
  pages = {50},
  publisher = {Multidisciplinary Digital Publishing Institute},
  issn = {2039-4349},
  doi = {10.3390/audiolres15030050},
  urldate = {2025-07-29},
  copyright = {http://creativecommons.org/licenses/by/3.0/},
  langid = {english}
}

@article{martinNoisePowerSpectral2001,
  title = {Noise Power Spectral Density Estimation Based on Optimal Smoothing and Minimum Statistics},
  author = {Martin, R.},
  year = 2001,
  month = jul,
  journal = {IEEE Transactions on Speech and Audio Processing},
  volume = {9},
  number = {5},
  pages = {504--512},
  issn = {1558-2353},
  doi = {10.1109/89.928915},
  urldate = {2025-06-26}
}

@article{mcdaidEstimatingGlobalCosts2021,
  title = {Estimating the Global Costs of Hearing Loss},
  author = {McDaid, David and Park, A-La and Chadha, Shelly},
  year = 2021,
  month = mar,
  journal = {International Journal of Audiology},
  volume = {60},
  number = {3},
  pages = {162--170},
  publisher = {Taylor \& Francis},
  issn = {1499-2027},
  doi = {10.1080/14992027.2021.1883197},
  urldate = {2025-09-04},
  pmid = {33590787}
}

@misc{miaoPesqPythonWrapper,
  title = {Pesq: {{Python Wrapper}} for {{PESQ Score}} (Narrow Band and Wide Band)},
  shorttitle = {Pesq},
  author = {Miao, Wang and Boeddeker, Christoph and Dantas, Rafael and Seelan, Ananda},
  urldate = {2025-11-25},
  copyright = {OSI Approved :: MIT License}
}

@techreport{murphySwitchingKalmanFilters1998,
  title = {Switching {{Kalman}} Filters},
  author = {Murphy, Kevin},
  year = 1998,
  pages = {21},
  institution = {University of California, Berkeley},
  urldate = {2025-07-22}
}

@article{nachtegaalHearingAbilityWorking2012,
  title = {Hearing Ability in Working Life and Its Relationship with Sick Leave and Self-Reported Work Productivity},
  author = {Nachtegaal, Janneke and Festen, Joost M. and Kramer, Sophia E.},
  year = 2012,
  journal = {Ear and hearing},
  volume = {33},
  number = {1},
  pages = {94--103},
  publisher = {LWW},
  doi = {10.1097/AUD.0b013e318228033e},
  urldate = {2025-07-29}
}

@article{orjiGlobalRegionalNeeds2020a,
  title = {Global and Regional Needs, Unmet Needs and Access to Hearing Aids},
  author = {Orji, Aislyn and Kamenov, Kaloyan and Dirac, Mae and Davis, Adrian and Chadha, Shelly and Vos, Theo},
  year = 2020,
  month = mar,
  journal = {International Journal of Audiology},
  volume = {59},
  number = {3},
  pages = {166--172},
  issn = {1708-8186},
  doi = {10.1080/14992027.2020.1721577},
  langid = {english},
  pmid = {32011190}
}

@inproceedings{pascualSEGANSpeechEnhancement2017,
  title = {{{SEGAN}}: {{Speech Enhancement Generative Adversarial Network}}},
  shorttitle = {{{SEGAN}}},
  booktitle = {Interspeech 2017},
  author = {Pascual, Santiago and Bonafonte, Antonio and Serr{\`a}, Joan},
  year = 2017,
  month = aug,
  pages = {3642--3646},
  publisher = {ISCA},
  doi = {10.21437/Interspeech.2017-1428},
  urldate = {2025-09-09},
  langid = {english}
}

@article{podusenkoAIDAActiveInferenceBased2022a,
  title = {{{AIDA}}: {{An Active Inference-Based Design Agent}} for {{Audio Processing Algorithms}}},
  shorttitle = {{{AIDA}}},
  author = {Podusenko, Albert and {van Erp}, Bart and Koudahl, Magnus and {de Vries}, Bert},
  year = 2022,
  month = mar,
  journal = {Frontiers in Signal Processing},
  volume = {2},
  pages = {842477},
  issn = {2673-8198},
  doi = {10.3389/frsip.2022.842477},
  urldate = {2022-09-16}
}

@inproceedings{reddyDnsmosP835NonIntrusive2022,
  title = {Dnsmos {{P}}.835: {{A Non-Intrusive Perceptual Objective Speech Quality Metric}} to {{Evaluate Noise Suppressors}}},
  shorttitle = {Dnsmos {{P}}.835},
  booktitle = {{{ICASSP}} 2022 - 2022 {{IEEE International Conference}} on {{Acoustics}}, {{Speech}} and {{Signal Processing}} ({{ICASSP}})},
  author = {Reddy, Chandan K A and Gopal, Vishak and Cutler, Ross},
  year = 2022,
  month = may,
  pages = {886--890},
  issn = {2379-190X},
  doi = {10.1109/ICASSP43922.2022.9746108},
  urldate = {2025-11-25}
}

@book{sarkkaBayesianFilteringSmoothing2013,
  title = {Bayesian {{Filtering}} and {{Smoothing}}},
  author = {S{\"a}rkk{\"a}, Simo},
  year = 2013,
  month = oct,
  publisher = {Cambridge University Press},
  isbn = {978-0-415-55809-9},
  langid = {english}
}

@inproceedings{scalartSpeechEnhancementBased1996,
  title = {Speech Enhancement Based on a Priori Signal to Noise Estimation},
  booktitle = {1996 {{IEEE International Conference}} on {{Acoustics}}, {{Speech}}, and {{Signal Processing Conference Proceedings}}},
  author = {Scalart, P. and Filho, J.V.},
  year = 1996,
  month = may,
  volume = {2},
  pages = {629-632 vol. 2},
  issn = {1520-6149},
  doi = {10.1109/ICASSP.1996.543199},
  urldate = {2025-10-29}
}

@inproceedings{scheiblerUniversalScorebasedSpeech2024,
  title = {Universal {{Score-based Speech Enhancement}} with {{High Content Preservation}}},
  booktitle = {Interspeech 2024},
  author = {Scheibler, Robin and Fujita, Yusuke and Shirahata, Yuma and Komatsu, Tatsuya},
  year = 2024,
  month = sep,
  pages = {1165--1169},
  publisher = {ISCA},
  doi = {10.21437/Interspeech.2024-138},
  urldate = {2025-09-05},
  langid = {english}
}

@article{senozVariationalMessagePassing2021,
  title = {Variational {{Message Passing}} and {{Local Constraint Manipulation}} in {{Factor Graphs}}},
  author = {{\c S}en{\"o}z, {\.I}smail and {van de Laar}, Thijs and Bagaev, Dmitry and {de Vries}, Bert},
  year = 2021,
  month = jul,
  journal = {Entropy},
  volume = {23},
  number = {7},
  pages = {807},
  publisher = {Multidisciplinary Digital Publishing Institute},
  issn = {1099-4300},
  doi = {10.3390/e23070807},
  urldate = {2022-03-15},
  copyright = {http://creativecommons.org/licenses/by/3.0/},
  langid = {english}
}

@article{smithBarkERBBilinear1999,
  title = {Bark and {{ERB}} Bilinear Transforms},
  author = {Smith, J.O. and Abel, J.S.},
  year = 1999,
  month = nov,
  journal = {IEEE Transactions on Speech and Audio Processing},
  volume = {7},
  number = {6},
  pages = {697--708},
  issn = {1558-2353},
  doi = {10.1109/89.799695},
  urldate = {2025-03-24}
}

@inproceedings{srinivasanComputationalAuditoryScene2006,
  title = {A Computational Auditory Scene Analysis System for Robust Speech Recognition},
  booktitle = {Interspeech 2006},
  author = {Srinivasan, Soundararajan and Shao, Yang and Jin, Zhaozhang and Wang, DeLiang},
  year = 2006,
  month = sep,
  pages = {paper 1547-Mon1WeS.1-0},
  publisher = {ISCA},
  doi = {10.21437/Interspeech.2006-19},
  urldate = {2025-10-05},
  langid = {english}
}

@article{upadhyaySingleChannelSpeech2016,
  title = {Single {{Channel Speech Enhancement}}: {{Using Wiener Filtering}} with {{Recursive Noise Estimation}}},
  shorttitle = {Single {{Channel Speech Enhancement}}},
  author = {Upadhyay, Navneet and Jaiswal, Rahul Kumar},
  year = 2016,
  month = jan,
  journal = {Procedia Computer Science},
  series = {Proceeding of the {{Seventh International Conference}} on {{Intelligent Human Computer Interaction}} ({{IHCI}} 2015)},
  volume = {84},
  pages = {22--30},
  issn = {1877-0509},
  doi = {10.1016/j.procs.2016.04.061},
  urldate = {2025-10-29}
}

@misc{valentini-botinhaoc.NoisySpeechDatabase2017,
  title = {Noisy Speech Database for Training Speech Enhancement Algorithms and {{TTS}} Models},
  author = {{Valentini-Botinhao, C.}},
  year = 2017,
  doi = {10.7488/ds/2117},
  langid = {english}
}

@article{viragSingleChannelSpeech1999,
  title = {Single Channel Speech Enhancement Based on Masking Properties of the Human Auditory System},
  author = {Virag, N.},
  year = 1999,
  month = mar,
  journal = {IEEE Transactions on Speech and Audio Processing},
  volume = {7},
  number = {2},
  pages = {126--137},
  issn = {1558-2353},
  doi = {10.1109/89.748118},
  urldate = {2025-06-26}
}

@incollection{wangIdealBinaryMask2005,
  title = {On {{Ideal Binary Mask As}} the {{Computational Goal}} of {{Auditory Scene Analysis}}},
  booktitle = {Speech {{Separation}} by {{Humans}} and {{Machines}}},
  author = {Wang, DeLiang},
  editor = {Divenyi, Pierre},
  year = 2005,
  pages = {181--197},
  publisher = {Springer US},
  address = {Boston, MA},
  doi = {10.1007/0-387-22794-6_12},
  urldate = {2025-10-03},
  isbn = {978-0-387-22794-8},
  langid = {english}
}

@article{wangTrainingTargetsSupervised2014,
  title = {On {{Training Targets}} for {{Supervised Speech Separation}}},
  author = {Wang, Yuxuan and Narayanan, Arun and Wang, DeLiang},
  year = 2014,
  month = dec,
  journal = {IEEE/ACM Transactions on Audio, Speech, and Language Processing},
  volume = {22},
  number = {12},
  pages = {1849--1858},
  issn = {2329-9304},
  doi = {10.1109/TASLP.2014.2352935},
  urldate = {2025-10-05}
}

@book{wienerExtrapolationInterpolationSmoothing1964,
  title = {Extrapolation, Interpolation, and Smoothing of Stationary Time Series},
  author = {Wiener, Norbert},
  year = 1964,
  publisher = {The MIT press},
  urldate = {2025-07-29}
}

@book{worldhealthorganizationWorldReportHearing2021,
  title = {World Report on Hearing},
  author = {{World Health Organization}},
  year = 2021,
  publisher = {World Health Organization},
  urldate = {2025-07-29},
  isbn = {92-4-002157-4},
  langid = {english}
}

@inproceedings{yangFspenUltraLightweightNetwork2024,
  title = {Fspen: An {{Ultra-Lightweight Network}} for {{Real Time Speech Enhancement}}},
  shorttitle = {Fspen},
  booktitle = {{{ICASSP}} 2024 - 2024 {{IEEE International Conference}} on {{Acoustics}}, {{Speech}} and {{Signal Processing}} ({{ICASSP}})},
  author = {Yang, Lei and Liu, Wei and Meng, Ruijie and Lee, Gunwoo and Baek, Soonho and Moon, Han-Gil},
  year = 2024,
  month = apr,
  pages = {10671--10675},
  issn = {2379-190X},
  doi = {10.1109/ICASSP48485.2024.10446016},
  urldate = {2025-09-05}
}

@article{yongOptimizationEvaluationSigmoid2013,
  title = {Optimization and Evaluation of Sigmoid Function with a Priori {{SNR}} Estimate for Real-Time Speech Enhancement},
  author = {Yong, Pei Chee and Nordholm, Sven and Dam, Hai Huyen},
  year = 2013,
  month = feb,
  journal = {Speech Communication},
  volume = {55},
  number = {2},
  pages = {358--376},
  issn = {0167-6393},
  doi = {10.1016/j.specom.2012.09.004},
  urldate = {2025-06-27}
}

@inproceedings{lukashchuk2025quotient,
  title     = {The Quotient Bayesian Learning Rule},
  author    = {Lukashchuk, Mykola and Tr{\'e}sor, Rapha{\"e}l and Nuijten, Wouter W. L. and {\.{I}}smail {\c{S}en\"{o}z} and de Vries, Bert},
  booktitle = {Advances in Neural Information Processing Systems},
  note      = {NeurIPS 2025 poster},
  year      = {2025},
  url       = {https://openreview.net/forum?id=XDisynd63Y}
}

\appendix

\section{THE WARPED-FREQUENCY FILTER BANK}
\label{app:wfb-details}

This appendix complements Section~\ref{sec:WFB}. We avoid restating the model and only record implementation details that are helpful for reproducing the front-end. Notation follows the main text: $x_k$ is the input, $y_t$ the output, $w_m$ the blockwise weight vector, and $z_{kj}$ the warped delay-line states generated by the all-pass cascade in \eqref{eq:all-pass-FB} and the output is produced by \eqref{eq:FIR-filter}.

\subsection{IMPLEMENTATION}
In the time domain, passing $z_{k,j-1}$ through $A(q^{-1})$ to generate the next tap $z_{kj}$ involves computing the following updates,
\begin{subequations}\label{eq:all-pass-recursive}
  \begin{align}
    z_{kj} &= v_{k-1,j} -\alpha z_{k,j-1} \\
    v_{kj}
 &= z_{k,j-1} + \alpha z_{kj} \,.
 \end{align}  
\end{subequations}

Here, each all-pass filter $A(q^{-1})$ maintains an internal state variable $v_{kj}$ that preserves a memory trace of the input signal. The delay line of cascaded all-pass filters has only one parameter, $0 \leq \alpha < 1$, which controls the phase shifts throughout the delay line. 

\subsection{SPECTRAL ANALYSIS \& SYNTHESIS}
The WFB front-end does not perform spectral analysis or synthesis. Those operations are part of the SEM (see Section~\ref{sec:model-specification}). The interface is as follows:
(i) the front-end exposes the warped state vector $z_m \in \mathbb{R}^J$ once per block to the SEM;
(ii) the SEM returns an updated weight vector $w_m \in \mathbb{R}^J$;
(iii) the front-end applies $w_m$ to the time-domain delay-line states to generate $y_t$ as defined in \eqref{eq:FIR-filter}.
This separation keeps the primary signal path strictly time-domain and causal, and leaves filter weight inference to the SEM.

\subsection{PARAMETER SELECTION}

Setting $\alpha=0$ reduces \eqref{eq:WFB} to a regular FIR filter. However, to achieve perceptually-motivated frequency warping, $\alpha$ is typically chosen to align the frequency transformation in $A(q^{-1})$ with the measured frequency warping of acoustic signals in the human cochlea. 

The warping parameter $\alpha$ determines the degree of frequency compression in the all-pass cascade and is crucial for achieving perceptually motivated frequency resolution. For auditory applications, Smith \textit{et al.} \cite{smithBarkERBBilinear1999} proposed a sampling-rate-dependent formula to approximate perceptual frequency scales such as the Bark scale

\begin{equation}\label{eq:alpha-formula}
\alpha = \frac{1.0674 \sqrt{\frac{2}{\pi} \arctan(0.06583 f_s)} - 0.1916}{1 + 1.0674 \sqrt{\frac{2}{\pi} \arctan(0.06583 f_s)}},
\end{equation}
where $f_s$ is the sampling frequency in Hz.
For a sampling rate of $f_s=16$ kHz, Smith \textit{et al.} \cite{smithBarkERBBilinear1999} showed that $\alpha=0.58$ gives a close match to Bark-scale frequency warping. For simplicity and computational ease, we choose $\alpha=0.5$.

\begin{figure*}[!htbp]
\centering
\scalebox{0.6}{\begin{tikzpicture}
    [node distance=15mm,auto,>=stealth',
every node/.append style={font=\Large}]
\node[smallbox](eq1_zk1){$=$};
\node[below=20mm of eq1_zk1](observation_x){$x_k$};
\draw[->, thick](observation_x)--(eq1_zk1);

\node[smallbox, above=15mm of eq1_zk1](eq_zk1_warping){$=$};
\draw[->, thick](eq1_zk1.north) -- (eq_zk1_warping.south);
\node[
draw, 
regular polygon, 
regular polygon sides=3, 
minimum size=0.2cm, 
rotate=0, above=10mm of eq_zk1_warping
] (gain_ap_1) {};
\node[left=2mm of gain_ap_1]{$-\alpha$};
\draw[->, thick](eq_zk1_warping.north) -- (gain_ap_1.south);

\node[smallbox,above=10mm of gain_ap_1](plus1){$+$};
\draw[->, thick](gain_ap_1.north) -- (plus1.south);

\node[smallbox, above=25mm of plus1](dot_a1){$=$};
\draw[->, thick](plus1.north) -- (dot_a1.south);
\node[box, left=10mm of plus1](q1){$q^{-1}$};
\node[smallbox, left=10mm of q1](plus2){$+$};

\node[
draw, 
regular polygon, 
regular polygon sides=3, 
minimum size=0.1cm, 
rotate=180, above=15mm of plus2
] (gain_ap_2) {};

\draw[->, thick](dot_a1.west) -| (gain_ap_2.south);
\node[right=5mm of gain_ap_2]{$\alpha$};

\draw[->, thick](gain_ap_2.north) -| (plus2.north);
\draw[->, thick](plus2.south) |- (eq_zk1_warping.west);

\draw[->, thick](plus2.east) -- (q1.west);
\draw[->, thick](q1.east) -- (plus1.west);

\draw[dashed]
([xshift=-10mm, yshift=23mm] gain_ap_2.west) rectangle 
([xshift=10mm, yshift=-13mm] eq_zk1_warping.south);
\node[align=center,xshift=-9mm, yshift=-22mm] at ($(gain_ap_2)!0.7!(eq_zk1_warping)$) {\shortstack{All-Pass \\Recursion \eqref{eq:all-pass-recursive}}};


\node[smallbox, above=15mm of dot_a1](eq1_zk2){$=$};
\draw[->, thick](dot_a1.north) -- node[left,yshift=3mm]{$z_{k2}$}(eq1_zk2.south);

\node[box,above=15mm of eq1_zk2](q2){$A(q^-1)$};
\draw[->, thick](eq1_zk2.north) -- (q2.south);

\node[smallbox, above=15mm of q2](eq1_zk3){$=$};
\draw[->, thick](q2.north) -- node[left]{$z_{k3}$} (eq1_zk3.south);

\node[above=15mm of eq1_zk3](eq1_zk3_inv){$\cdots$};
\draw[->, thick](eq1_zk3.north) -- (eq1_zk3_inv.south);

\node[above=15mm of eq1_zk3_inv](eq1_zk3_inv2){};

\node[box, above=15mm of eq1_zk3_inv](q3){$A(q^-1)$};
\draw[->, thick](eq1_zk3_inv) -- (q3);

\node[smallbox, right=15mm of eq1_zk1](eq2_zk1){$=$};
\node[smallbox, right=32.5mm of eq1_zk2](eq2_zk2){$=$};
\node[smallbox, right=50mm of eq1_zk3](eq2_zk3){$=$};
\node[smallbox, above right=15 mm and 61.5mm of q3](eq1_zkJ){$=$};

\node[above=230mm of eq2_zk1](zk1_to_SEM){};
\node[right=15mm of zk1_to_SEM](zk2_to_SEM){};
\node[right=15mm of zk2_to_SEM](zk3_to_SEM){};
\node[right=15mm of zk3_to_SEM](zkJ_to_SEM){};
\draw[->, thick](eq2_zk1.north) -- (zk1_to_SEM)node[right]{$z_{k1}$};
\draw[->, thick](eq2_zk2.north) --  (zk2_to_SEM)node[right]{$z_{k2}$};
\draw[->, thick](eq2_zk3.north) -- (zk3_to_SEM)node[right]{$z_{k3} \cdots$};
\draw[->, thick](eq1_zkJ.north) -- (zkJ_to_SEM)node[right]{$z_{kJ}$};

\node[above=16mm of q3](q3_inv){};

\draw[-, thick](q3.north)--node[left]{$z_{kJ}$}(q3_inv.center);

\draw[->, thick]
  (q3_inv.center)
  to[
    jump=(eq2_zk1)--(zk1_to_SEM),
    jumps=left
  ] ($ (q3_inv.center)!0.33!(eq1_zkJ.west) $)
  to[
    jump=(eq2_zk2)--(zk2_to_SEM),
    jumps=left
  ] ($ (q3_inv.center)!0.66!(eq1_zkJ.west) $)
  to[
    jump=(eq2_zk3)--(zk3_to_SEM),
    jumps=left
  ] (eq1_zkJ.west);


\draw[dashed]
([xshift=10mm, yshift=10mm] eq1_zkJ.north) rectangle 
([xshift=-50mm, yshift=-10mm] eq1_zk1.south);
\node[align=center,xshift=-57mm, yshift=110mm] at ($(eq1_zkJ)!0.5!(eq1_zk1)$) {All-Pass Filter Bank \eqref{eq:all-pass-FB}};

\node[smallbox, right=70mm of eq2_zk1](mul_zk1){$\times$};
\node[smallbox,right=70mm of eq2_zk2](mul_zk2){$\times$};
\node[smallbox, right=70mm of eq2_zk3](mul_zk3){$\times$};
\node[smallbox, right=70mm of eq1_zkJ](mul_zkJ){$\times$};

\node[above=230mm of mul_zk1](wk1_to_WFB){};
\node[right=15mm of wk1_to_WFB](wk2_to_WFB){};
\node[right=15mm of wk2_to_WFB](wk3_to_WFB){};
\node[right=15mm of wk3_to_WFB](wkJ_to_WFB){};
\draw[<-, thick](mul_zk1.north) -- (wk1_to_WFB)node[right]{$w_{m1}$};
\draw[<-, thick](mul_zk2.north) --(wk2_to_WFB)node[right]{$w_{m2}$};
\draw[<-, thick](mul_zk3.north) -- (wk3_to_WFB)node[right]{$w_{m3} \cdots$};
\draw[<-, thick](mul_zkJ.north) --(wkJ_to_WFB)node[right]{$w_{mJ}$};

\node[smallbox, right=30mm of mul_zk3](plus_w_k3){$+$};
\node[smallbox,right=47.5mm of mul_zk2](plus_w_k2){$+$};
\node[smallbox, right=65mm of mul_zk1](plus_w_k1){$+$};

\draw[->, thick](eq1_zk2.east)to [jump=(eq2_zk1)--(zk1_to_SEM), jumps=left](eq2_zk2.west);

\draw[->, thick]
  (eq1_zk3.east)
  to[
    jump=(eq2_zk1)--(zk1_to_SEM),
    jumps=left
  ] ($ (eq1_zk3.east)!0.5!(eq2_zk3.west) $) 
  to[
    jump=(eq2_zk2)--(zk2_to_SEM),
    jumps=left
  ] (eq2_zk3.west);

\draw[->, thick](eq1_zk1.east)--(eq2_zk1.west)node[midway,below,yshift=-3mm]{$z_{k1} =x_{k}$};
\draw[->, thick](eq2_zk1.east)--(mul_zk1.west);

\draw[->, thick](eq2_zk2.east)to[
    jump=(mul_zk1)--(wk1_to_WFB),
    jumps=left
  ](mul_zk2.west);

\draw[->, thick](eq2_zk3.east)to[
    jump=(mul_zk1)--(wk1_to_WFB),
    jumps=left
  ] ($ (eq2_zk3.east)!0.6!(mul_zk3.west) $) 
  to[
    jump=(mul_zk2)--(wk2_to_WFB),
    jumps=left
  ](mul_zk3.west);
  
\draw[->, thick](eq1_zkJ.east)to[
    jump=(mul_zk1)--(wk1_to_WFB),
    jumps=left
  ] ($ (eq1_zkJ.east)!0.3!(mul_zkJ.west) $) 
  to[
    jump=(mul_zk2)--(wk2_to_WFB),
    jumps=left
  ]($ (eq1_zkJ.east)!0.7!(mul_zkJ.west) $)
  to[
    jump=(mul_zk3)--(wk3_to_WFB),
    jumps=left
  ] (mul_zkJ.west);

\draw[->, thick](mul_zk3)--(plus_w_k3);
\draw[->, thick](mul_zkJ)-|(plus_w_k3);
\draw[->, thick](plus_w_k3)--(plus_w_k2);

\draw[->, thick](mul_zk2)--(plus_w_k2);

\draw[->, thick](plus_w_k2)--(plus_w_k1);

\draw[->, thick](mul_zk1)--(plus_w_k1);

\node[below=20mm of plus_w_k1](y){$y_k$};
\draw[->, thick](plus_w_k1)--(y);

\draw[dashed]
([xshift=57mm, yshift=10mm] mul_zkJ.north) rectangle 
([xshift=-10mm, yshift=-10mm] mul_zk1.south);
\node[align=center,xshift=55mm, yshift=110mm] at ($(mul_zk1)!0.5!(mul_zkJ)$) {Warped FIR synthesis
 \eqref{eq:FIR-filter}};

\end{tikzpicture}}
\caption{
Block diagram of the Warped-Frequency Filter Bank (WFB) architecture. 
The input signal $x_k$ passes through a cascade of first-order all-pass filters $A(q^{-1})$, producing warped delay-line signals $z_{kj}$ with internal states $v_{kj}$. 
A time-domain FIR structure with weights $w_m$ generates the output $y_k$. 
In parallel, the warped signals $z_{kj}$ are provided to the Spectral Enhancement Model (SEM), which infers and synthesizes the time-domain coefficients $w_m$, enabling perceptually aligned, low-latency enhancement.}
\label{fig:front-end-architecture}
\end{figure*}
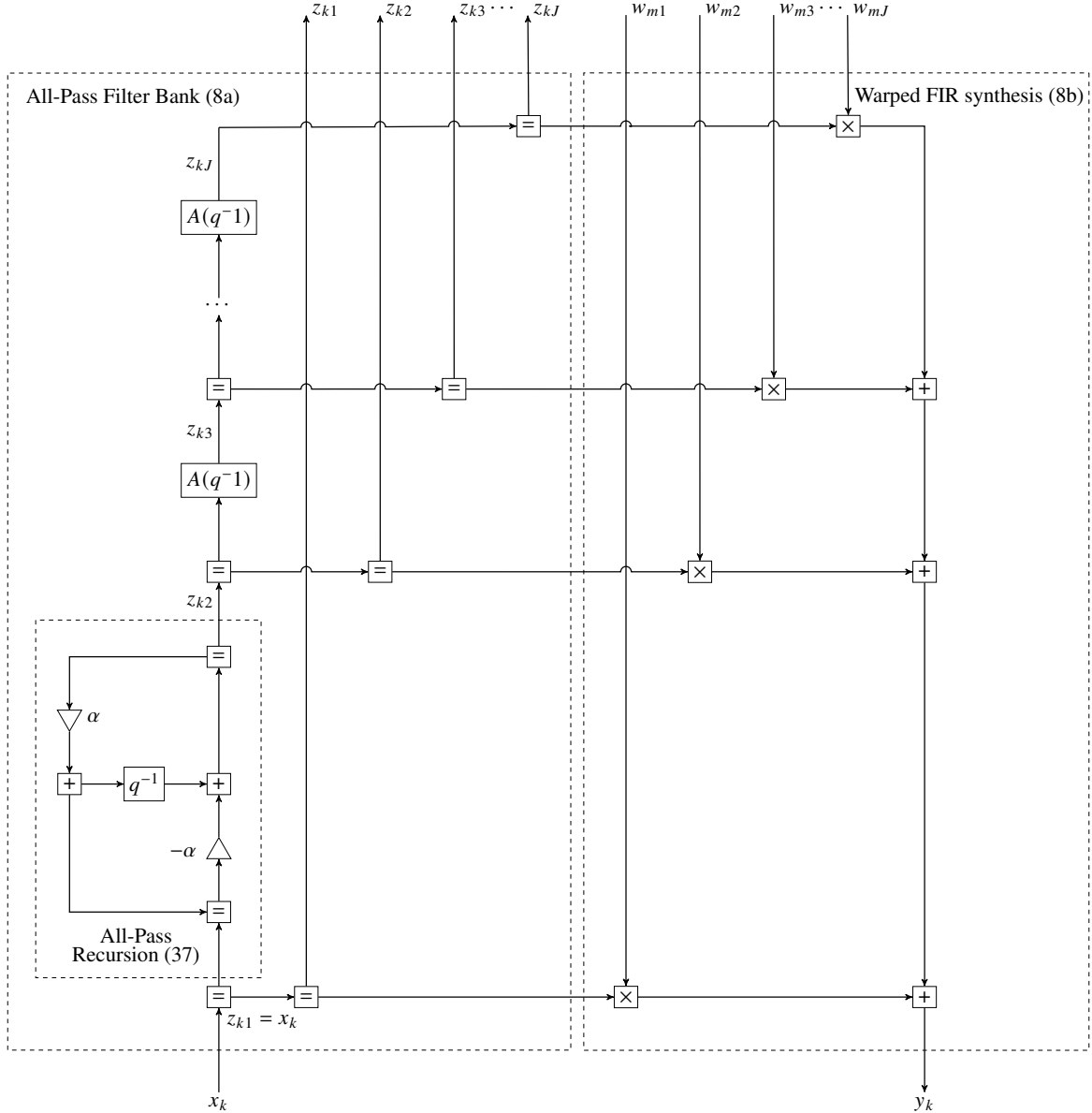

The WFB algorithm is summarized in Algorithm~\ref{alg:WFB} and illustrated in Figure~\ref{fig:front-end-architecture}.

\vspace{0.5em}
\begin{algorithm}[ht!]
\caption{Warped All-Pass Cascade}
\label{alg:WFB}
\begin{algorithmic}[1]
\Require 
Input sample $x_t$; 
previous states $\{v_{k-1,j}\}_{j=2}^{J}$; 
block index $m=\lfloor k/M \rfloor$; 
filter weights (from SEM) $w_m = [w_{m1},\ldots,w_{mJ}]$; 
warping coefficient $\alpha \in [0,1)$.
\Ensure 
Output sample $y_k$; 
updated states $\{v_{k,j}\}_{j=2}^{J}$.
\State $z_{k,1} \gets x_k$
\For{$j=2$ \textbf{to} $J$} \Comment{first-order all-pass cascade}
  \State $z_{k,j} \gets v_{k-1,j} - \alpha\, z_{k,j-1}$
  \State $v_{k,j} \gets z_{k,j-1} + \alpha\, z_{k,j}$
\EndFor
\State $y_k \gets \sum_{j=1}^{J} w_{m j}\, z_{k,j}$ \Comment{weights held constant for $k \in \{mM,\ldots,(m+1)M-1\}$}
\end{algorithmic}
\end{algorithm}
\vspace{0.5em}

\section{THE BAYESIAN LEAKY INTEGRATOR} \label{app:bayesian-leaky-integrator}

This appendix provides the derivation and implementation details of the Bayesian Leaky Integrator (BLI) used for signal-to-noise ratio (SNR) tracking SEM. The BLI corresponds to the speech and noise tracking component described in Subsection~\ref{sec:snr-tracking}. The following derivation shows how the classical leaky integrator can be expressed in discrete time and how its recursive form arises naturally as the mean update of a steady-state Kalman filter.

\subsection{THE DISCRETE-TIME LEAKY INTEGRATOR}
A continuous-time leaky integrator filter is defined by 
\begin{equation}\label{eq:cont-time-leaky-integrator}
    \tau \frac{\mathrm{d}y(t)}{\mathrm{d}t} + y(t) = x(t)\,,
\end{equation}
where $x(t)$ and $y(t)$ are the input and output signals respectively, and $\tau$ is the characteristic time constant for this filter. A leaky integrator is a first-order low-pass filter, ideal for monitoring the moving average of a signal $x(t)$.

We introduce a discrete time series $y_k \triangleq y(kT)$ where $k = 0,1,2,\ldots$ and $T$ is the sampling period. Let us approximate the continuous-time differential by
\begin{equation}
\frac{\mathrm{d}y(t)}{\mathrm{d}t} \approx \frac{y_k - y_{k-1}}{T} \,.
\end{equation}

Using the Backward Euler method  \cite[Sec. 20.4, pp. 68–70]{butcherNumericalMethodsOrdinary2016} , the discrete-time leaky integrator is then given by
\begin{align}
    \tau \frac{y_k-y_{k-1}}{T} + y_k = x_k \,, 
\end{align}
which can be rewritten as
\begin{align}\label{eq:leaky-integrator}
    y_k &= y_{k-1} + \gamma (x_k - y_{k-1})\,,
\end{align}
where 
\begin{equation}\label{eq:forgetting-factor_backward}
 \gamma = \frac{T}{T+\tau}   
\end{equation} 
is a dimensionless \emph{forgetting factor}.

\subsection{SETTING THE FORGETTING FACTOR}
When developing a signal processing system incorporating a leaky integrator, it is essential to choose the forgetting factor $\gamma$ to align with our goals. We will now establish a connection between the forgetting factor and the temporal characteristics of the leaky integrator. Consider setting $x(t)$ in \eqref{eq:cont-time-leaky-integrator} as a step function shifting from $0$ to $1$ at $t=0$. For the continuous system in \eqref{eq:cont-time-leaky-integrator}, the response to this step unit is given by

\begin{equation}
    \label{eq:unit_step_responce}
    y(t) = 1 - e^{-t/\tau}
\end{equation}

Thus, after a time $t=\tau$, the output $y(t)$ reaches the value $1-1/e \approx 0.63$ and approaches approximately $0.9$ after $2.3\tau$ seconds. This interval, indicated as $\tau_{90} = 2.3\tau$, is known as the $90\%$ settling time. The parameter $\tau_{90}$ holds meaningful significance: in the context of the leaky integrator \eqref{eq:cont-time-leaky-integrator}, it takes about $\tau_{90}$ for the response to catch up with the input.

We are interested in relating the forgetting factor $\gamma$ to $\tau_{90}$, since we want to expose $\tau_{90}$ as a parameter to the engineer. From equation \eqref{eq:forgetting-factor_backward} and the relationship $\tau_{90} = 2.3\tau$, we can substitute $\tau = \tau_{90}/2.3$ to obtain:

\begin{equation}
\label{eq:tau_to_lambda}
\gamma = \frac{T}{T+ (\tau_{90}/2.3)}\,,
\end{equation}
where $\tau_{90}$ is the parameter that is exposed to an end user.

\subsection{THE BAYESIAN LEAKY INTEGRATOR} 
Consider a probabilistic generative model given by
\begin{subequations}\label{eq:ssm-leaky-integrator}
 \begin{align}
    p(s_k|s_{k-1}) &= \mathcal{N}(s_k | s_{k-1}, \sigma^2_s) \\
    p(x_k|s_k) &= \mathcal{N}(x_k | s_k, \sigma^2_x)\,,
\end{align}   
\end{subequations}
where $x_k \in \mathbb{R}$ is an observed input signal, $s_k \in \mathbb{R}$ is a latent state signal, and $\sigma^2_s > 0$ and $\sigma^2_x > 0$ are given process and observation noise variances, respectively. Next, we describe how to choose the parameters $\sigma^2_s$ and $\sigma^2_x$ such that the posterior mean estimator of \eqref{eq:ssm-leaky-integrator} behaves as a leaky integrator filter in \eqref{eq:leaky-integrator}. We will refer to such a filter as a Bayesian Leaky Integrator (BLI). 

Because \eqref{eq:ssm-leaky-integrator} is a linear Gaussian state-space system, the filtering posterior $p(s_k| x_{1:k})=\mathcal{N}(s_k | \mu_k,\sigma_k^{2})$ can be updated recursively by the Kalman filter \cite[Theorem 4.2]{sarkkaBayesianFilteringSmoothing2013}, given by

\begin{subequations}\label{eq:kalman-filter}
\begin{align}
    \sigma_{k|k-1}^2 &= \sigma^2_{k-1} + \sigma^2_s \label{eq:predicted-variance} \\
    K_k &= \frac{\sigma_{k|k-1}^2}{\sigma_{k|k-1}^2 + \sigma^2_x} \label{eq:Kalman-gain} \\
    \mu_k &= \mu_{k-1} + K_k (x_k - \mu_{k-1}) \label{eq:kalman-mean-upd} \\
    \sigma_k^2 &= (1-K_k) \sigma_{k|k-1}^2 \,. \label{eq:posterior-variance}
\end{align}    
\end{subequations}

The update equation \eqref{eq:kalman-mean-upd} coincides with the classical leaky–integrator recursion \eqref{eq:leaky-integrator} when the Kalman gain $K_k$ equals the forgetting factor $\lambda$. However, the Kalman formulation additionally tracks the posterior variance $\sigma_k^{2}$, providing a principled measure of the estimator's uncertainty.

To relate the Kalman gain in~\eqref{eq:Kalman-gain} to the forgetting factor $\lambda$ in \eqref{eq:forgetting-factor_backward}, we examine the steady state. At steady state, the posterior variance converges to a constant value, i.e., $\sigma_k^2 = \sigma_{k-1}^2 = \sigma^2_{\infty}$ for sufficiently large $k$.

Under steady-state conditions, the Kalman filter recursions \eqref{eq:kalman-filter} reduce to

\begin{subequations}\label{eq:kalman-filter-steady-state}
\begin{align}
    \sigma_{\infty|k-1}^2 &= \sigma^2_{\infty} + \sigma^2_s \label{eq:ss-predicted-variance} \\
    K_{\infty} &= \frac{\sigma_{\infty|k-1}^2}{\sigma_{\infty|k-1}^2 + \sigma^2_x}  \label{eq:k_infty} \\
    \mu_k &= \mu_{k-1} + K_{\infty} (x_k - \mu_{k-1})  \label{eq:ss-mean-update} \\
    \sigma^2_{\infty} &= (1-K_{\infty}) \sigma_{\infty|k-1}^2 \,.\label{eq:sigma_infty}
\end{align}    
\end{subequations}

From \eqref{eq:sigma_infty} and \eqref{eq:ss-predicted-variance}, we have
\begin{equation}
\sigma^2_{\infty} = (1-K_{\infty})(\sigma^2_{\infty} + \sigma^2_s) \,.
\end{equation}

Solving for $\sigma^2_{\infty}$,

\begin{equation}
\sigma^2_{\infty} = \frac{(1-K_{\infty})\sigma^2_s}{K_{\infty}} \,.
\end{equation}

Substituting this into \eqref{eq:k_infty} and simplifying yields

\begin{equation}
\label{eq:sigma_s-constraint}
    \frac{\sigma^2_s}{\sigma^2_x} = \frac{K_{\infty}^2}{1-K_{\infty}} \,.
\end{equation}

To satisfy the constraint in \eqref{eq:sigma_s-constraint}, we set the parameters in our BLI filter as
\begin{subequations}\label{eq:sigma_s-constraint_bli}
   \begin{align}
\sigma^2_x &= 1 \\
\sigma^2_s &= \frac{\lambda^{2}}{1-\lambda}\,.
\end{align} 
\end{subequations}

Following the above discussion, we define the BLI as follows: 
\begin{definition}[Bayesian Leaky Integrator]
For a given forgetting factor $\lambda \in (0,1)$, the Bayesian Leaky Integrator (BLI) is defined as the Kalman filter for the state-space model \eqref{eq:ssm-leaky-integrator} with noise variances satisfying \eqref{eq:sigma_s-constraint_bli}.
\end{definition}
\section{THE WIENER GAIN}
\label{app:wiener}

The Wiener filter is a statistical estimator that minimizes mean-square error (MSE) using knowledge of signal and noise statistics. The gain depends on the ratio of speech and noise.

We consider an additive mixture model in the short-time Fourier transform (STFT) domain. For each time frame $m$ and frequency bin $j$,
\begin{equation}
\label{eq:lin-mixture}
X_{mj} = S_{mj} + N_{mj},
\end{equation}
where $X_{mj}$ is the observed (noisy) STFT coefficient, $S_{mj}$ denotes the speech coefficient, and $N_{mj}$ denotes the background-noise coefficient. For notational convenience, we drop the indices $(m,j)$ in the derivations that follow.

We assume the speech and noise coefficients are zero-mean, independent, circularly symmetric (proper) complex Gaussians with variances $\sigma_S^2$ and $\sigma_N^2$, respectively,

\begin{equation}
S \sim \mathcal{CN}(0,\sigma_S^2), \qquad
N \sim \mathcal{CN}(0,\sigma_N^2), 
\end{equation}

Here $\mathcal{CN}(0,\sigma^2)$ denotes a zero-mean, circular complex Gaussian with variance $\sigma^2$ and density
\begin{equation}
p(z) = \frac{1}{\pi\sigma^2}\exp\!\Big(-\frac{|z|^2}{\sigma^2}\Big).    
\end{equation}

The joint generative model is given by
\begin{equation}
p(X,S,N) = p(S)\,p(N)\,\delta\big(X-(S+N)\big).
\end{equation}
Marginalizing $N$ yields

\begin{align}
p(S,X)&= \int p(S)p(N)\,\delta(X-(S+N))\mathrm{d}N\\
&= p(S)p(X-S),\notag
\end{align}
using $\delta(X-(S+N))=\delta(N-(X-S))$ and the sifting property. Hence the likelihood is the shifted noise density,

\begin{align}
p(X\,|\,S) &= \frac{p(S,X)}{p(S)}\\ &= p(X-S)\notag\\
&= \frac{1}{\pi\sigma_N^2}\exp\Big(-\frac{|X-S|^2}{\sigma_N^2}\Big).\notag
\end{align}

By Bayes’ rule, the posterior is

\begin{equation}
p(S\,|\,X) = \frac{p(X\,|\,S)\,p(S)}{p(X)}
\propto
\exp\!\left(-\frac{|X-S|^2}{\sigma_N^2}-\frac{|S|^2}{\sigma_S^2}\right).
\end{equation}

Expanding $|X-S|^2=|X|^2-2\,\mathbb{R}\{X^*S\}+|S|^2$ and collecting terms gives

\begin{equation}
\frac{|X-S|^2}{\sigma_N^2}+\frac{|S|^2}{\sigma_S^2}
= \frac{|X|^2}{\sigma_N^2}
-\frac{2\,\mathbb{R}\{X^*S\}}{\sigma_N^2}
+\Big(\frac{1}{\sigma_N^2}+\frac{1}{\sigma_S^2}\Big)|S|^2.
\end{equation}

Let the precision be  $\Lambda \triangleq \tfrac{1}{\sigma_N^2}+\tfrac{1}{\sigma_S^2}>0$. Completing the square via
\begin{equation}
\Lambda\,|S-\mu|^2 = \Lambda|S|^2 - 2\,\mathbb{R}\{(\Lambda\mu)^*S\} + \Lambda|\mu|^2,
\end{equation}
and matching the cross term $-\tfrac{2\,\mathbb{R}\{X^*S\}}{\sigma_N^2}$ yields $(\Lambda\mu)^*=\tfrac{X^*}{\sigma_N^2}$, hence
\begin{equation}
\mu = \Lambda^{-1}\frac{X}{\sigma_N^2}
= \frac{\sigma_S^2}{\sigma_S^2+\sigma_N^2}\,X,
\qquad
\Sigma = \Lambda^{-1} = \frac{\sigma_S^2\,\sigma_N^2}{\sigma_S^2+\sigma_N^2}.
\end{equation}

Therefore $p(S\,|\,X)=\mathcal{CN}\!\big(\mu,\,\Sigma\big)$, the MMSE estimator equals the posterior mean,
\begin{equation}
\hat S = \mathbb{E}[S\,|\,X] = \mu = \frac{\sigma_S^2}{\sigma_S^2+\sigma_N^2}\,X.
\end{equation}

Under zero means the variances equal the per-bin powers,
\begin{equation}
\sigma_S^2=\mathbb{E}[|S|^2]\triangleq P^{(s)}, \qquad
\sigma_N^2=\mathbb{E}[|N|^2]\triangleq P^{(n)}.
\end{equation}
The spectral Bayesian Wiener filter is
\begin{equation}
\hat S = \mathbb{E}[S\,|\,X]
= \underbrace{\frac{P^{(s)}}{P^{(s)}+P^{(n)}}}_{\tilde{w}}\,X,
\end{equation}
with Wiener gain $\tilde{w}\in[0,1]$ determined by the speech–noise power ratio. The gain $\tilde{w}$ can also be expressed in terms of logarithmic power densities. Let $\gamma\triangleq P^{(s)}/P^{(n)}$ and 
$\xi\triangleq\ln\gamma$. Then
\begin{equation}
\tilde{w}=\frac{\gamma}{1+\gamma}
=\frac{1}{1+\mathrm{e}^{-\xi}} =\sigma(\xi),
\end{equation}
where $\sigma(\xi)$ is the logistic function.

\section{VARIATIONAL DERIVATIONS FOR THE LOGIT NODE}
\label{app:logit}
The logit node relates a continuous variable $x\in\mathbb{R}$ to a binary variable $y\in\{0,1\}$ as
\begin{equation}
\label{eq:logit-lik}
f(y,x)\;=\;\mathrm{Ber}\!\left(y\,|\,\sigma(x)\right),
\qquad
\sigma(x)=\frac{1}{1+\exp(-x)}.
\end{equation}
We work within the mean-field variational message passing (VMP) framework.  Throughout the appendix, for any random quantity $z$ with variational marginal $q(z)$, we denote expectations by an overbar:
\begin{equation}
\bar{z}\ \triangleq\ \mathbb{E}_{q(z)}[z],
\qquad
\overline{z^k}\ \triangleq\ \mathbb{E}_{q(z)}[z^k],
\qquad
\Var[z]\ \triangleq\ \overline{z^2}-\bar{z}^{\,2}.
\end{equation}

To handle the nonlinearity of the logit likelihood, we approximate it using the Jaakkola--Jordan variational bound~\cite{bishopPatternRecognitionMachine2006}, expressed in~\eqref{eq:logit-lik}.

\begin{align}
\label{eq:boundsig}
    \sigma(x) &\;\ge\; \sigma(\zeta)\,
\exp\!\left(
    \tfrac{x - \zeta}{2}
    - \lambda(\zeta)\bigl(x^{2} - \zeta^{2}\bigr)
\right),
\quad \text{where} \\
\lambda(\zeta) &= \frac{1}{2\zeta}\left(\sigma(\zeta) - \tfrac{1}{2}\right).
\end{align}

\subsection{FACTORIZATION}
We impose the mean-field factorization
\begin{equation}
\label{eq:mf-factorization}
q(y,x,\zeta)=q(y)\,q(x)\,q(\zeta),
\end{equation}
and use the standard VMP local update (suppressing constants $C$ that do not depend on the message argument)
\begin{equation}
\label{eq:vmp-update}
\vec{\nu}_j(s_j)\ \propto\
\exp\!\Big(\mathbb{E}_{q(s_{a\setminus j})}\big[\ln f_a(s_a)\big]\Big).
\end{equation}

\subsection{VARIATIONAL MESSAGES}
\subsubsection{Forward message to y}
Using~\eqref{eq:vmp-update} and~\eqref{eq:boundsig},

\begin{align}
\ln \vec{v}(y) 
&= \mathbb{E}_{x,\zeta}\!\bigl[\ln f(y,x;\zeta)\bigr] + C \\[6pt]
&\approx \mathbb{E}_{x,\zeta}\!\left[ xy + \ln \sigma(\zeta) - \frac{x+\zeta}{2}
     - \lambda(\zeta)\bigl(x^2 - \zeta^2\bigr) \right] + C \notag\\[6pt]
&= \bar{x}y + C. \notag
\end{align}
\begin{align}
\Rightarrow \; \vec{v}(y) &\propto \exp(\bar{x}y)\\
&= \exp(\bar{x})^y \, 1^{\,1-y} \notag\\[6pt]
&\propto 
   \left(\frac{\exp(\bar{x})}{1+\exp(\bar{x})}\right)^{y}
   \left(\frac{1}{1+\exp(\bar{x})}\right)^{1-y} \notag\\[6pt]
&= \sigma(\bar{x})^y \,\bigl(1-\sigma(\bar{x})\bigr)^{1-y} \notag\\
&= \operatorname{Ber}\!\bigl(y \,|\,\sigma(\bar{x})\bigr).\notag
\end{align}

\subsubsection{Backward message to x}
Similarly,
\begin{align}
\ln \cev{v}(x) 
&= \mathbb{E}_{y,\zeta}\!\bigl[\ln f(y,x;\zeta)\bigr] + C \\[6pt]
&\approx \mathbb{E}_{y,\zeta}\!\left[ xy + \ln \sigma(\zeta) 
   - \frac{x+\zeta}{2} - \lambda(\zeta)\bigl(x^2 - \zeta^2\bigr) \right] + C \notag\\[6pt]
&= \bar{y}x - \tfrac{x}{2} - \lambda(\hat{\zeta})x^2 + C \notag\\[6pt]
&= -\lambda(\hat{\zeta})\left( x - \frac{\bar{y}-\tfrac{1}{2}}{2\lambda(\hat{\zeta})} \right)^{\!2} + C.\notag
\end{align}

\begin{equation}
\Rightarrow\; \cev{v}(x) \;\propto\;
\mathcal{N}\!\left(x \,\Bigg|\,
\left(\bar{y} - \tfrac{1}{2}\right)(2\lambda(\hat{\zeta}))^{-1},
\ (2\lambda(\hat{\zeta}))^{-1}\right).
\end{equation}

In practice, we set $q(\zeta)$ to a point mass at its maximizer $\hat{\zeta}$ (derived below).

\subsubsection{Update for the variational parameter \texorpdfstring{$\zeta$}{zeta}}
\begin{align}
\ln \cev{v}(\zeta) 
&= \mathbb{E}_{y,x}\!\bigl[\ln f(y,x;\zeta)\bigr] + C \\[6pt]
&\approx \mathbb{E}_{y,x}\!\left[ xy + \ln \sigma(\zeta) 
   - \frac{x+\zeta}{2} - \lambda(\zeta)\bigl(x^2 - \zeta^2\bigr) \right] + C .\notag
\end{align}

\noindent
We absorb the terms independent of $\zeta$ into $C$, and define
\begin{equation}
g(\zeta) = \ln \sigma(\zeta) - \frac{\zeta}{2} 
- \lambda(\zeta)\!\left(\,\overline{x^2} - \zeta^2\right).
\end{equation}

\medskip
We then require the maximum:
\begin{align}
\frac{d g}{d\zeta} 
&= \frac{\sigma'(\zeta)}{\sigma(\zeta)} - \tfrac{1}{2}
   - \overline{x^2}\,\lambda'(\zeta) 
   + \lambda'(\zeta)\zeta^2 + 2\lambda(\zeta)\zeta \\[6pt]
&= \tfrac{1}{2} - \sigma(\zeta) + \lambda'(\zeta)\bigl(\zeta^2 - \overline{x^2}\bigr) 
   + 2\lambda(\zeta)\zeta \notag\\[6pt]
&= \lambda'(\zeta)\bigl(\zeta^2 - \overline{x^2}\bigr).\notag
\end{align}

\noindent
Because $\lambda(\zeta)$ is monotone, the maximum is unaffected by $\lambda'(\zeta)$.  
Setting the derivative to zero yields
\begin{equation}
    \zeta^2 = \overline{x^2}
\quad\;\;\Rightarrow\;\;\quad
\hat{\zeta} = \sqrt{\,\overline{x}^2 + \Var[x]}\,.
\end{equation}

\subsubsection{Average energy}
\begin{align}
\mathcal{U}&[f(y,x;\zeta)] \\
&= -\,\mathbb{E}_{y,x,\zeta}\!\bigl[\ln f(y,x;\zeta)\bigr] \notag\\[6pt]
&= -\,\mathbb{E}_{y,x,\zeta}\!\left[
    xy + \ln \sigma(\zeta) - \frac{x+\zeta}{2}
    - \lambda(\zeta)\bigl(x^{2}-\zeta^{2}\bigr)
  \right] \notag\\[6pt]
&= - \overline{xy} - \ln \sigma(\hat{\zeta})
   + \frac{\bar{x}+\hat{\zeta}}{2}
   + \lambda(\hat{\zeta})\Bigl(\overline{x}^{2} + \Var[x] - \hat{\zeta}^{2}\Bigr).\notag
\end{align}
The point-mass update $\zeta=\hat{\zeta}$   simplifies during inference.

\section{RESULTS}
\label{app:results}

Evaluation results across multiple acoustic environments 
(\textsc{Bus}, \textsc{Cafe}, \textsc{Living}, \textsc{Public Square}, and \textsc{Office}) and input 
SNR conditions (2.5, 7.5, 12.5, 17.5~dB). We present standard perceptual quality and intelligibility 
metrics, namely PESQ, SIG, BAK, and OVL, to compare the Unprocessed speech,  to the proposed SEM module. The following tables provide a detailed breakdown of scores 
for each noise type and SNR level.

\begin{table*}[hb!]
\centering
\caption{Objective scores by SNR and noise location.}
\label{tab:metrics-quadrants}

\subfloat[PESQ]{%
\begin{minipage}{0.5\textwidth}\centering\scriptsize
\begin{tabular}{l|rrrr}
\hline
\multicolumn{5}{c}{BUS} \\ \hline
System & 2.5 dB & 7.5 dB & 12.5 dB & 17.5 dB \\ \hline
Unprocessed & 1.75 & 2.26 & 2.69 & 3.19 \\
SEM (ours) & 2.01 & 2.58 & 2.88 & 3.33 \\
\hline
\multicolumn{5}{c}{CAFE} \\ \hline
System & 2.5 dB & 7.5 dB & 12.5 dB & 17.5 dB \\ \hline
Unprocessed & 1.15 & 1.33 & 1.54 & 1.92 \\
SEM (ours) & 1.20 & 1.44 & 1.74 & 2.24 \\
\hline
\multicolumn{5}{c}{LIVING} \\ \hline
System & 2.5 dB & 7.5 dB & 12.5 dB & 17.5 dB \\ \hline
Unprocessed & 1.18 & 1.40 & 1.67 & 2.20 \\
SEM (ours) & 1.28 & 1.58 & 1.92 & 2.36 \\
\hline
\multicolumn{5}{c}{PSQUARE} \\ \hline
System & 2.5 dB & 7.5 dB & 12.5 dB & 17.5 dB \\ \hline
Unprocessed & 1.24 & 1.45 & 1.82 & 2.35 \\
SEM (ours) & 1.40 & 1.68 & 2.00 & 2.63 \\
\hline
\multicolumn{5}{c}{OFFICE} \\ \hline
System & 2.5 dB & 7.5 dB & 12.5 dB & 17.5 dB \\ \hline
Unprocessed & 1.76 & 2.29 & 2.78 & 3.22 \\
SEM (ours) & 2.14 & 2.65 & 2.94 & 3.39 \\
\hline
\end{tabular}
\end{minipage}}
\hfill
\subfloat[SIG]{%
\begin{minipage}{0.5\textwidth}\centering\scriptsize
\begin{tabular}{l|rrrr}
\hline
\multicolumn{5}{c}{BUS} \\ \hline
System & 2.5 dB & 7.5 dB & 12.5 dB & 17.5 dB \\ \hline
Unprocessed & 3.31 & 3.51 & 3.52 & 3.49 \\
SEM (ours) & 3.33 & 3.39 & 3.38 & 3.39 \\
\hline
\multicolumn{5}{c}{CAFE} \\ \hline
System & 2.5 dB & 7.5 dB & 12.5 dB & 17.5 dB \\ \hline
Unprocessed & 2.25 & 3.07 & 3.52 & 3.58 \\
SEM (ours) & 2.56 & 3.26 & 3.38 & 3.40 \\
\hline
\multicolumn{5}{c}{LIVING} \\ \hline
System & 2.5 dB & 7.5 dB & 12.5 dB & 17.5 dB \\ \hline
Unprocessed & 2.40 & 3.32 & 3.39 & 3.54 \\
SEM (ours) & 2.85 & 3.28 & 3.37 & 3.34 \\
\hline
\multicolumn{5}{c}{PSQUARE} \\ \hline
System & 2.5 dB & 7.5 dB & 12.5 dB & 17.5 dB \\ \hline
Unprocessed & 3.21 & 3.46 & 3.53 & 3.56 \\
SEM (ours) & 3.37 & 3.40 & 3.42 & 3.40 \\
\hline
\multicolumn{5}{c}{OFFICE} \\ \hline
System & 2.5 dB & 7.5 dB & 12.5 dB & 17.5 dB \\ \hline
Unprocessed & 3.49 & 3.46 & 3.48 & 3.45 \\
SEM (ours) & 3.31 & 3.30 & 3.38 & 3.39 \\
\hline
\end{tabular}
\end{minipage}}

\subfloat[BAK]{%
\begin{minipage}{0.5\textwidth}\centering\scriptsize
\begin{tabular}{l|rrrr}
\hline
\multicolumn{5}{c}{BUS} \\ \hline
System & 2.5 dB & 7.5 dB & 12.5 dB & 17.5 dB \\ \hline
Unprocessed & 3.07 & 3.50 & 3.68 & 3.77 \\
SEM (ours) & 3.56 & 3.80 & 3.85 & 3.91 \\
\hline
\multicolumn{5}{c}{CAFE} \\ \hline
System & 2.5 dB & 7.5 dB & 12.5 dB & 17.5 dB \\ \hline
Unprocessed & 1.66 & 2.26 & 2.87 & 3.30 \\
SEM (ours) & 2.17 & 2.98 & 3.29 & 3.57 \\
\hline
\multicolumn{5}{c}{LIVING} \\ \hline
System & 2.5 dB & 7.5 dB & 12.5 dB & 17.5 dB \\ \hline
Unprocessed & 1.76 & 2.73 & 2.99 & 3.38 \\
SEM (ours) & 2.46 & 3.21 & 3.48 & 3.58 \\
\hline
\multicolumn{5}{c}{PSQUARE} \\ \hline
System & 2.5 dB & 7.5 dB & 12.5 dB & 17.5 dB \\ \hline
Unprocessed & 2.50 & 2.92 & 3.27 & 3.52 \\
SEM (ours) & 3.17 & 3.41 & 3.62 & 3.68 \\
\hline
\multicolumn{5}{c}{OFFICE} \\ \hline
System & 2.5 dB & 7.5 dB & 12.5 dB & 17.5 dB \\ \hline
Unprocessed & 3.40 & 3.69 & 3.85 & 3.98 \\
SEM (ours) & 3.77 & 3.92 & 3.99 & 4.02 \\
\hline
\end{tabular}
\end{minipage}}
\hfill
\subfloat[OVL]{%
\begin{minipage}{0.5\textwidth}\centering\scriptsize
\begin{tabular}{l|rrrr}
\hline
\multicolumn{5}{c}{BUS} \\ \hline
System & 2.5 dB & 7.5 dB & 12.5 dB & 17.5 dB \\ \hline
Unprocessed & 2.62 & 2.93 & 3.03 & 3.07 \\
SEM (ours) & 2.82 & 2.98 & 3.01 & 3.05 \\
\hline
\multicolumn{5}{c}{CAFE} \\ \hline
System & 2.5 dB & 7.5 dB & 12.5 dB & 17.5 dB \\ \hline
Unprocessed & 1.62 & 2.18 & 2.64 & 2.90 \\
SEM (ours) & 1.92 & 2.52 & 2.73 & 2.90 \\
\hline
\multicolumn{5}{c}{LIVING} \\ \hline
System & 2.5 dB & 7.5 dB & 12.5 dB & 17.5 dB \\ \hline
Unprocessed & 1.69 & 2.51 & 2.66 & 2.91 \\
SEM (ours) & 2.14 & 2.64 & 2.80 & 2.84 \\
\hline
\multicolumn{5}{c}{PSQUARE} \\ \hline
System & 2.5 dB & 7.5 dB & 12.5 dB & 17.5 dB \\ \hline
Unprocessed & 2.33 & 2.65 & 2.87 & 3.01 \\
SEM (ours) & 2.66 & 2.80 & 2.94 & 2.94 \\
\hline
\multicolumn{5}{c}{OFFICE} \\ \hline
System & 2.5 dB & 7.5 dB & 12.5 dB & 17.5 dB \\ \hline
Unprocessed & 2.88 & 2.98 & 3.09 & 3.14 \\
SEM (ours) & 2.90 & 2.96 & 3.07 & 3.10 \\
\hline
\end{tabular}
\end{minipage}}

\end{table*}

\end{document}